\documentclass{article}
\usepackage{arxiv}

\usepackage[utf8]{inputenc} 
\usepackage[T1]{fontenc}    
\usepackage{hyperref}       
\usepackage{url}            
\usepackage{booktabs}       
\usepackage{amsfonts}       
\usepackage{nicefrac}       
\usepackage{microtype}      
\usepackage[numbers]{natbib}
\usepackage{doi}

\usepackage{graphicx}  
\usepackage{dcolumn}
\usepackage{caption}  
\usepackage{subcaption}  

\usepackage{multirow}  
\usepackage{booktabs}

\usepackage{bm}
\usepackage{amsmath}
\usepackage{amssymb}

\usepackage{xcolor}  
\usepackage{color, colortbl}  
\definecolor{green}{RGB}{0,100,0} 
\definecolor{red}{RGB}{200,0,0} 
\usepackage{hyperref}
\usepackage{cleveref}       

\title{A Hybrid Generative Reduced-Order Model for the Minimal Flow Unit} 

\date{}

\newif\ifuniqueAffiliation

\ifuniqueAffiliation 
\author{ \href{https://orcid.org/0009-0003-0164-5601}{\includegraphics[scale=0.06]{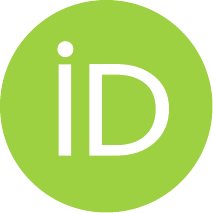}\hspace{1mm}Niccolò Tonioni}\thanks{https://niccolotonioni.github.io/} \\
	Pprime Institute,\\
	CNRS, Université de Poitiers, ISAE-ENSMA,\\
    Chasseneuil-du-Poitou, 86360, France.\\
	\texttt{niccolo.tonioni@univ-poitiers.fr} \\
	\And
	Lionel Agostini \\
	Pprime Institute,\\
	CNRS, Université de Poitiers, ISAE-ENSMA,\\
    Chasseneuil-du-Poitou, 86360, France.\\
    \And
    Marcial Sanchis-Agudo, \\
    FLOW, Engineering Mechanics,\\
    KTH Royal Institute of Technology,\\
    Stockholm, SE-100 44, Sweden.
	\And
	Mohammad Umair, \\
	Laboratoire des Écoulements Géophysiques et Industriels, \\
    BP 53, 38041, Grenoble CEDEX 09, France\\
    \And 
    FLOW, Engineering Mechanics,\\
    KTH Royal Institute of Technology,\\
    Stockholm, SE-100 44, Sweden.
	\And
	Franck Kerhervé \\
	Pprime Institute,\\
	CNRS, Université de Poitiers, ISAE-ENSMA,\\
    Chasseneuil-du-Poitou, 86360, France.\\
	\And
	Laurent Cordier \\
	Pprime Institute,\\
	CNRS, Université de Poitiers, ISAE-ENSMA,\\
    Chasseneuil-du-Poitou, 86360, France.\\
	\And
	Ricardo Vinuesa\\
	Department of Aerospace Engineering,\\
	University of Michigan,\\ 
    Ann Arbor, MI 48109, United States\\
    and\\
    FLOW, Engineering Mechanics,\\
    KTH Royal Institute of Technology,\\
    Stockholm, SE-100 44, Sweden.
}
\else
\usepackage{authblk}

\newbox{\orcid}\sbox{\orcid}{\includegraphics[scale=0.06]{orcid.pdf}} 
\author[1]{%
	\href{https://orcid.org/0009-0003-0164-5601}{\usebox{\orcid}\hspace{1mm}Niccolò Tonioni\thanks{\href{https://niccolotonioni.github.io}{niccolotonioni.github.io},\hspace{1mm}\texttt{niccolo.tonioni@univ-poitiers.fr}}}%
}
\author[1]{Lionel Agostini\thanks{\texttt{lionel.agostini@cnrs.fr}}}
\author[2]{Marcial Sanchis-Agudo}
\author[2]{Mohammad Umair}
\author[1]{Franck Kerhervé}
\author[1]{Laurent Cordier}
\author[4,2]{Ricardo Vinuesa\thanks{\href{https://www.vinuesalab.com/}{vinuesalab},\hspace{1mm}\texttt{rvinuesa@umich.edu}}}

\affil[1]{Pprime Institute, CNRS, Université de Poitiers, ISAE-ENSMA,Chasseneuil-du-Poitou, 86360, France}
\affil[2]{FLOW, Engineering Mechanics, KTH Royal Institute of Technology, Stockholm, SE-100 44, Sweden}
\affil[3]{Laboratoire des Écoulements Géophysiques et Industriels, BP 53, 38041, Grenoble CEDEX 09, France}
\affil[4]{Department of Aerospace Engineering,University of Michigan, Ann Arbor, MI 48109, United States}
\fi

\renewcommand{\headeright}{A Preprint}
\renewcommand{\shorttitle}{A Hybrid Generative Reduced-Order Model}

\hypersetup{
pdftitle={A Hybrid Generative Reduced-Order Model for the Minimal Flow Unit},
pdfsubject={reduced-order-modeling},
pdfauthor={Niccolò Tonioni, Lionel Agostini, Marcial Sanchis-Agudo, Mohammad Umair, Franck Kerhervé, Laurent Cordier, Ricardo Vinuesa},
pdfkeywords={Reduced-Order Models, Minimal Flow Unit, Near-Wall Turbulence, Turbulent Flows, Variational Autoencoders, Generative Adversarial Networks, Transformers},
}

\begin{document}
\maketitle

\begin{abstract}
    A data-driven reduced-order modelling framework is proposed for wall-bounded turbulent flows to forecast the intermittent near-wall dynamics over extended time horizons from sparse sensor measurements. The approach combines a $\beta$-VAE-GAN, which compresses high-dimensional flow fields into a low-dimensional latent space, with a sensor-conditioned Transformer that forecasts the evolution of the latent variables. The temporal module employs Easy Attention, a static time-mixing operator that replaces the learnable query-key mechanism of standard self-attention at reduced computational cost, combined with an adapted AdaLN-Zero modulation mechanism for sensor-based conditioning. Evaluated on the Minimal Flow Unit ($Re_\tau = 200$) at $y^+ = 14$, the compression stage recovers $87\%$ of the turbulent kinetic energy within a four-dimensional latent space, exceeding the standard $\beta$-VAE baseline by more than $10\%$. The latent dimensions autonomously encode the characteristic timescales of the flow, with specific coordinates capturing the low-frequency signature of the near-wall regeneration cycle ($T^+ \approx 1724$), establishing the physical interpretability of the learnt representation. The sensor-conditioned Transformer maintains accurate forecasts over $17{,}288\,t^+$ from an initialisation window of only $128\,t^+$, whilst end-to-end inference reconstructs $82\%$ of the turbulent kinetic energy. The principal limitation is the attenuation of rare, extreme-amplitude events, a consequence of the encoder prioritising the most statistically recurrent flow states within the low-dimensional bottleneck. Nevertheless, the framework accurately reproduces the alternating active and quiescent phases of the regeneration cycle, demonstrating its suitability as a surrogate model for the intermittent dynamics of wall-bounded turbulence.
\end{abstract}


\label{sec:01_intro}
Wall-bounded turbulent flows play a central role in applications ranging from energy and aerospace technologies to atmospheric and environmental sciences. The economic and ecological implications of these flows are profound: it is estimated that approximately 15\% of total global energy consumption is dissipated by turbulence in the boundary layers of transport vehicles \cite{jimenez2013near}. Despite this significance, accurate prediction and control remain elusive \cite{agostini2014influence}.

This modeling complexity arises from the presence of a solid surface. Whilst turbulence far from boundaries exhibits the self-similar behavior and small-scale universality described by Kolmogorov \cite{kolmogorov1941local}, the presence of a wall fundamentally breaks these symmetries, introducing strong anisotropy and inhomogeneity \cite{Biferale2003}.  In the resulting wall-induced shear layer, two overlapping regions can be distinguished: an inner region, immediately adjacent to the wall, where viscous effects dominate, and an outer region, governed primarily by inertial forces. In both regions, the flow self-organizes into three-dimensional structures with a broad range of characteristic spatial arrangements and temporal dynamics \cite{robinson1991coherent}. The resulting mosaic of multiscale interactions gives rise to non-linear coupling and chaotic dynamics \cite{mathis2009large, marusic2010high, marusic2010predictive}, which constitute a central obstacle for both experimental characterisation and numerical modelling.

To decipher this mosaic, Reduced-Order Modelling (ROM) offers a viable path by constructing simplified surrogates that capture the dominant flow dynamics. These models rely on the conjecture of Ruelle and Takens \cite{ruelle1971nature} that the dissipative nature of the Navier-Stokes equations (NSE) confines the system trajectory to a lower-dimensional strange attractor. The construction of such low-dimensional representations generally follows one of two paradigms: an \textit{inductive} approach, where elementary structures are proposed based on physical insight, or a \textit{deductive} approach, where dominant patterns are extracted directly from data \cite{panton1999self}.

The inductive approach has been particularly fruitful in studies of plane Couette flow, where the relatively simple geometry and forcing have enabled researchers to identify and model fundamental regeneration mechanisms \cite{hamilton1995regeneration}. For example, Waleffe and collaborators derived low-order dynamical systems by projecting the NSE onto \textit{a priori} defined functions, such as streaks and rolls \cite{waleffe1997self, waleffe1998streamwise}. Whilst these models reproduce the self-sustaining cycle, they do not fully capture the stochastic intermittency of bursting events. Extensions by Moehlis \textit{et al.} \cite{moehlis2004low} addressed this by incorporating additional modes to capture subcritical transitions. However, generalizing these models to pressure-driven channel flows is non-trivial. The sinusoidal forcing and simplified boundary conditions of the Couette configuration do not translate directly to channel flow, where the mean shear itself can sustain streak instabilities through transient growth \cite{schoppa1997genesis}.

Consequently, research in channel flow has favoured the deductive, data-driven paradigm. The foundational method in this approach is Proper Orthogonal Decomposition (POD) \cite{Lumley1967Structure}, which defines turbulent structures as the orthogonal spatial modes that capture the maximum variance of velocity fluctuations around the mean flow. These modes are identified as the eigenfunctions of the Karhunen-Lo\`{e}ve decomposition of velocity field observations \cite{loeve1955probability, cordier2008proper}, and their temporal dynamics are subsequently recovered through Galerkin projection of the NSE onto the retained mode basis \cite{Holmes_Lumley_Berkooz_Rowley_2012, sirovich1987turbulence, cordier2008two}. Early applications to channel flows \cite{aubry1988dynamics, podvin1998low} demonstrated that severely truncated POD reduced systems can exhibit intermittent dynamics, although the physical fidelity of these truncations remains debated \cite{zhou1992coherence}.

A more fundamental limitation of linear decomposition methods, such as POD, when applied to turbulence, is the slow decay of the Kolmogorov $n$-width \cite{ohlberger2015reduced}. In advection-dominated systems with strong non-linearities, the singular values decay slowly, requiring a prohibitively large number of linear modes to approximate the solution manifold accurately \cite{greif2019decay}. Indeed, linear methods must represent advecting structures by superposing many stationary basis functions. By contrast, non-linear representations can intrinsically encode translational invariance, tracking these coherent structures more efficiently \cite{ahmed2020breaking, peherstorfer2022breaking}. This theoretical advantage, combined with the increasing availability of high-fidelity simulation databases, has driven the widespread adoption of non-linear dimensionality reduction techniques based on autoencoders (AEs).

AEs are machine learning architectures that address these limitations by learning a non-linear mapping from high-dimensional flow fields to a compact latent manifold \cite{goodfellow2016deep}.  In this framework, the definition of a turbulent `structure' is extended from a spatial mode to an information-rich pattern \cite{shannon1959coding} that maximises reconstructive fidelity. Unlike POD, which provides explicit, orthogonal spatial modes, AEs produce latent variables that act as coordinates on a curved manifold \cite{magri2022interpretability}. Although the relationship between latent coordinates and physical flow features remains implicit, the superior compression efficiency has driven the adoption of AEs across a broad spectrum of turbulence research, from state prediction \cite{brunton2020machine, agostini2020exploration} to flow control \cite{vinuesa2022enhancing, vinuesa2024perspectives}. For instance, Agostini and Leschziner \cite{agostini2022auto} utilized autoencoders to isolate the non-linear modulation of inner-layer stresses by outer-layer structures in a high-Reynolds channel flow. Similarly, Balasubramanian \textit{et al.} \cite{balasubramanian2023predicting} employed Convolutional Neural Networks (CNNs) to map off-wall channel flow measurements to wall-shear stress, demonstrating the utility of non-linear models for wall modeling in Large-Eddy Simulations (LES).

Beyond spatial compression, machine learning architectures have been applied to model temporal dynamics. Linot and Graham \cite{linot2023dynamics} introduced a framework based on AEs and Neural Ordinary Differential Equations (Neural ODEs) to model a minimal Couette flow, successfully capturing the system trajectory and long-term statistics. More recently, Solera-Rico \textit{et al.} \cite{solera2024beta} proposed an architecture combining a $\beta$-Variational Autoencoder ($\beta$-VAE) with a Transformer, demonstrating effective temporal forecasting for chaotic flow between two collinear flat plates. However, this architecture has yet to be assessed in the context of wall-bounded turbulence, where the distinctive physics of the near-wall region impose two demanding requirements. First, regarding spatial representation, the $\beta$-VAE bottleneck is known to attenuate high-wavenumber features \cite{higgins2017beta}. This smoothing effect is particularly detrimental in the near-wall region, where fine-scale streaks and quasi-streamwise vortices are critical for Reynolds stress production. Second, concerning temporal forecasting, an autoregressive rollout conditioned solely on its own predictions is inherently susceptible to error accumulation and distributional drift. This limitation is especially consequential in a flow governed by an intermittent regeneration cycle, as the model must demonstrates to reproduce rare quiescent phases.

\begin{figure}[h]
\centering
    \includegraphics[width=0.98\linewidth, trim=0cm 3cm 0cm 7cm, clip]{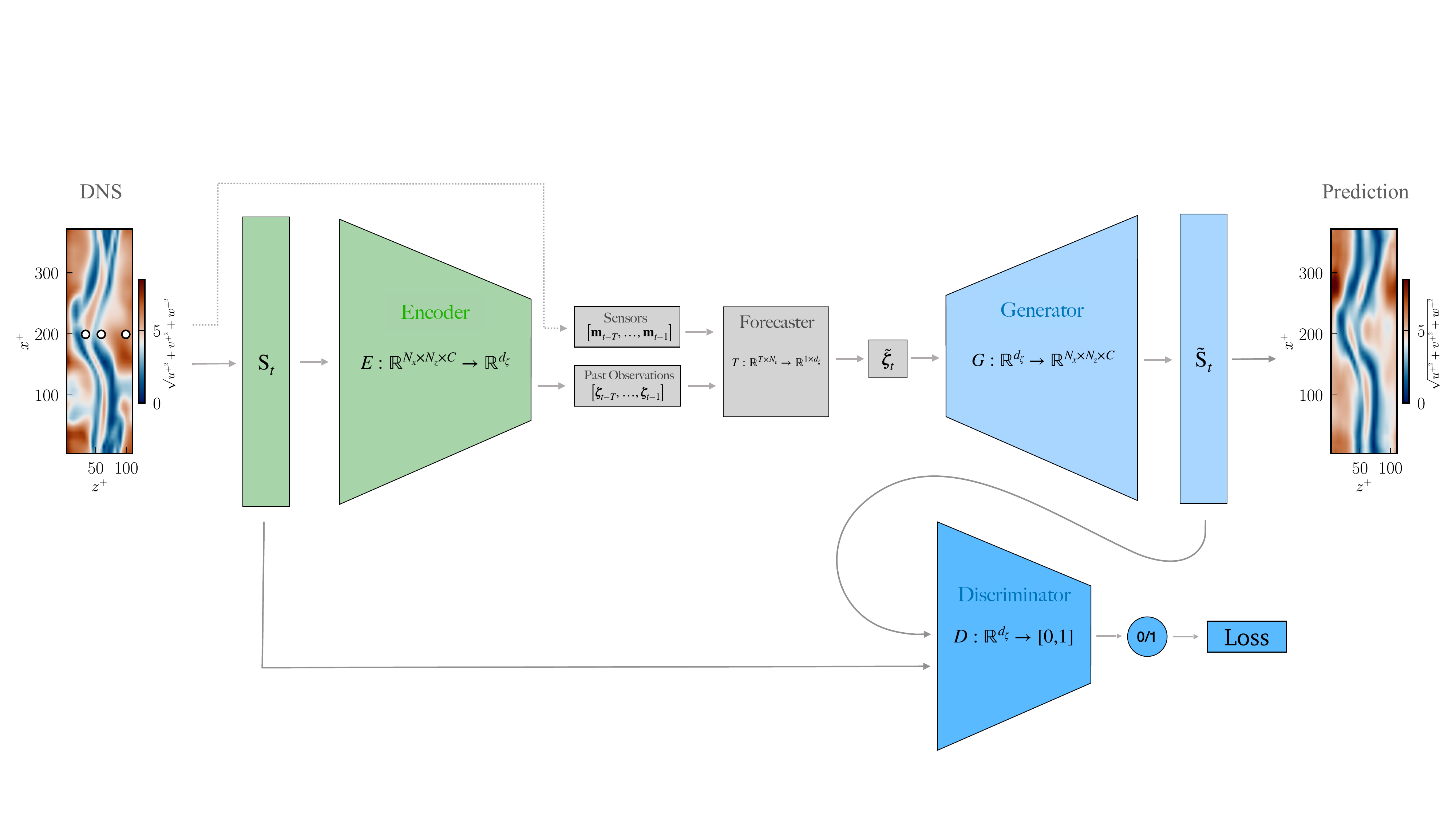}
    \caption{\protect Schematic representation of the proposed model framework. \textit{Training phase:} The $\beta$-variational autoencoder ($\beta$-VAE) and discriminator are trained simultaneously in a generative adversarial framework, where the decoder serves as generator and receives evaluative feedback from the discriminator. A conditional transformer model is then trained to forecast latent variable evolution from minimal sensors $m_i$ as input. \textit{Inference phase:} Only the transformer and decoder are retained to forecast flow field predictions from minimal sensors signals.}
\label{fig:intro/data-driven-framework}
\end{figure}

Building on these foundations, the present study proposes a reduced-order modelling framework illustrated in Figure \ref{fig:intro/data-driven-framework} that addresses the two aforementioned limitations. First, to mitigate the attenuation of high-wavenumber features, the standard $\beta$-VAE \cite{higgins2017beta} is augmented with a Generative Adversarial Network (GAN) discriminator \cite{goodfellow2014generative}. As recently demonstrated by the authors in the context of fluid-structure interaction \cite{tonioni2026vivaldy}, this $\beta$-VAE-GAN formulation preserves fine-scale spatial structures with greater fidelity than the baseline $\beta$-VAE. Second, to prevent distributional drift during the autoregressive rollout, the Transformer is conditioned on sparse sensor measurements. This is achieved via an adapted AdaLN-Zero modulation mechanism \cite{Peebles_2023_ICCV}, which provides a physical constraint at each forecasting step to correct the predicted trajectory. Finally, computational overhead is minimised by employing Easy-Attention \cite{easy} in place of standard Self-Attention \cite{vaswani2023attentionneed}.

.

The principal contributions of the present study are : (i) the first application of a $\beta$-VAE-GAN architecture to wall-bounded turbulence, demonstrating improved preservation of high-wavenumber spectral features relative to standard $\beta$-VAE approaches; (ii) an adaptation of the AdaLN-Zero modulation mechanism for sensor-based conditioning of the Transformer, enabling stable autoregressive forecasting over extended time horizons; and (iii) a comprehensive evaluation of the proposed framework on the Minimal 
Flow Unit \cite{jimenez1991minimal} ($Re_\tau = 200$) at $y^+ = 14$, demonstrating that the learnt latent dimensions autonomously encode the characteristic timescales of the near-wall regeneration cycle and that the framework reproduces intermittent ejection, sweep, and quiescent events.

The remainder of the paper is organised as follows. The numerical setup and dataset preparation are detailed in \S \ref{sec:02_dataset}. Next, \S \ref{sec:03_methods} outlines the architecture of the reduced-order model alongside the proposed training strategy. The predictive capabilities of the framework are analysed and discussed in \S \ref{sec:04_results}. Finally, concluding remarks are provided in \S \ref{sec:05_conclusion}.


\section{Direct Numerical Simulation and Data Preparation}
\label{sec:02_dataset}

This section details the generation and preprocessing of the turbulent channel flow dataset. The high-fidelity Direct Numerical Simulation (DNS) used to generate the raw flow fields are first described (\S\ref{sec:02_dataset/dns}). Subsequently, the data processing pipeline is outlined, including the extraction of two-dimensional snapshots, the train-test splitting strategy, and the normalization procedures required for the learning framework (\S\ref{sec:02_dataset/preprocessing}).

\subsection{Direct Numerical Simulation of Minimal Flow Unit}
\label{sec:02_dataset/dns}

\begin{figure}[htpb]
    \centering
    \includegraphics[width=0.5\linewidth]{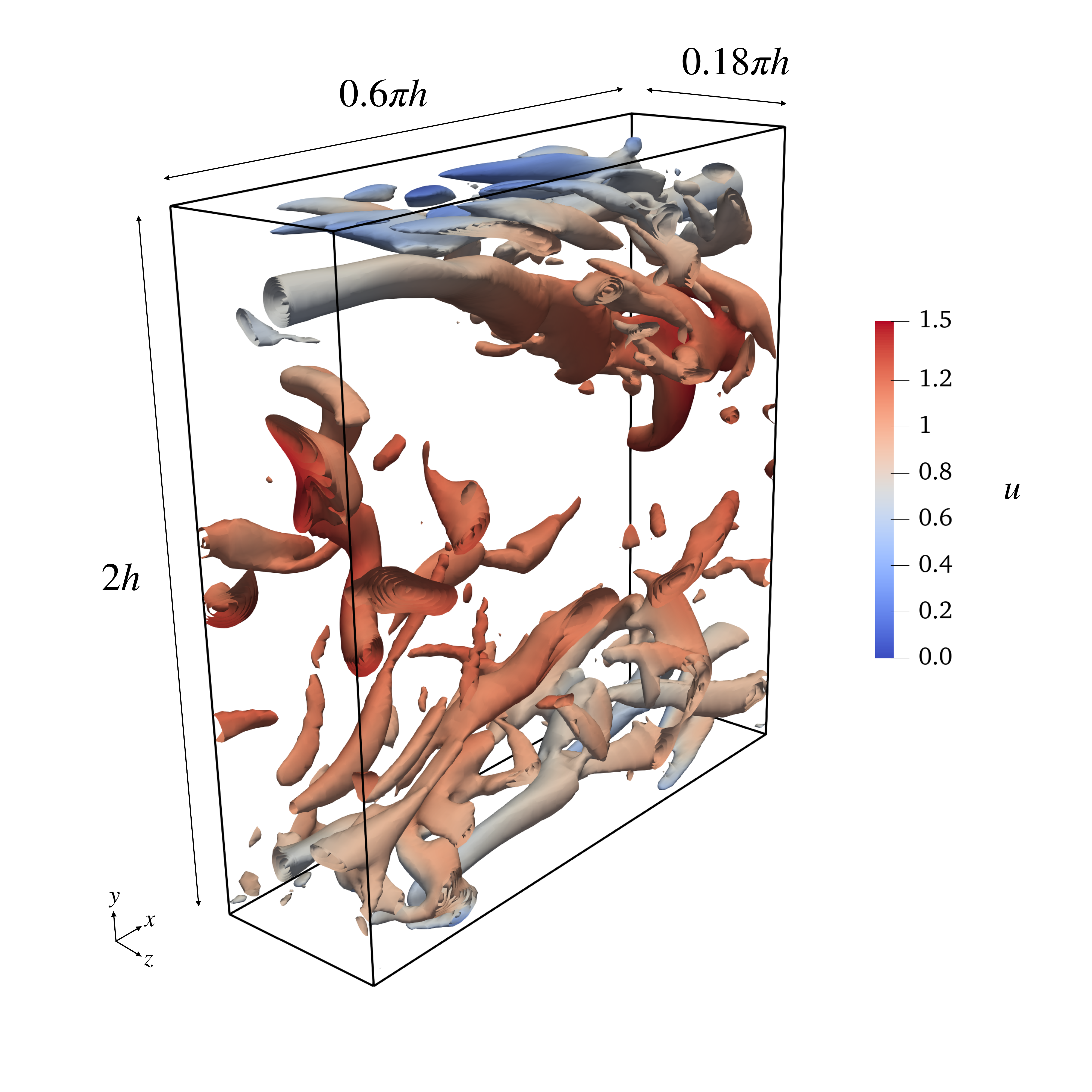}
    \caption{Instantaneous visualization of the MFU computational domain. Iso-surfaces of the Q-criterion \citep{hunt1988eddies} are colored by the streamwise velocity $u$. The domain dimensions are $L_x^+ \approx 377$, $L_y^+ = 400$, and $L_z^+ \approx 113$ in wall units, considering $h=1$.}
    \label{fig:02_dataset/dataset_visualization}
\end{figure}

The dataset was generated via DNS of a plane turbulent channel flow at a friction Reynolds number $Re_\tau = u_\tau h / \nu = 200$, where $u_\tau$ is the friction velocity, $h$ is the channel half-height, and $\nu$ is the kinematic viscosity. To maximize computational efficiency while preserving the essential self-sustaining mechanisms of wall-bounded turbulence, a Minimal Flow Unit domain is selected \citep{jimenez1991minimal}. The computational domain size is set to $L_x \times L_y \times L_z = 0.6\pi h \times 2h \times 0.18\pi h$, as illustrated in Figure \ref{fig:02_dataset/dataset_visualization}. Boundary conditions are periodic in the streamwise ($x$) and spanwise ($z$) directions, with no-slip conditions imposed at the walls ($y = \pm h$).

Simulations were performed using \texttt{SOD2D} \citep{gasparino2024sod2d}, a GPU-accelerated spectral-element solver. The spatial discretization utilizes Gauss-Lobatto-Legendre (GLL) nodes within elements of polynomial order five. The domain is discretized into $12\times36\times12$ elements, along the $x$, $y$, and $z$ directions, respectively. The resulting uniform grid spacings are $\Delta x^+ = 8.9$ and $\Delta z^+ = 2.7$. In the wall-normal direction, elements follow a power-law distribution, resulting in a resolution varying from $\Delta y^+ = 0.4$ at the wall to $\Delta y^+ = 4.4$ at the centerline. Time integration was performed using a fourth-order Runge-Kutta scheme with a fixed time-step $\Delta t^* = 10^{-3}$, where time is normalized by the convective time scale as $t^* = t u_b / h$, with $u_b$ denoting the bulk velocity.

\subsection{Data Extraction and Preprocessing}
\label{sec:02_dataset/preprocessing}

To ensure a rigorous evaluation of the model's generalization capabilities, two separate DNS runs were performed. The first run ($0 \leq t^* \leq 1380$) was utilized for training and validation, while the second run ($1380 < t^* \leq 2769$), initialized from the final state of the first, served exclusively as a test set. This temporal separation prevents data leakage and ensures the test set encompasses a distinct realization of the flow's chaotic evolution.

\begin{figure}[htbp]
  \centering
  \begin{minipage}[b]{0.45\textwidth}
    \centering
    \includegraphics[width=\linewidth]{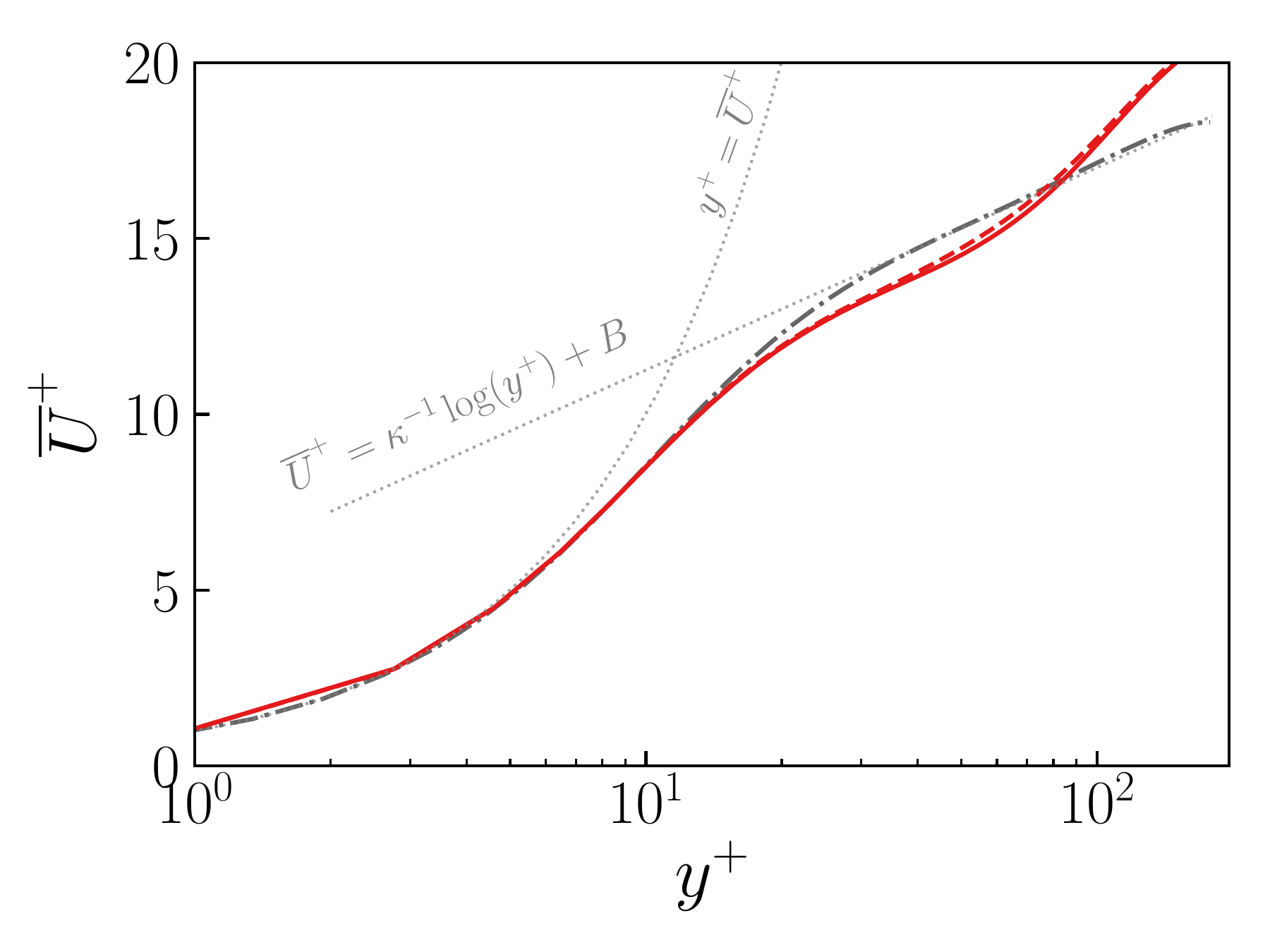}
  \end{minipage}
  \hfill
  \begin{minipage}[b]{0.45\textwidth}
    \centering
    \includegraphics[width=\linewidth]{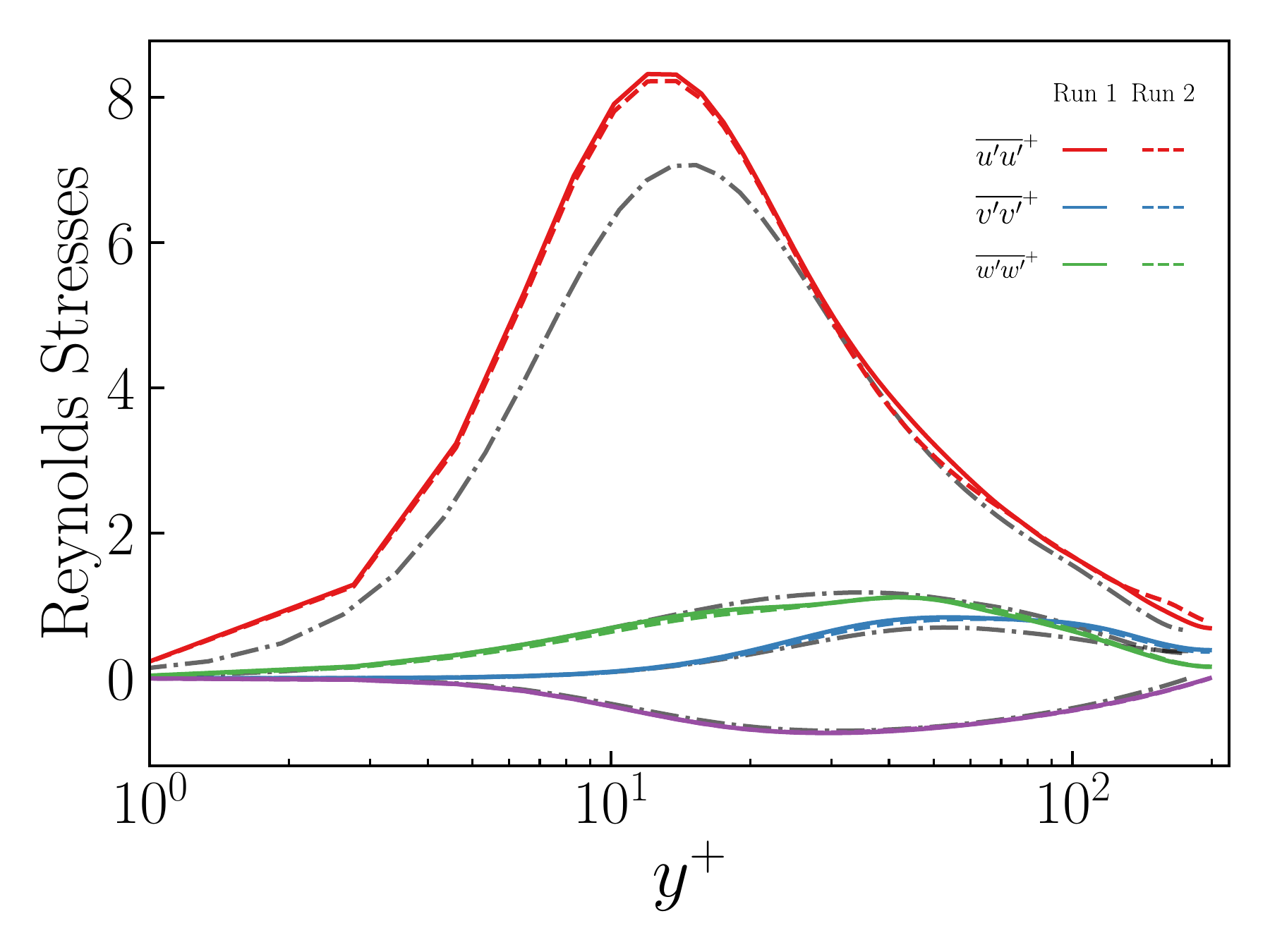}
  \end{minipage}
  \caption{Comparison of the datasets first- and second-order statistics against reference DNS data.  Left panel: streamwise mean velocity profile. Right panel: Reynolds stress profiles. Colored lines represent results from the first and second DNS runs at $Re_\tau = 200$, while dashed black lines show reference data from the full channel simulation of Kim \textit{et al.} \cite{kim1987turbulence} at $Re_\tau = 180$. Velocity is normalized by the friction velocity $u_\tau$, stresses by $u_\tau^2$, and wall-normal distance $y$ in wall units ($y^+ = y u_\tau/\nu$).}
  \label{fig:02_dataset/comparison_full_channel}
\end{figure}

Snapshots were extracted at intervals of $\Delta t^*_{save} = 0.08$, yielding a total of 34,615 fields. Since the raw DNS data resides on a grid with non-uniform GLL node distribution, each snapshot was interpolated onto a uniform Cartesian grid of size $N_x \times N_y \times N_z = 48\times216\times24$ using fifth-order spectral interpolation. This preprocessing step ensures compatibility with standard convolutional neural network architectures. The resulting Cartesian grid spacings are $\Delta x^+ = 7.8$, $\Delta y^+ = 1.8$, and $\Delta z^+ = 4.7$. As shown in Figure \ref{fig:02_dataset/comparison_full_channel}, the first- and second-order statistics show good agreement with the reference full-channel DNS of Kim and Moin \cite{kim1987turbulence}, with the small discrepancy in $\overline{u^\prime u^\prime}$ ascribable to the constrained domain of the MFU relative to the full channel, and to the slight difference in Reynolds number ($Re_\tau = 200$ versus $Re_\tau = 180$ for the reference). Furthermore, the statistical profiles computed independently for the first and second runs overlap almost perfectly. This mutual consistency confirms that the sampling duration of each dataset is sufficient to obtain converged turbulence statistics.

Since the learning framework operates on two-dimensional turbulent fluctuations, a Reynolds decomposition is first applied to the velocity fields:

\begin{equation}
    u = \overline{u} + u^\prime, \quad
    v = \overline{v} + v^\prime, \quad
    w = \overline{w} + w^\prime
\end{equation}

\noindent where the mean flow $\overline{\left(\cdot\right)}$ is computed by averaging over time. Then two-dimensional flow snapshots were extracted from the $x$-$z$ plane at $y^+ = 14$, corresponding to the peak of turbulent kinetic energy. The snapshots from the first run ($N_t^{\mathrm{train/val}} = 17{,}246$ snapshots) were used to construct the training and validation datasets, while those from the second run ($N_t^{\mathrm{test}} = 17{,}369$ snapshots) formed a separate test dataset. Each dataset thus spans approximately $17{,}300\,t^+$ in wall units, with a temporal resolution of $\Delta t^+ \approx 1$. 

To verify that the planes extracted from the two runs exhibit comparable spatio-temporal complexity, a global entropy measure based on the Proper Orthogonal Decomposition is adopted, following \cite{aubry1991spatiotemporal}. This verification ensures that the evaluated reconstruction errors reflect the model's true generalisation capability rather than a systematic disparity in dynamical complexity between the training and test sets. Each dataset, initially of size $(N_t, N_x, N_z, C)$ where $C = 3$ represents the velocity components, is reshaped into a matrix of dimensions $(N, M)$ with $N = N_t$ snapshots and $M = N_x N_z C$ spatial degrees of freedom. Since $M < N$, the POD decomposition yields $M$ eigenvalues $\lambda_i$ corresponding to the energy content of each spatial mode. The global entropy is then defined as:

\begin{equation}
    H = - \frac{1}{\log{M}}\sum_{i=1}^M p_i \log p_i
    \label{eq:02.2_global_entropy}
\end{equation}

\noindent where $p_i$ represents the relative energy contribution of the $i$-th POD mode:

\begin{equation}
    p_i = \frac{\lambda_i}{\sum_{j=1}^M \lambda_j}.
\end{equation}

\noindent This entropy measure quantifies the degree of energy distribution across the POD modes. A high entropy ($H \to 1$) indicates energy distributed across many modes (disordered flow), while low entropy ($H \to 0$) suggests energy concentration in few structures. The entropy values computed are:

\begin{equation}
    H^{\text{train/val}} = 0.6974, \quad H^{\text{test}} = 0.6975.
\end{equation}

The nearly identical values confirm comparable spatio-temporal complexity between the two sets. The specific subdivision of the first run snapshots into training and validation sets will be described in the respective sections for the $\beta$-VAE-GAN (\S\ref{sec:03.1_beta_vae_gan}) and conditional transformer (\S\ref{sec:03.2_cond-transf}), as the split strategy differ between the two models.

Finally, to facilitate training stability while preserving the physical anisotropy of the flow, the fluctuation fields from each dataset were normalized using a reference scaling parameter, computed separately for each run. The global standard deviation of the streamwise velocity fluctuation is utilized scaled by a factor of $3$. This scaling is selected to bound the majority of turbulent events within the interval $[-1, 1]$ while preserving the relative magnitudes of the transverse components:

\begin{equation}
    u^\prime_\mathrm{norm} = \frac{u^\prime}{3\sigma_u}, \quad
    v^\prime_\mathrm{norm} = \frac{v^\prime}{3\sigma_u}, \quad
    w^\prime_\mathrm{norm} = \frac{w^\prime}{3\sigma_u}
\end{equation}

\noindent where $\sigma_u$ is the global root-mean-square (RMS) of the streamwise fluctuations $u^\prime$, computed separately for the training and test datasets to prevent data leakage between the two sets.


\section{Reduced-Order Modeling Approach}
\label{sec:03_methods}
This section presents the proposed machine learning framework for reduced-order modelling. The architecture comprises two primary components: a $\beta$-VAE-GAN that learns a compact, non-linear latent representation of the spatial flow structures (\S\ref{sec:03.1_beta_vae_gan}), and a transformer that models the temporal evolution of these latent variables conditioned on sparse sensor inputs (\S\ref{sec:03.2_cond-transf}). 

The two models are trained sequentially rather than end-to-end to ensure that the $\beta$-VAE-GAN focuses exclusively on capturing the essential spatial structures of turbulence, without being influenced by the transformer prediction objective. In an end-to-end approach, the latent representation may be biased towards features that facilitate prediction rather than those that best represent the underlying flow physics. By decoupling the compression and prediction tasks, the framework maintains a latent space optimised on the spatial structures while preserving modelling flexibility.

\subsection{Latent Feature Extraction via $\beta$-VAE-GAN}
\label{sec:03.1_beta_vae_gan}

For the latent feature extraction, a hybrid generative architecture combining $\beta$-Variational Autoencoders \cite{higgins2017beta} and Generative Adversarial Networks \cite{goodfellow2014generative} is proposed. Drawing upon advancements in learned image compression \cite{tschannen2018deep, agustsson2019generative, mentzer2020high}, this formulation augments the 
autoencoder with an adversarial objective to better preserve the topology of the input distribution. Such structural fidelity is achieved because the adversarial loss evaluates the global statistical consistency of the generated fields. In contrast, relying solely on point-wise metrics like Mean Squared Error (MSE) inherently applies a low-pass filter to the data, attenuating the high-wavenumber features of the flow. In the following, the $\beta$-VAE formulation is presented first, followed by the GAN component and their hybrid coupling.

Let $\mathbf{S}_t \in \mathbb{R}^{N_x \times N_z \times C}$ denote a flow snapshot at time $t$, and let
$\boldsymbol{\zeta}_t \in \mathbb{R}^{d_\zeta}$ be a set of latent variables acting as a coordinate system of a curved manifold, with $d_\zeta \ll N_x N_z C$. An autoencoder is defined as the architecture that learns the non-linear mapping from the high-dimensional snapshots $\mathbf{S}_t$ to the low-dimensional latent variables $\boldsymbol{\zeta}_t$ (encoding) and its inverse (decoding). While standard autoencoders learn deterministic mappings, VAEs are designed to learn \textit{probabilistic} representations \cite{kingma2013auto}. By modeling the snapshots $\mathbf{S}_t$ as independent realizations of a random variable $\mathbf{S}$ drawn from an underlying probability distribution $p_{\mathbf{S}}$, and assuming the prior $p_{\boldsymbol{\zeta}}$ follows a standard multivariate normal, the encoder $E$ approximates the intractable posterior $p_{\boldsymbol{\zeta}\vert\mathbf{S}}$ via a variational distribution $q_{\mathbf{W}_E}(\boldsymbol{\zeta}\vert\mathbf{S})$, parametrized by weights $\mathbf{W}_E$. The decoder $G$, parametrized by weights $\mathbf{W}_G$, serves as a generative model. It maps latent variables back to the reconstructed flow field, which parametrizes the mean of the likelihood $p(\mathbf{S}\vert\boldsymbol{\zeta})$. The $\beta$-VAE objective function modifies the standard evidence lower bound, used in VAE, by introducing a scaling factor $\beta$:

\begin{equation}
    \mathcal{L}_{EG}^{\beta-\mathrm{VAE}} = \mathbb{E}_{\mathbf{S} \sim p_{\mathbf{S}}}\bigl[d(\mathbf{S}, \tilde{\mathbf{S}})\bigr] + \beta\,D_{\mathrm{KL}}\bigl(q_{\mathbf{W}_E}(\boldsymbol{\zeta}\vert\mathbf{S}) \,\|\, p_{\boldsymbol{\zeta}}\bigr),
\label{eq:03_methods/loss_function_beta_VAE}
\end{equation}

\noindent where $\mathbb{E}[\cdot]$ denotes the expectation,  $\tilde{\mathbf{S}} = G(\boldsymbol{\zeta})$ represents the reconstruction, $d(\cdot, \cdot)$ is a distortion metric (e.g., MSE), and $D_{\mathrm{KL}}$ denotes the Kullback--Leibler divergence. The parameter $\beta$ regulates the trade-off between reconstruction fidelity and the disentanglement of the latent space. A higher $\beta$ enforces stricter independence among the latent variables, ensuring that distinct dimensions capture distinct features, but also increases the tolerable distortion \cite{burgess2018understanding, jacobsen2022disentangling}. This increased distortion, combined with the inherent low-pass filtering of point-wise distortion metrics, is what attenuates the high-wavenumber features of the flow. To counteract this effect, an adversarial component is integrated into the objective function. The GAN formulation underlying this component is presented next before introducing the hybrid coupling.

A standard GAN comprises a generator $G$ and a discriminator $D$, trained in an adversarial game. The generator maps latent vectors from a prior $\boldsymbol{\zeta} \sim p_{\boldsymbol{\zeta}}$ to synthesized samples $\tilde{\mathbf{S}}$, while the discriminator attempts to distinguish between input samples $\mathbf{S} \sim p_{\mathbf{S}}$ and generated samples. The non-saturating adversarial objective is given by:

\begin{align}
    \mathcal{L}_{G}^\mathrm{GAN} &= \mathbb{E}_{\boldsymbol{\zeta} \sim p_{\boldsymbol{\zeta}}} \Big[-\,\log\bigl(D\bigl(G(\boldsymbol{\zeta})\bigr)\bigr)\Big], \label{eq:methods/gan_loss_gen}\\
    \mathcal{L}_{D}^\mathrm{GAN} &= \mathbb{E}_{\mathbf{S} \sim p_{\mathbf{S}}} \Big[-\,\log\bigl(D(\mathbf{S})\bigr)\Big] + \mathbb{E}_{\tilde{\mathbf{S}} \sim p_{\tilde{\mathbf{S}}}} \Big[-\,\log\bigl(1 - D\bigl(\tilde{\mathbf{S}}\bigr)\bigr) \Big]. 
\label{eq:03_methods/gan_loss_disc}
\end{align} 

\noindent The generator loss, $\mathcal{L}_{G}^\mathrm{GAN}$, represents the negative log probability that the discriminator assigns to the generated data being real. Conversely, the discriminator loss, $\mathcal{L}_{D}^\mathrm{GAN}$, minimizes the negative log probability assigned to input data being real (first term) and generated data being fake (second term). This formulation drives the generator to produce outputs that are statistically indistinguishable from the true data distribution $p_{\mathbf{S}}$. However, in a standard GAN formulation, the latent vector is sampled from a fixed prior distribution (typically standard Gaussian noise) that is agnostic to the underlying physics. Consequently, the resulting latent space lacks structural organisation and does not inherently encode the spatial features of the flow data.

In the proposed framework, the GAN generator is coupled with a VAE encoder, creating a structured low-dimensional manifold whilst exploiting the adversarial loss to recover spectral fidelity. The total training objective combines the $\beta$-VAE and GAN formulations into a hybrid adversarial loss:

\begin{align}
    \mathcal{L}_{EG}^\mathrm{Hybrid}  &= \underbrace{\mathcal{L}_{EG}^{\beta-\mathrm{VAE}}}_{\text{Reconstruction + Regularization}} \;-\;\gamma\,\underbrace{\mathbb{E}_{\mathbf{S} \sim p_{\mathbf{S}}}\Bigl[\mathbb{E}_{\boldsymbol{\zeta} \sim q_{\mathbf{W}_E}(\cdot|\mathbf{S})} \bigl[\log\bigl(D(G(\boldsymbol{\zeta}))\bigr)\bigr]\Bigr]}_{\text{Adversarial term}}, 
    \label{eq:adversarial_loss_formulation/enc_gen_loss}\\
    \mathcal{L}_{D}^\mathrm{Hybrid} &= \mathbb{E}_{\mathbf{S} \sim p_{\mathbf{S}}} \Bigl[-\,\log\bigl(D(\mathbf{S})\bigr)\Bigr] + \mathbb{E}_{\mathbf{S} \sim p_{\mathbf{S}}} \Bigl[\mathbb{E}_{\boldsymbol{\zeta} \sim q_{\mathbf{W}_E}(\cdot|\mathbf{S})} \bigl[-\,\log\bigl(1 - D\bigl(G(\boldsymbol{\zeta})\bigr)\bigr)\bigr]\Bigr],
    \label{eq:adversarial_loss_formulation/dsc_loss}
\end{align}

\noindent where the hyperparameter $\gamma$ weights the adversarial contribution. Here, the discriminator evaluates reconstructions sampled from the approximate posterior $\boldsymbol{\zeta} \sim q_{\mathbf{W}_E}(\cdot|\mathbf{S})$, rather than the uninformative fixed prior of standard GANs. This ensures that the adversarial penalty enforces statistical consistency on the reconstructed flow fields drawn from the structured latent-space.

\begin{figure}[htbp]
    \centering
    
    \includegraphics[width=0.9\linewidth, trim=0.5cm 8cm 1.0cm 8cm, clip]{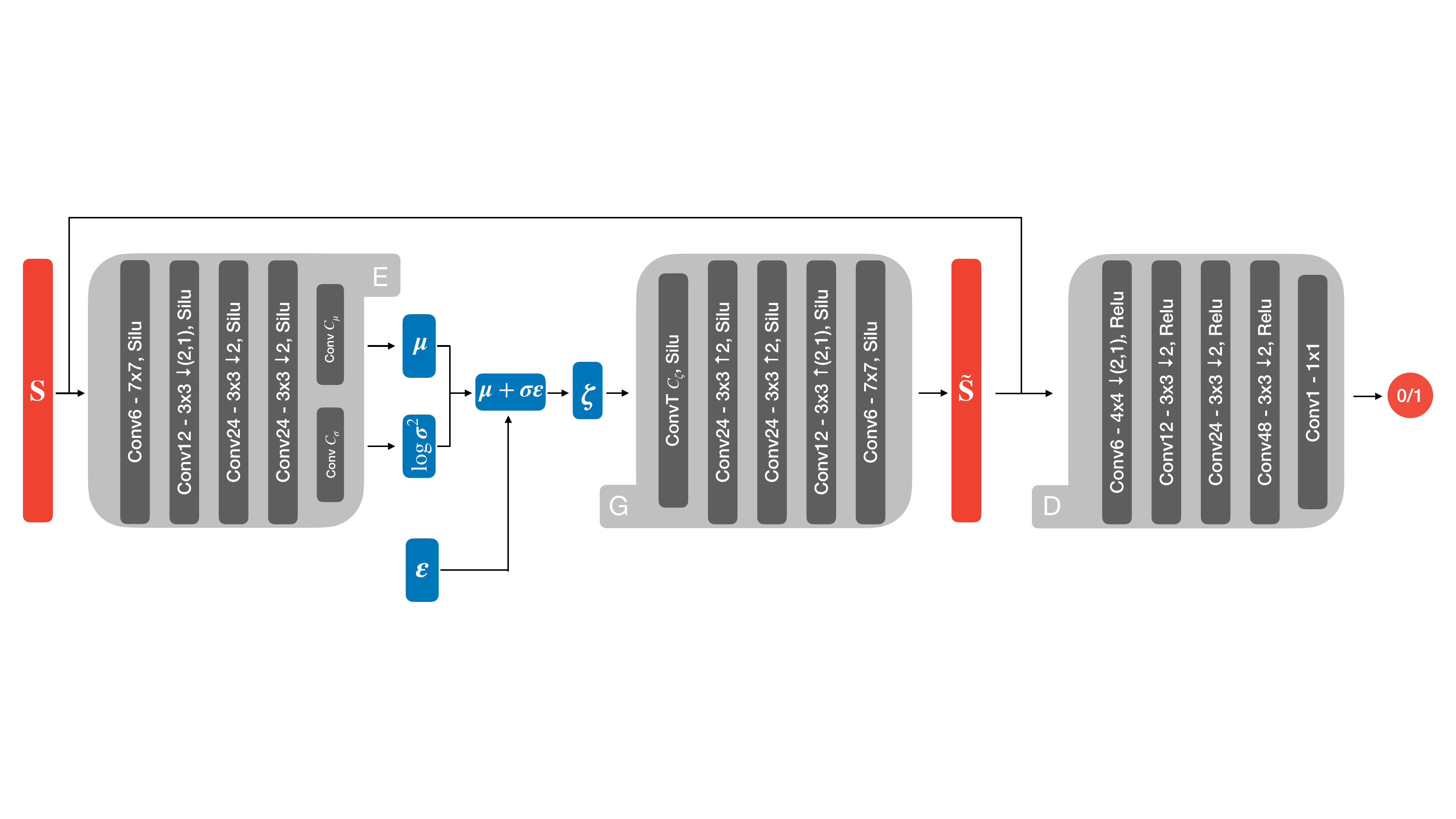}
    \caption{
    Schematic of the $\beta$-VAE-GAN architecture. The encoder ($E$), generator ($G$), and discriminator ($D$) utilize convolutional layers. Each Conv$C_f$ layer applies $k \times k$ filters with $C_f$ output channels. Downward arrows ($\downarrow(2,1)$) denote strided downsampling in the streamwise and spanwise directions, respectively, while upward arrows ($\uparrow(2,1)$) denote Lanczos upsampling. Layers labeled Conv~$C_\mu$ and Conv~$C_\sigma$ project the encoder output to the latent mean and variance, while ConvT~$C_{\zeta}$ projects the sampled latent variable $\boldsymbol{\zeta}$ back to the spatial domain.
    }
    \label{fig:03_methods/advae_arch}
\end{figure}

Figure~\ref{fig:03_methods/advae_arch} presents the $\beta$-VAE-GAN architecture used in this study. Circular padding is employed across all convolutional layers to enforce the periodic boundary conditions in the streamwise and spanwise directions. The encoder comprises $N_{L_e} = 4$ convolutional layers with SiLU activations. Strided convolutions are applied from the second layer onwards to progressively compress the spatial dimensions, with the final layer acting as a fully connected layer, projecting the output to the latent mean and variance. The generator (decoder) follows a symmetric architecture: a transpose convolution first maps the sampled latent vector $\boldsymbol{\zeta}$ to a low-resolution spatial representation, after which $N_{L_g} = 4$ convolutional layers with SiLU activations and upsampling progressively restore the spatial resolution. Upsampling is performed using Lanczos interpolation to mitigate checkerboard artefacts \cite{odena2016deconvolution}. The discriminator comprises $N_{L_d} = 4$ strided convolutional layers with ReLU activations, each normalised via spectral normalisation \cite{miyato2018spectral} to enforce a Lipschitz constraint on the critic, thereby stabilising adversarial training. A final $1 \times 1$ convolution, following the PatchGAN formulation \cite{isola2017image}, projects the feature maps to a single-channel spatially resolved map of real/fake logits, from which the adversarial loss is computed directly.  No final activation function is applied, as the adversarial loss is computed directly from these raw logits.

The model was trained on the training/validation dataset described in \S\ref{sec:02_dataset/preprocessing}, randomly partitioned into 90\% training and 10\% validation sets; random partitioning being appropriate since spatial compression is invariant to temporal ordering. Training was performed for 2,000 epochs using the Adam optimiser \cite{kingma2014adam} on a single NVIDIA A100 (80\,GB) GPU, with a cosine decay learning rate schedule and a linear warmup period applied to both the encoder-decoder and discriminator networks. To stabilise training, the discriminator weights were updated every two epochs and remained frozen in the interim. The regularisation parameter $\beta$ followed a cyclical annealing schedule \cite{fu2019cyclical} with two cycles. The adversarial weight $\gamma$ was held at zero for the initial 50 epochs, linearly ramped up to a target value $\gamma_{\max}$ over the subsequent 150 epochs, and kept constant thereafter. The validation loss was monitored every 20 epochs to trigger early stopping if necessary. All hyperparameters are summarised in Table~\ref{tab:3.1_hyperparameters}.

\begin{table}[htbp]
    \centering
    \small
    \begin{tabular}{l l p{0.5cm} l l}
        \toprule
        \multicolumn{2}{c}{\textbf{Loss}} & & \multicolumn{2}{c}{\textbf{Training}} \\
        \cmidrule{1-2} \cmidrule{4-5} 
        $\beta$ Schedule          & Cyclical (2 cyc)      & & Hardware                & NVIDIA A100 \\
        $\beta_{min}$ / $\beta_{max}$ & $0.0$ / $10^{-3}$ & & Train / Val split       & 90\% / 10\% \\
        Cycle ratio               & 0.5                   & & Optimizer               & Adam \\
        $\gamma$ Schedule         & Linear Ramp           & & Batch size              & 32 \\
        $\gamma$ Delay            & 50 epochs             & & Total epochs            & 2000 \\
        $\gamma$ Ramp             & 150 epochs            & & Warmup steps            & 500 \\
        $\gamma_{max}$            & $1\times 10^{-3}$     & & Initial LR              & $1\times 10^{-3}$ \\
                                  &                       & & Final LR                & $1\times 10^{-5}$ \\
                                  &                       & & Schedule type           & Cosine Decay \\
        \bottomrule
    \end{tabular}
    \caption{Summary of hyperparameters and training configuration for the $\beta$-VAE-GAN.}
    \label{tab:3.1_hyperparameters}
\end{table} 

\subsection{Temporal Forecasting via Sensor-Conditioned Transformer}
\label{sec:03.2_cond-transf}

Having established a compact latent representation of the spatial flow structures, the second phase of the framework is concerned with modelling their temporal evolution. To this end, a transformer-based architecture is employed to autoregressively forecast the future trajectory of the latent state $\boldsymbol{\zeta}_t$. Transformer architectures are preferred over recurrent (LSTM, RNN) and convolutional alternatives, having been shown to outperform them in capturing non-linear, chaotic dynamics \cite{solera2024beta, Eiximeno_Sanchis-Agudo_Miró_Rodriguez_Vinuesa_Lehmkuhl_2025}. To extend the forecasting horizon, the proposed transformer is explicitly conditioned on sparse sensor measurements, directly coupling the latent dynamics prediction to observable physical inputs and constraining the trajectories to the underlying flow state.

The forecasting task is formulated as a sequence-to-sequence prediction problem. Retaining the notation of \S\ref{sec:03.1_beta_vae_gan}, $\boldsymbol{\zeta}_t \in \mathbb{R}^{d_{\zeta}}$ denotes the latent representation of snapshot $\mathbf{S}_t$ produced by the pre-trained $\beta$-VAE-GAN encoder. Sparse sensor measurements are extracted via a measurement operator $\mathcal{H}$ acting on the physical field:

\begin{equation}
    \mathbf{m}_t = \mathcal{H}(\mathbf{S}_t) \in \mathbb{R}^{d_m},
\end{equation}

\noindent where $d_m$ is the number of sensors. Given a context window of length $T$, the time-aligned sequences of past latent states and sensor readings are defined as:

\begin{equation}
    \mathbf{Z}_{t} = [\boldsymbol{\zeta}_{t-T+1},\dots,\boldsymbol{\zeta}_{t}]^\top \in \mathbb{R}^{T\times d_\zeta},
    \quad
    \mathbf{M}_{t} = [\mathbf{m}_{t-T+1},\dots,\mathbf{m}_{t}]^\top \in \mathbb{R}^{T\times d_m}.
\end{equation}

\noindent Then, the transformer architecture parameterizes a temporal predictor $f_\theta$, with trainable weights $\theta$, which performs one-step-ahead forecasting in the latent space:

\begin{equation}
    \hat{\boldsymbol{\zeta}}_{t+1} = f_\theta(\mathbf{Z}_{t}, \mathbf{M}_{t}).
    \label{eq:03_methods/temporal_objective}
\end{equation}

\noindent Multi-step rollouts are then obtained auto-regressively by appending the predicted state $\hat{\boldsymbol{\zeta}}_{t+1}$ to the context window and shifting the sequence forward.

\begin{figure}[htbp]
    \centering
    \includegraphics[width=0.75\linewidth, trim=6.5cm 0.5cm 0cm 0.5cm, clip]{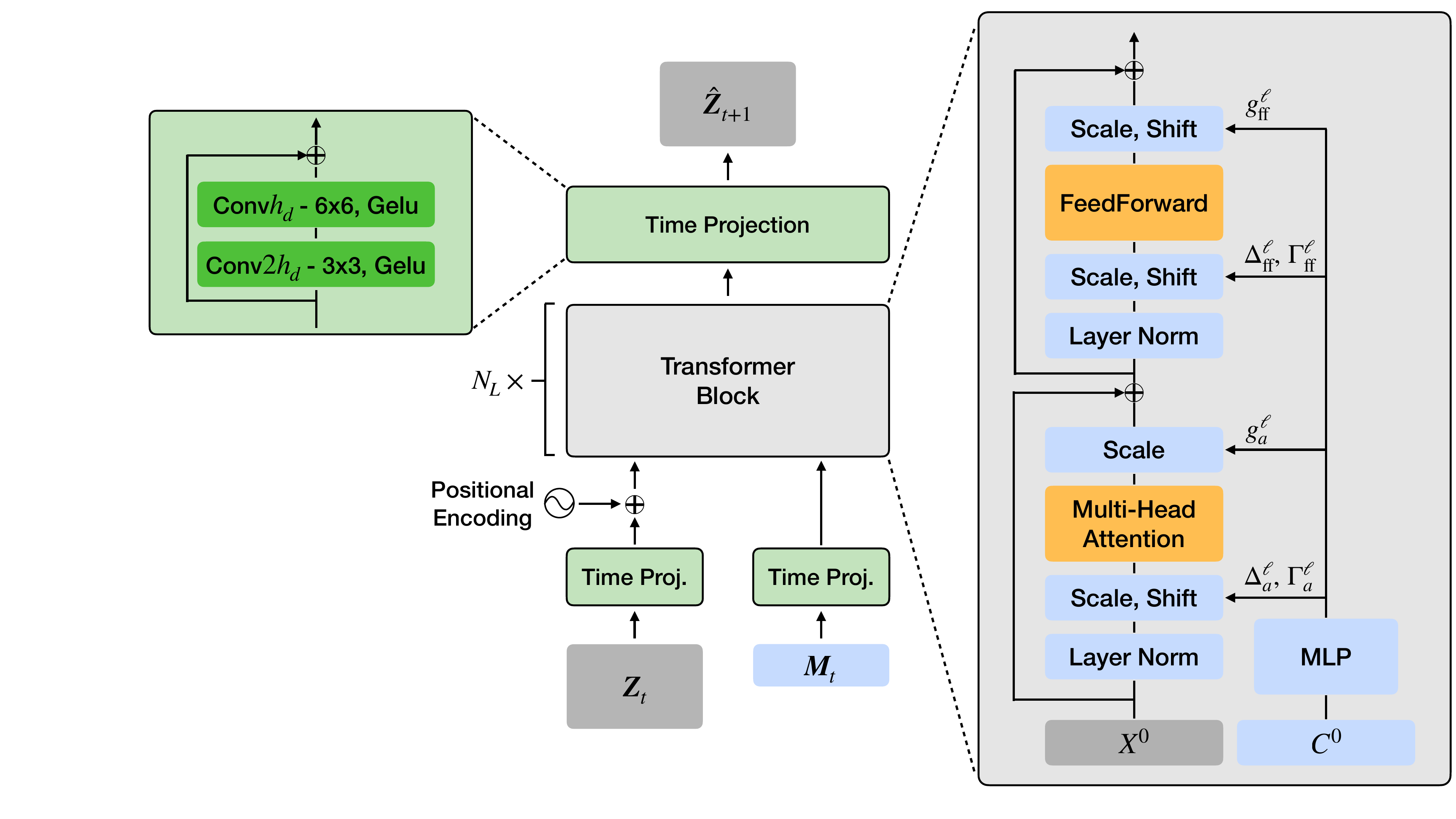}
    \caption{Schematic of the sensor-conditioned Transformer architecture. The framework processes sequences of latent states ($\mathbf{Z}_t$) and sensor measurements ($\mathbf{M}_t$) through parallel convolutional embeddings, utilizing AdaLN-Zero \cite{Peebles_2023_ICCV} to modulate the transformer blocks.}
    \label{fig:03_methods/trsf_arch}
\end{figure}

\newpage
Figure~\ref{fig:03_methods/trsf_arch} illustrates the Transformer backbone architecture used in this study. The input latent sequence is first mapped to an embedding of dimension $d_e$ through a time projection layer ($\mathrm{TP}$), followed by the addition of an absolute positional encoding:

\begin{equation}
    X^0 = \mathrm{TP}_{d_e}(\mathbf{Z}_{t}) + P_{\mathrm{sin}} \in \mathbb{R}^{T\times d_e},
\end{equation}

\noindent where $P_{\mathrm{sin}}$ represents the standard sinusoidal positional encoding. The time projection module itself is designed as a two-layer convolutional network featuring Gelu activations and a residual connection. To enhance feature extraction, this module employs an inverse bottleneck architecture: the intermediate hidden dimension is temporarily expanded to twice the target dimension ($2h_d$) before being projected back down. This expansion provides a higher-dimensional representation, enhancing the network's expressivity without inflating the computational cost of the subsequent attention mechanism. For this initial input projection, the final target dimension is $h_d = d_e$. To capture both local and broader temporal trends, the convolutional kernel size increases from the first to the second layer, enlarging the network's temporal receptive field.

In parallel, the sensor sequence is projected to a conditioning dimension $d_c$,

\begin{equation}
    C^0 = \mathrm{TP}_{d_c}(\mathbf{M}_{t}) \in \mathbb{R}^{T\times d_c},
\end{equation}

\noindent and is injected into every Transformer block through Adaptive Layer Normalization modulation with zero weights initialisation (AdaLN-Zero, described below). The Transformer itself comprises $N_L=4$ conditioned blocks with an embedding dimension $d_e=64$, $N_h=4$ attention heads, and a feed-forward network hidden dimension of $d_{\mathrm{ff}}=256$. The maximum context window of past latent states is set to $T = 128$ steps, corresponding to approximately $128 \, t^+$ in wall units. 

The conditioning signal is provided by $d_m = 3$ spatial sensors, each measuring the three fluctuating velocity components ($u^\prime, v^\prime, w^\prime$). Sensor locations are determined via a two-point spatial correlation analysis of the streamwise velocity fluctuations $u^\prime$. A primary reference sensor is placed at the centre of the domain to maximise the distance from the periodic boundarie and serves as correlation reference. The remaining two sensors are positioned at the points of maximum anti-correlation (phase opposition) and zero correlation (phase quadrature), respectively, along the same streamwise coordinate $x^+$.

The AdaLN-Zero (Diffusion Transformer-style) modulation mechanism is selected for conditioning due to its superior performance over cross-attention and in-context conditioning at a comparable computational cost \cite{Peebles_2023_ICCV}. A distinctive feature of the proposed architecture is the adaptation of this modulation to process physical sensor measurements. While AdaLN-Zero is traditionally driven by discrete class labels or text embeddings, the present framework utilises the aforementioned one-dimensional convolutional projection to map the sensor time series into compatible conditioning signals. These signals directly modulate the LayerNorm statistics. Specifically, within each transformer block $\ell$, a multilayer perceptron (MLP) maps the sensor conditioning tokens $\mathbf{C}^0$ to shift ($\Delta$), scale ($\Gamma$), and residual gate ($g$) parameters for both the attention and feed-forward branches:

\begin{equation}
    (\Delta_a^\ell,\Gamma_a^\ell,g_a^\ell)=\psi_a^\ell(C^0),
    \qquad
    (\Delta_\mathrm{ff}^\ell,\Gamma_\mathrm{ff}^\ell,g_\mathrm{ff}^\ell)=\psi_\mathrm{ff}^\ell(C^0).
\end{equation}

\noindent Given the intermediate latent tokens $X$, the adaptive LayerNorm (AdaLN) operator is defined as:

\begin{equation}
    \mathrm{AdaLN}(X;\Delta,\Gamma) = \mathrm{LN}(X)\odot(1+\Gamma)+\Delta,
\end{equation}

\noindent where $\odot$ denotes element-wise multiplication and $\mathrm{LN}(\cdot)$ the layer-norm operation. The final linear layer of each modulation MLP ($\psi^\ell$) is initialized to zero, forcing the residual gates ($g^\ell$) and modulation parameters ($\Delta^\ell, \Gamma^\ell$) to zero at the start of training. Consequently, each transformer block initially behaves as an identity mapping (due to the residual connections), ensuring a stable optimization procedure. As training advances, the network progressively learns to integrate the continuous sensor stream \cite{Peebles_2023_ICCV, goyal2017accurate}.

Three variants of the attention operator $\mathrm{Attn}$ are considered in the present study. Easy Attention \cite{easy} is employed in the proposed architecture: a simplification of standard self-attention \cite{vaswani2023attentionneed}, which replaces the traditional query-key mechanism by a static, learnable weight matrix. This eliminates the $\mathcal{O}(d_e T^2)$ overhead of dynamically computing pairwise similarities for every input, instead acting as a fixed temporal mixing operator that encodes the invariant dynamics of the system. Two alternatives are evaluated for comparison: standard Self-Attention and a hybrid Self + Easy formulation. These are defined as follows:

\begin{itemize}

\item \textbf{Self-Attention (baseline SA)}
Standard causal multi-head self-attention is computed by concatenating the outputs of $N_h$ independent attention heads:
\begin{align}
    \mathrm{head}_h &= \mathrm{softmax}\!\left(\frac{Q_h K_h^\top}{\sqrt{d_h}} + M\right)V_h, \\
    \mathrm{SelfAttn}(X) &= \mathrm{Concat}(\mathrm{head}_1, \dots, \mathrm{head}_{N_h})W_O,
\end{align}
where $Q_h=XW_{Q,h}$, $K_h=XW_{K,h}$, $V_h=XW_{V,h}$, $d_h=d_e/N_h$, and $M$ is the causal mask.

\item \textbf{Easy Attention (EA)}
Content-independent causal temporal mixing, where the attention weights are learned parameters rather than similarity scores:
\begin{equation}
\mathrm{EasyAttn}(X)=\mathrm{Concat}_{h=1}^{N_h}\!\left[\boldsymbol{\alpha}_h V_h\right]W_O,
\qquad
\boldsymbol{\alpha}_h\in\mathbb{R}^{T\times T},
\label{eq:03_methods/easyattn}
\end{equation}
with $\boldsymbol{\alpha}_h$ causally masked (upper-triangular entries suppressed). This operator learns a global temporal mixing of latent modes that is shared across samples.

\item \textbf{Self + Easy (hybrid SA + EA)}
A hybrid operator using content-dependent and content-independent mixing in parallel:
\begin{equation}
\mathrm{Attn}(X)=\mathrm{SelfAttn}(X)+\mathrm{EasyAttn}(X).
\end{equation}
\end{itemize}

With the attention variants defined, the forward pass through each conditioned transformer block ($\ell=1,\dots,N_L$) sequentially updates the latent representation $X^\ell$ as follows:

\begin{align}
    \hat{X}^\ell_a &= \mathrm{AdaLN}(X^\ell;\Delta_a^\ell,\Gamma_a^\ell), \nonumber\\
    X^{\ell+\frac{1}{2}} &= X^\ell + \mathrm{Drop}\!\Big(g_a^\ell \odot \mathrm{Attn}(\hat{X}^\ell_a)\Big), \nonumber\\
    \hat{X}^\ell_\mathrm{ff} &= \mathrm{AdaLN}(X^{\ell+\frac{1}{2}};\Delta_\mathrm{ff}^\ell,\Gamma_\mathrm{ff}^\ell), \nonumber\\
    X^{\ell+1} &= X^{\ell+\frac{1}{2}} + \mathrm{Drop}\!\Big(g_\mathrm{ff}^\ell \odot \mathrm{MLP}_\mathrm{ff}(\hat{X}^\ell_\mathrm{ff})\Big),
    \label{eq:03_methods/cond_block}
\end{align}

\noindent where $\mathrm{MLP}_\mathrm{ff}$ is a position-wise feed-forward network, and $\mathrm{Drop}$ denotes dropout regularization (active only during training). 

The final block output $X^{N_L}\in\mathbb{R}^{T\times d_e}$ is mapped back to the latent space using a time projection layer with target dimension is $h_d = d_\zeta$:

\begin{equation}
    \hat{\mathbf{Z}}_{t+1} = \mathrm{TP}_{d_\zeta}(X^{N_L}) \in \mathbb{R}^{T\times d_\zeta},
    \label{eq:03_methods/decoder}
\end{equation}

\noindent where the output sequence, $\hat{\mathbf{Z}}_{t+1} = [\hat{\boldsymbol{\zeta}}_{t-T+2}, \dots, \hat{\boldsymbol{\zeta}}_{t+1}]^\top$, consists of the latent states shifted one step into the future.  During training, this full shifted sequence is employed to compute the forecasting loss efficiently across all time steps within the window (teacher forcing). During inference, the network simply extracts the final element of this sequence, $\hat{\boldsymbol{\zeta}}_{t+1}$, which serves as the one-step-ahead prediction for the subsequent autoregressive rollout.

The model and its variants were trained on the latent representations of the training/validation dataset described in \S\ref{sec:02_dataset/preprocessing}, as encoded by the pre-trained $\beta$-VAE-GAN. Given the causal nature of the forecasting problem, the dataset was partitioned sequentially rather than randomly: the first 90\% of the temporal sequence was allocated for training and the final 10\% reserved for validation. Training was performed over 2,000 epochs using Adam optimiser \cite{kingma2014adam} on the same hardware as the $\beta$-VAE-GAN (a single NVIDIA A100 GPU), with optimisation governed by a cosine decay learning rate schedule with a linear warmup period. The validation loss was evaluated every 20 epochs to trigger early stopping if necessary.

During training, a scheduled sampling strategy, following Bengio \textit{et al.} \cite{Bengio2015ScheduledSampling}, was adopted to bridge the gap between training and inference conditions. In standard sequence training, the true encoded latent states are provided as inputs throughout; at inference time, however, these states are unavailable and the model must propagate its own predictions autoregressively. In chaotic systems, early prediction errors can amplify rapidly, a phenomenon known as exposure bias. To mitigate this, the adopted training process gradually transitions from a fully guided scheme towards one in which a dynamically increasing fraction of the input tokens is replaced by the model's own previous-step predictions, training the model to recover from accumulating errors. In the present work, the replacement probability is linearly ramped from 0\% to 50\% over the first half of the training epochs.

To further stabilise the long-horizon autoregressive rollouts, two complementary modifications to the loss function are incorporated alongside the scheduled sampling strategy. First, a time-weighted forecasting loss was employed: 

\begin{equation}
    \mathcal{L} = \sum_{\tau=1}^{T} \delta^{T-\tau} \, l\!\left(\hat{\boldsymbol{\zeta}}_{\tau}, \boldsymbol{\zeta}_{\tau}\right),
    \label{eq:03_methods/time_weighted_loss}
\end{equation}

\noindent where $l(\cdot)$ denotes the standard step-wise loss metric (e.g., mean squared error) and $\delta \in (0, 1]$ is a temporal weighting factor. By setting $\delta = 0.98$, the objective exponentially assigns higher penalty weights to predictions made near the end of the context window ($\tau \approx T$). Because the scheduled sampling strategy randomly swaps tokens throughout the input sequence, predictions at later time steps must be formulated using a history that contains a higher absolute accumulation of these predicted tokens. Heavily weighting the end of the window thus forces the model to effectively integrate the full context and recover from these errors, prioritizing stability when the context window is fully saturated.

Second, additive Gaussian noise is injected following the scheduled sampling step, acting as a regulariser by forcing the network to map perturbed latent states back to the true future trajectory. Each mixed input token $\boldsymbol{\zeta}^{\mathrm{ss}}_{\tau}$ is perturbed to yield the final network input $\boldsymbol{\zeta}^{\mathrm{mix}}_{\tau}$ as:

\begin{equation}
    \boldsymbol{\boldsymbol{\zeta}}^{\mathrm{mix}}_{\tau} = \boldsymbol{\zeta}^{\mathrm{ss}}_{\tau} + \boldsymbol{\epsilon}, \qquad \boldsymbol{\epsilon} \sim \mathcal{N}(0, \sigma_n^2 \mathbf{I}),
    \label{eq:03_methods/ss_noise}
\end{equation}

\noindent where $\sigma_n = 0.01$ is the standard deviation of the injected noise.

To explicitly isolate and quantify the impact of these two regularisation strategies, models trained with the time-weighted loss and additive noise will be evaluated as distinct variants in the subsequent results section (\S \ref{sec:04.2_latent_prediction}). These models will be denoted with an asterisk, specifically $\text{EA}^*$ and $\text{SA+EA}^*$, to distinguish them from their standard-training counterparts.

Table \ref{tab:3.2_hyper_transformer} summarises all architecture and training hyperparameters.

\begin{table}[htbp]
    \centering
    \small
    \begin{tabular}{@{} l l @{\hspace{2em}} l l @{}}
        \toprule
        \multicolumn{2}{@{}c}{\textbf{Architecture}} & \multicolumn{2}{c@{}}{\textbf{Training}} \\
        \cmidrule(r){1-2} \cmidrule(l){3-4} 
        Context length ($T$)       & 128                     & Hardware           & NVIDIA A100 \\
        Input emb. dim. ($d_e$)  & 64                      & Train / Val split  & 90\% / 10\% \\
        Sensor emb. dim. ($d_c$) & 18                      & Optimizer          & Adam \\
        Layers ($N_L$)             & 4                       & Batch size         & 32 \\
        Heads ($N_H$)              & 4                       & Total epochs       & 2000 \\
        FFN hidden dim. ($d_{\mathrm{ff}}$) & 256            & Warmup steps       & 500 \\
        \addlinespace
        Time projection            & Conv. Net.     & Initial LR         & $3\times10^{-4}$ \\
        Positional embed.          & Sinusoidal              & Final LR           & $2.5\times10^{-5}$ \\
        Conditioning               & AdaLN-Zero              & Schedule type      & Cosine Decay \\
        Dropout                    & 0.1                     & Loss ($\mathcal{L}$)& MSE \\
                                   &                         & Scheduled sampling & Enabled \\
        \bottomrule
    \end{tabular}
    \caption{Hyperparameters and training configuration for the sensor-conditioned transformer.}
    \label{tab:3.2_hyper_transformer}
\end{table}

\newpage


\section{Results}
\label{sec:04_results}
This section assesses the efficacy of the proposed reduced-order modelling framework using the test dataset detailed in \S\ref{sec:02_dataset/preprocessing}. To ensure physical interpretability, all quantitative evaluations are performed on flow fields re-dimensionalised to their original physical units, rather than on the normalised data used for training stability. The analysis is structured in three stages: first, baseline reconstruction fidelity is established by benchmarking the $\beta$-VAE-GAN against standard $\beta$-VAE and POD approaches (\S\ref{sec:04.1_compression_performance}); second, the predictability of the latent dynamics is examined via the sensor-conditioned Transformer (\S\ref{sec:04.2_latent_prediction}); and finally, the forecasting capabilities of the full framework are comprehensively evaluated (\S\ref{sec:04.3_full_framework}).

\subsection{Reconstruction Quality: Compression Fidelity and Disentanglement}
\label{sec:04.1_compression_performance}
The compression performance of the $\beta$-VAE-GAN is quantified on the test set using two distinct metrics: statistical fidelity and latent disentanglement.

The statistical fidelity is assessed by the model's ability to reconstruct the turbulent kinetic energy (TKE) of the flow. Let $k^+$ denote the TKE of the input fields in wall units, defined as:

\begin{equation}
    k^+ = \frac{1}{2} \left\langle\left(\overline{u^\prime u^\prime}^+ + \overline{v^\prime v^\prime}^+ + \overline{w^\prime w^\prime}^+\right)\right\rangle_{x^+,z^+},
\end{equation}

\noindent where the overbar $\overline{(\cdot)}$ denotes a temporal average and $\langle \cdot \rangle_{x^+,z^+}$ denotes spatial averaging over the domain. To quantify the global reconstruction accuracy, the reconstructed energy coefficient, $E_k^+$, is defined as:

\begin{equation}
    E_{k^+} = 1 - \frac{|{k^+} - \tilde{k}^+|}{k^+},
\label{eq:04.1_tke_reconstruction_coefficient}
\end{equation}

\noindent where $\tilde{k}^+$ is the TKE of the model reconstruction in wall units, and $|\cdot|$ represents the absolute value.

Second, the latent disentanglement is quantified via the mean absolute Pearson correlation coefficient of the off-diagonal latent variable pairs. The Pearson correlation coefficient between two scalar sequences $a$ and $b$ is defined as:

\begin{equation}
    \rho(a,b) = \frac{\mathrm{cov}(a,b)}{\sigma_a \, \sigma_b},
    \label{eq:04.1_pearson_general}
\end{equation}

\noindent where $\mathrm{cov}(\cdot,\cdot)$ denotes the covariance and $\sigma_{(\cdot)}$ the standard deviation. Employing this definition, the mean absolute off-diagonal Pearson correlation of the latent space is expressed as:

\begin{equation}
     \overline{r}_\zeta = \frac{2}{d_{\zeta}(d_{\zeta} - 1)} \sum_{i=1}^{d_\zeta} \sum_{j > i}^{d_\zeta} |\rho(\zeta_i, \zeta_j)|,
\label{eq:04.1_mean_correlation}
\end{equation}

\noindent where $d_{\zeta}$ is the dimension of the latent space. A lower value of $\overline{r}_\zeta$ indicates a more disentangled, orthogonal latent representation.

\begin{figure}[htbp]
  \centering
  \begin{minipage}[b]{0.45\textwidth}
    \centering
    \includegraphics[width=\linewidth]{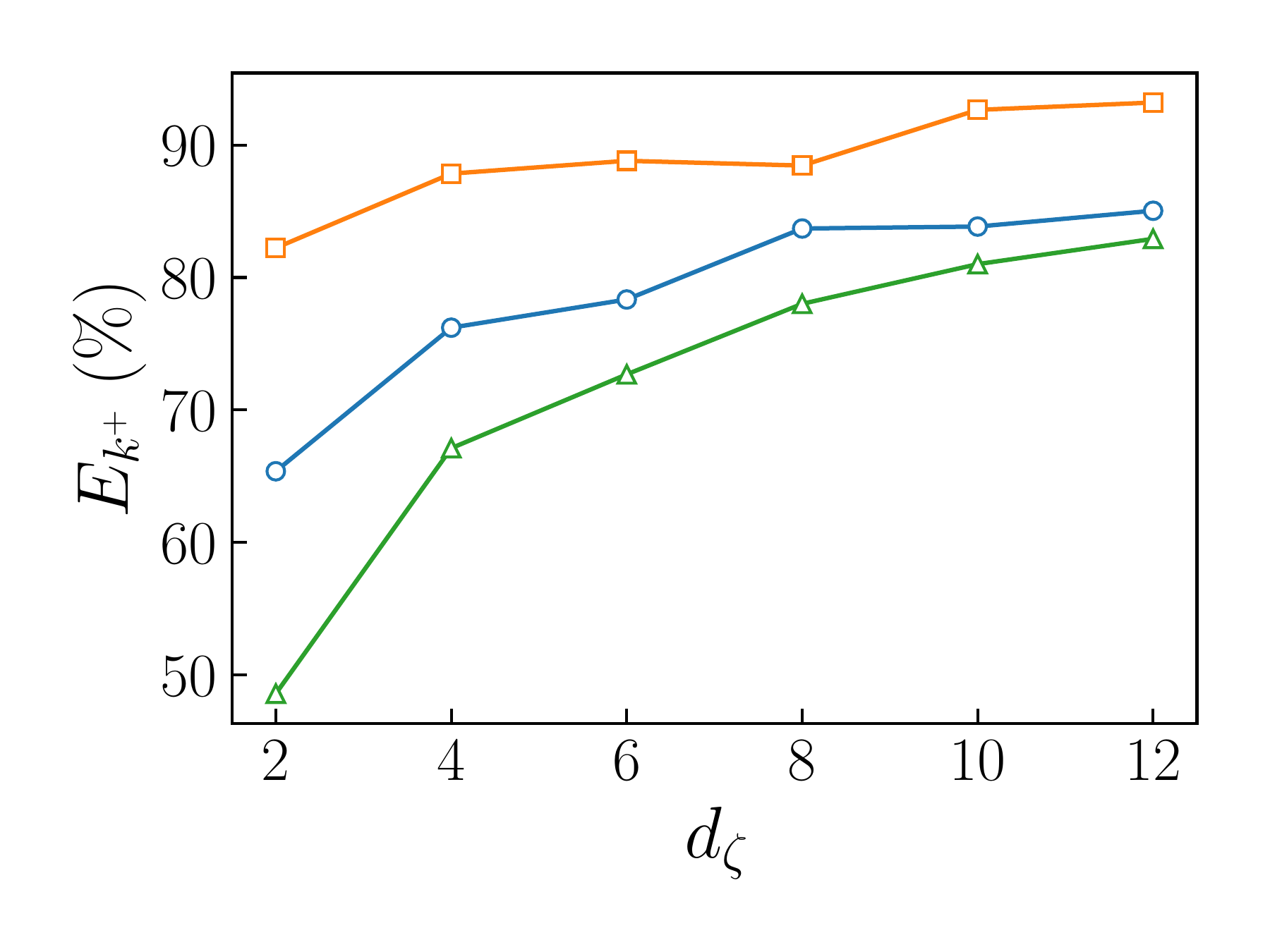}
  \end{minipage}
  \hfill
  \begin{minipage}[b]{0.45\textwidth}
    \centering
    \includegraphics[width=\linewidth]{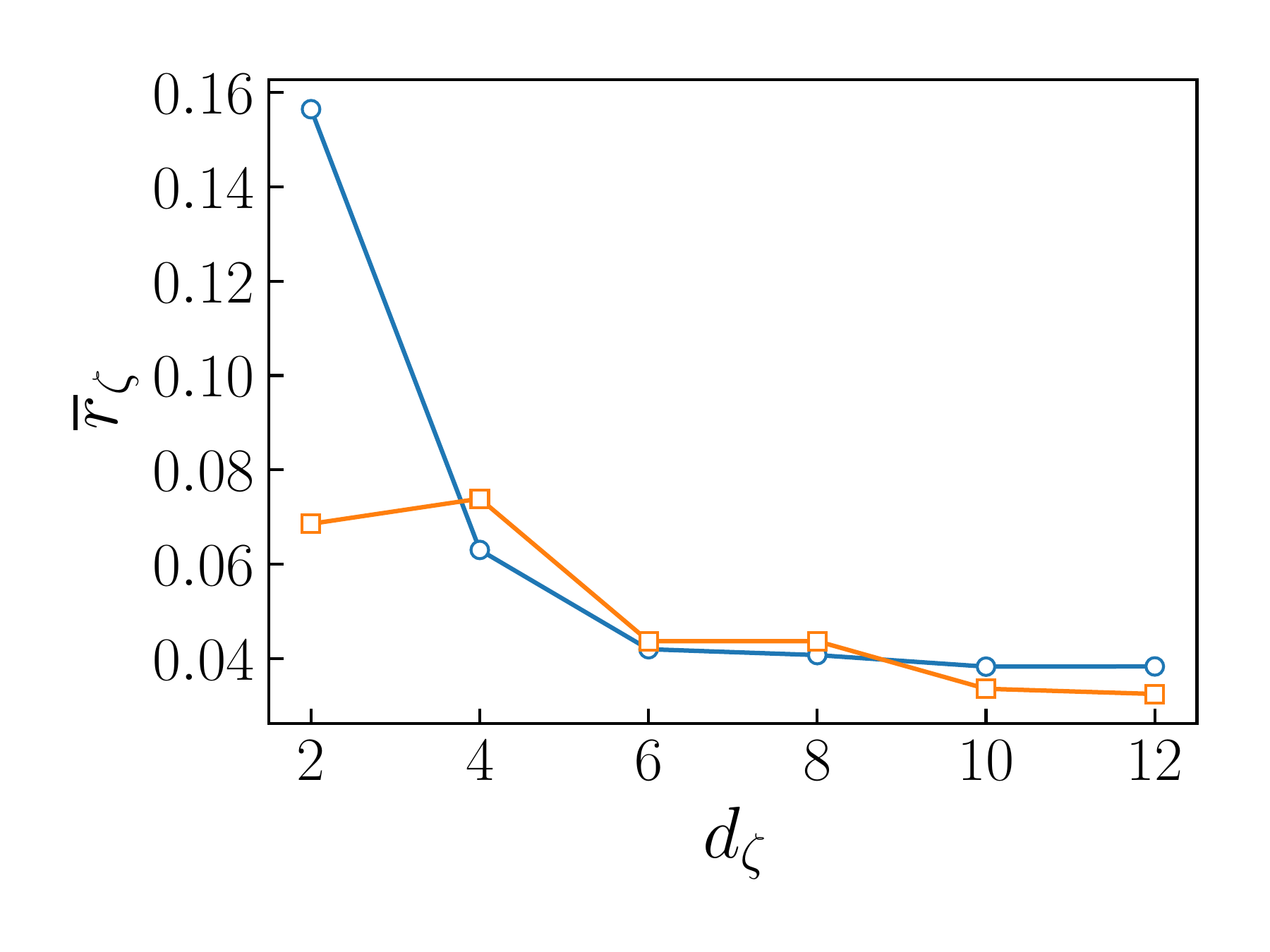}
  \end{minipage}
  \caption{Comparison of compression performance on the test set across different latent dimensions $d_{\zeta}$. Left: turbulent kinetic energy reconstruction coefficient $E_k$ (Eq.~\ref{eq:04.1_tke_reconstruction_coefficient}). Right: mean absolute correlation between latent variables $\overline{r}_\zeta$ (Eq.~\ref{eq:04.1_mean_correlation}). Colors denote: POD (green); $\beta$-VAE (blue); $\beta$-VAE-GAN (orange). Note that POD achieves zero correlation by construction (orthogonal modes) and is therefore omitted from the right panel.}
  \label{fig:04.1_compression_metrics}
\end{figure}

Figure~\ref{fig:04.1_compression_metrics} presents both metrics evaluated across varying latent dimensions $d_{\zeta} \in \{2, 4, 6, 8, 10, 12\}$, comparing $\beta$-VAE-GAN (orange line) against standard $\beta$-VAE (blue) and POD (green) baselines. The non-linear methods consistently outperform POD across all latent dimensions. Moreover, the proposed $\beta$-VAE-GAN formulation achieves better performance compared to the standard $\beta$-VAE, particularly at very low latent dimensions, demonstrating the benefit of the adversarial training component. For instance, at $d_{\zeta} = 2$, POD reconstructs less than 60\% of the test set TKE and the $\beta$-VAE less than 70\%, while the $\beta$-VAE-GAN reconstructs over 80\%. As expected, the performance gap between methods diminishes as the latent dimension increases. Importantly, the improved statistical fidelity of the $\beta$-VAE-GAN does not compromise latent disentanglement; it achieves comparable or higher decorrelation than the standard $\beta$-VAE.

To further analyse the quality of the spatial reconstruction, the two-point spatial autocorrelation of the streamwise velocity fluctuations is evaluated. For a spatial separation $r_z^+$ along the spanwise direction $z^+$, the normalised autocorrelation $R_{u^{\prime}u^{\prime}}^+(r_z^+)$ in wall units is defined as a specific application of Eq.~(\ref{eq:04.1_pearson_general}), evaluated between spatially shifted realisations of the velocity field:

\begin{equation}
    R_{u^{\prime}u^{\prime}}^+(r_z^+) = \frac{\langle \overline{u^{\prime +}(z^+) \, u^{\prime +}(z^+ + r_z^+)} \rangle_{x^+}}{\langle \overline{(u^{\prime +}(z^+))^2} \rangle_{x^+}},
\label{eq:04.1_autocorrelation}
\end{equation}

\begin{figure}[htbp]
    \centering
    \begin{minipage}[b]{0.48\textwidth}
        \centering
        \includegraphics[width=\linewidth]{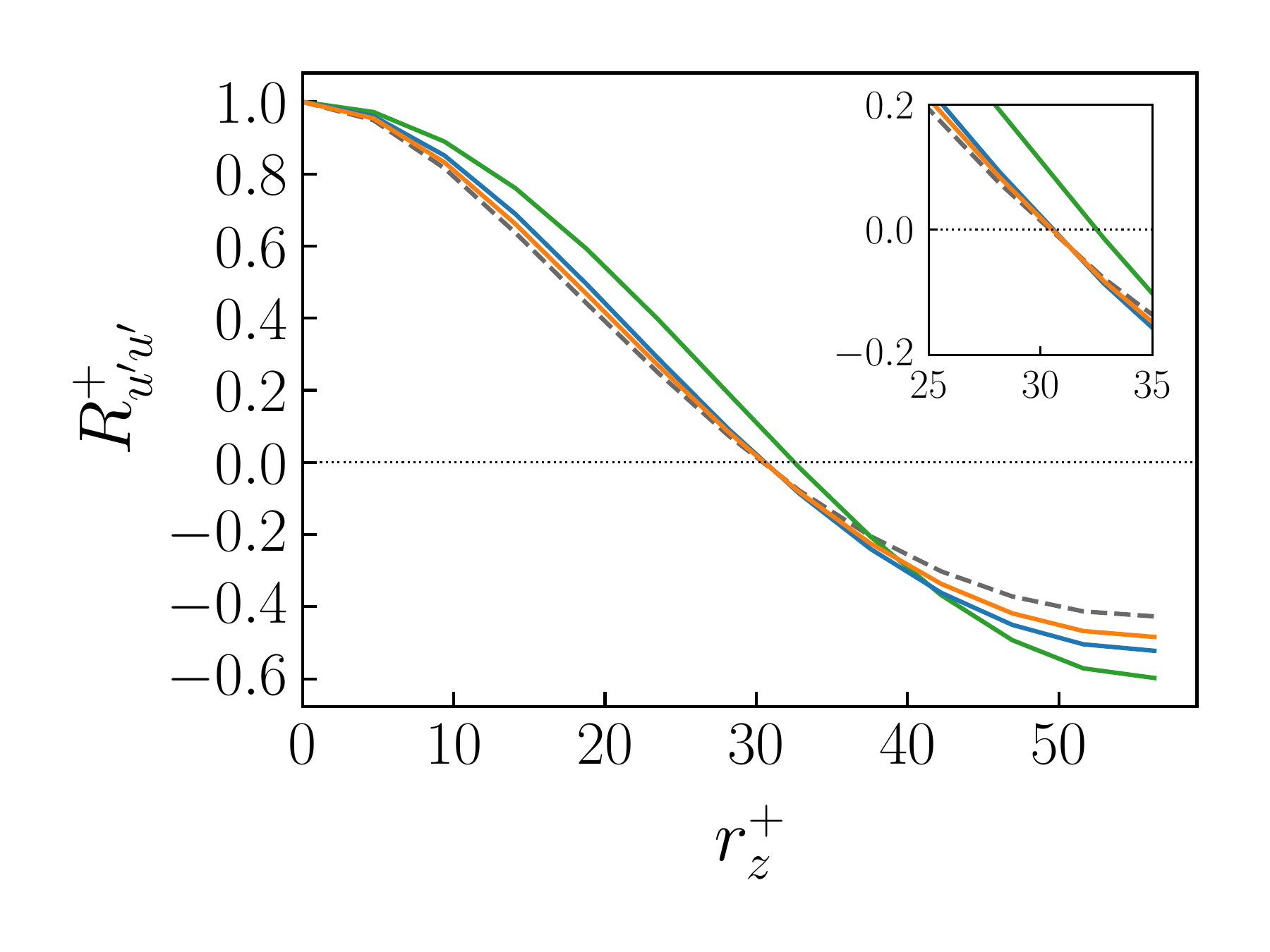}
    \end{minipage}
    \hfill
    \begin{minipage}[b]{0.48\textwidth}
        \centering
        \includegraphics[width=\linewidth]{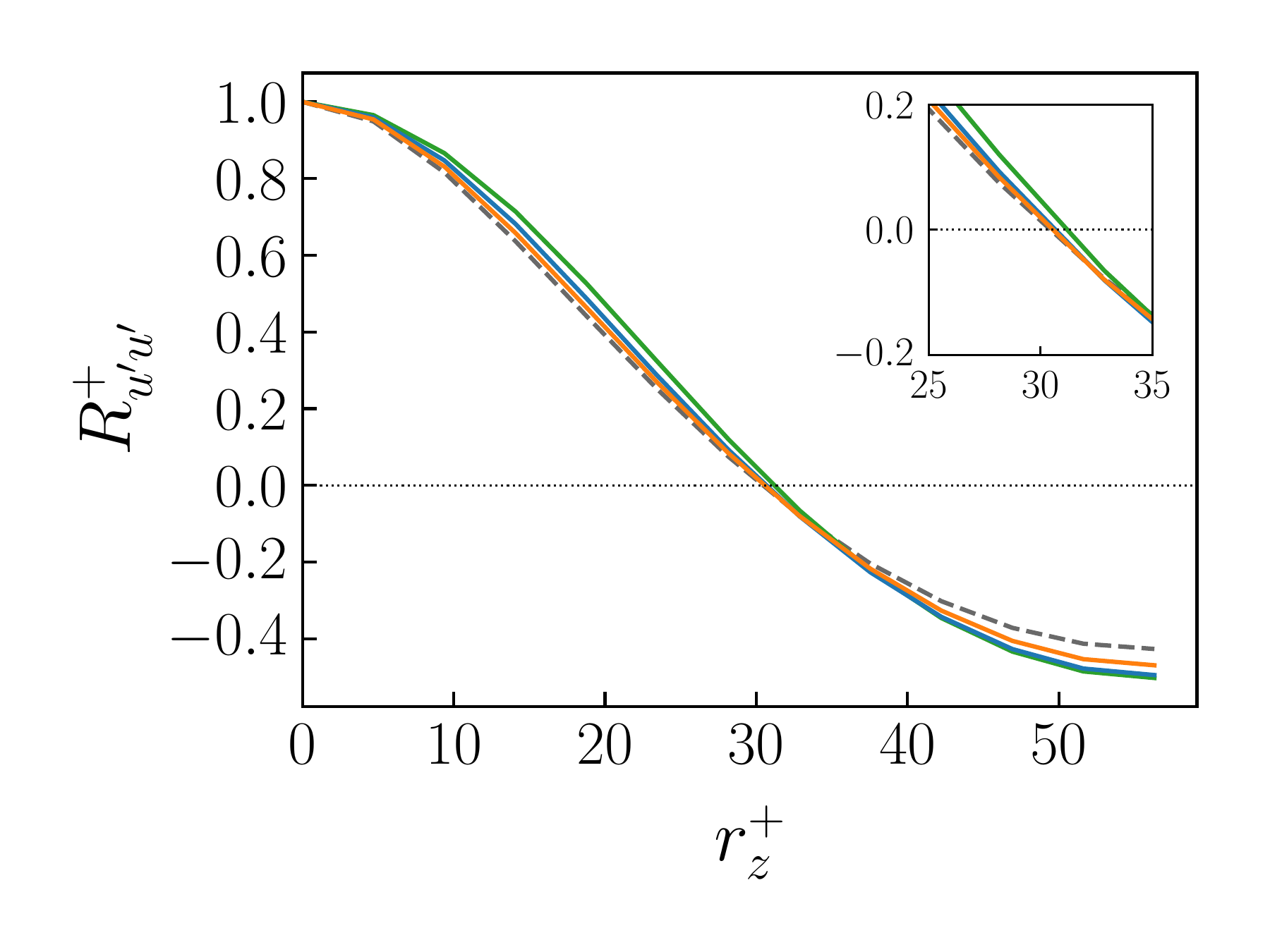}
    \end{minipage}
    \caption{Two-point spatial autocorrelation of the streamwise velocity fluctuations, $R_{u^{\prime}u^{\prime}}^+$, compared across latent space dimensions $d_{\zeta} = 4$ (left) and $d_{\zeta} = 12$ (right). The inset in the upper right of each panel provides a magnified view of the region near the zero-crossing. Profiles correspond to the DNS reference (gray), POD (green), $\beta$-VAE (blue), and $\beta$-VAE-GAN (orange).}
    \label{fig:04.1_1d_correlation}
\end{figure}

Figure~\ref{fig:04.1_1d_correlation} conveys the autocorrelation profiles for latent dimensions $d_{\zeta} = 4$ (left) and $d_{\zeta} = 12$ (right), computed with respect to a reference spanwise location of $z^+ = 0$. The $\beta$-VAE-GAN profile (orange) demonstrates closer agreement with the DNS reference (grey) than the baseline models. Notably, the curves are nearly coincident up to the zero-crossing, indicating that the framework successfully preserves the spanwise integral length scale of the flow, even when heavily compressed to a latent dimension of just $d_{\zeta} = 4$. Beyond the zero-crossing, in the region of negative correlation, the $\beta$-VAE-GAN continues to track the DNS most accurately, although a slightly larger discrepancy is observed here compared to the positive correlation regime.

Based on these results, the $\beta$-VAE-GAN model with latent dimension $d_\zeta = 4$ is selected for the remainder of the study, as it provides a compromise between preserving the turbulent kinetic energy of the flow and minimising the latent dimensionality for the Transformer.

\newpage
\subsection{Temporal Forecasting in Latent Space}
\label{sec:04.2_latent_prediction}

The second stage of the evaluation assesses the Transformer's ability to forecast the evolution of the latent state learnt by the $\beta$-VAE-GAN. To this end, the similarity between the predicted latent trajectories, denoted as the set of states $\{\hat{\boldsymbol{\zeta}}_t\}$, and the reference encoded trajectories, $\{\boldsymbol{\zeta}_t\}$, is quantified using three complementary families of metrics. The first family comprises pointwise accuracy metrics, which measure average trajectory deviations. The second captures temporal alignment through correlation analysis. The third characterises the geometric structure of the predicted manifold relative to the reference.

Pointwise accuracy is assessed using the Normalised Root Mean Square Error (NRMSE) and the Normalised Mean Absolute Error (NMAE). Based on the global average, denoted by the operator $\langle \cdot \rangle$, these are defined as:

\begin{equation}
    \text{NRMSE} = \sqrt{ \frac{\langle (\boldsymbol{\zeta}_t - \hat{\boldsymbol{\zeta}}_t) \odot (\boldsymbol{\zeta}_t - \hat{\boldsymbol{\zeta}}_t) \rangle}{\langle \boldsymbol{\zeta}_t \odot \boldsymbol{\zeta}_t \rangle} }, \qquad 
    \text{NMAE} = \frac{\langle |\boldsymbol{\zeta}_t - \hat{\boldsymbol{\zeta}}_t| \rangle}{\langle |\boldsymbol{\zeta}_t| \rangle} \times 100\%,
\label{eq:04.2_errors}
\end{equation}

\noindent where the errors are rendered dimensionless by normalising against the root-mean-square magnitude and the mean absolute magnitude of the reference latent variables, respectively. Temporal alignment is subsequently assessed via the Pearson correlation coefficient (Eq.~\ref{eq:04.1_pearson_general}), which is here evaluated between the predicted and reference latent trajectories::

\begin{equation}
    \rho = \frac{\text{cov}(\boldsymbol{\zeta}_t, \hat{\boldsymbol{\zeta}}_t)}{\sigma_{\boldsymbol{\zeta}} \sigma_{\hat{\boldsymbol{\zeta}}}}.
\label{eq:04.2_pearson}
\end{equation}

The structural alignment between the predicted and reference manifolds is evaluated using the Hausdorff distance ($d_H$), which defines the `worst-case' geometric discrepancy between the predicted and target trajectories:

\begin{equation}
    d_H = \max \left\{ \sup_{\boldsymbol{\zeta} \in \{\boldsymbol{\zeta}_t\}} \inf_{\hat{\boldsymbol{\zeta}} \in \{\hat{\boldsymbol{\zeta}}_t\}} \|\boldsymbol{\zeta} - \hat{\boldsymbol{\zeta}}\|_2, \sup_{\hat{\boldsymbol{\zeta}} \in \{\hat{\boldsymbol{\zeta}}_t\}} \inf_{\boldsymbol{\zeta} \in \{\boldsymbol{\zeta}_t\}} \|\boldsymbol{\zeta} - \hat{\boldsymbol{\zeta}}\|_2 \right\}.
\label{eq:04.2_hausdorff}
\end{equation}

Table~\ref{tab:comprehensive_metrics} summarises these performance metrics across all evaluated temporal models. The results indicate that the Easy Attention variant equipped with the proposed training modifications (time-weighted loss and additive Gaussian noise injection) outperforms the alternative attention mechanisms and standard training across nearly all statistical and geometric criteria. Notably, the $\text{EA}^*$ model achieves the lowest overall trajectory deviation (NRMSE of $0.7932$) whilst simultaneously maintaining strong structural alignment with the reference manifold ($d_H = 2.0380$ and $\rho = 0.6337$).

Although the hybrid $\text{SA+EA}^*$ model exhibits marginally better absolute error (NMAE), the $\text{EA}^*$ architecture is selected since it achieves the best results across the remaining criteria while eliminating the $\mathcal{O}(d_e T^2)$ computational overhead of standard self-attention (see \S\,\ref{sec:03.2_cond-transf}). Consequently, the subsequent analysis focuses on a detailed analysis of the forecasting quality for the $\text{EA}^*$ architecture and provides a physical interpretation of the learnt latent dynamics.

\begin{table}[htbp]
    \centering
    \small
    \begin{tabular}{l c c c c c c}
        \toprule
        \textbf{Model} & \textbf{NRMSE} $\downarrow$ & \textbf{NMAE} (\%) $\downarrow$ & $\boldsymbol{\rho}$ $\uparrow$ & $\boldsymbol{d_H}$ $\downarrow$  \\
        \midrule
        SA         & 1.1353 & 108.47 & 0.3317 & 2.5066 \\
        EA         & 0.9228 & 90.46  & 0.5160 & 2.6964 \\
        EA$^*$     & \textbf{0.7932} & 75.44  & \textbf{0.6337} & \textbf{2.0380} \\
        SA+EA      & 0.8652 & 81.76  & 0.5721 & 2.5099 \\
        SA+EA$^*$  & 0.7977 & \textbf{73.84}  & 0.6015 & 2.4109 \\
        \bottomrule
    \end{tabular}
    \caption{Quantitative comparison of model performance metrics, as defined in equations (\ref{eq:04.2_errors})--(\ref{eq:04.2_hausdorff}). Arrows ($\uparrow$, $\downarrow$) indicate whether higher or lower values correspond to better performance. The evaluated attention mechanisms are abbreviated as SA (Self-Attention), EA (Easy-Attention), and SA+EA (parallel integration). Models marked with an asterisk ($*$) incorporate the time-weighted loss and noise-injection training strategies detailed in \S\,\ref{sec:03.2_cond-transf}.}
    \label{tab:comprehensive_metrics}
\end{table}

Before examining the prediction quality, it is instructive to establish a consistent ordering of the latent variables. Unlike POD, which inherently ranks modes by kinetic energy, the latent dimensions of the $\beta$-VAE-GAN carry no \textit{a priori} ordering, and their indexing may vary across training runs. Consequently, the latent variables are sorted during post-processing according to their latent activity \cite{burda2015importance}, a variance-based measure defined as:

\begin{equation}
    A_j = \frac{1}{N_t} \sum_{i=1}^{N_t} \left( \zeta_{i,j} - \mu_j \right)^2,
\label{eq:04.2_latent_activity}
\end{equation}

\noindent where $\mu_j$ is the temporal mean of the $j$-th latent variable. Dimensions encoding active information about the flow exhibit large variance (high activity) in response to changes in the observations, whilst uninformative dimensions remain approximately static. The variables are hereafter re-indexed in descending order of activity ($A_1 \ge A_2 \ge A_3 \ge A_4$), ensuring that $\zeta_1$ corresponds to the dimension with the highest variance across the test dataset.

\begin{figure}[htbp]
    \centering
    \begin{minipage}[c]{\textwidth}
        \centering
        \includegraphics[width=0.55\linewidth]{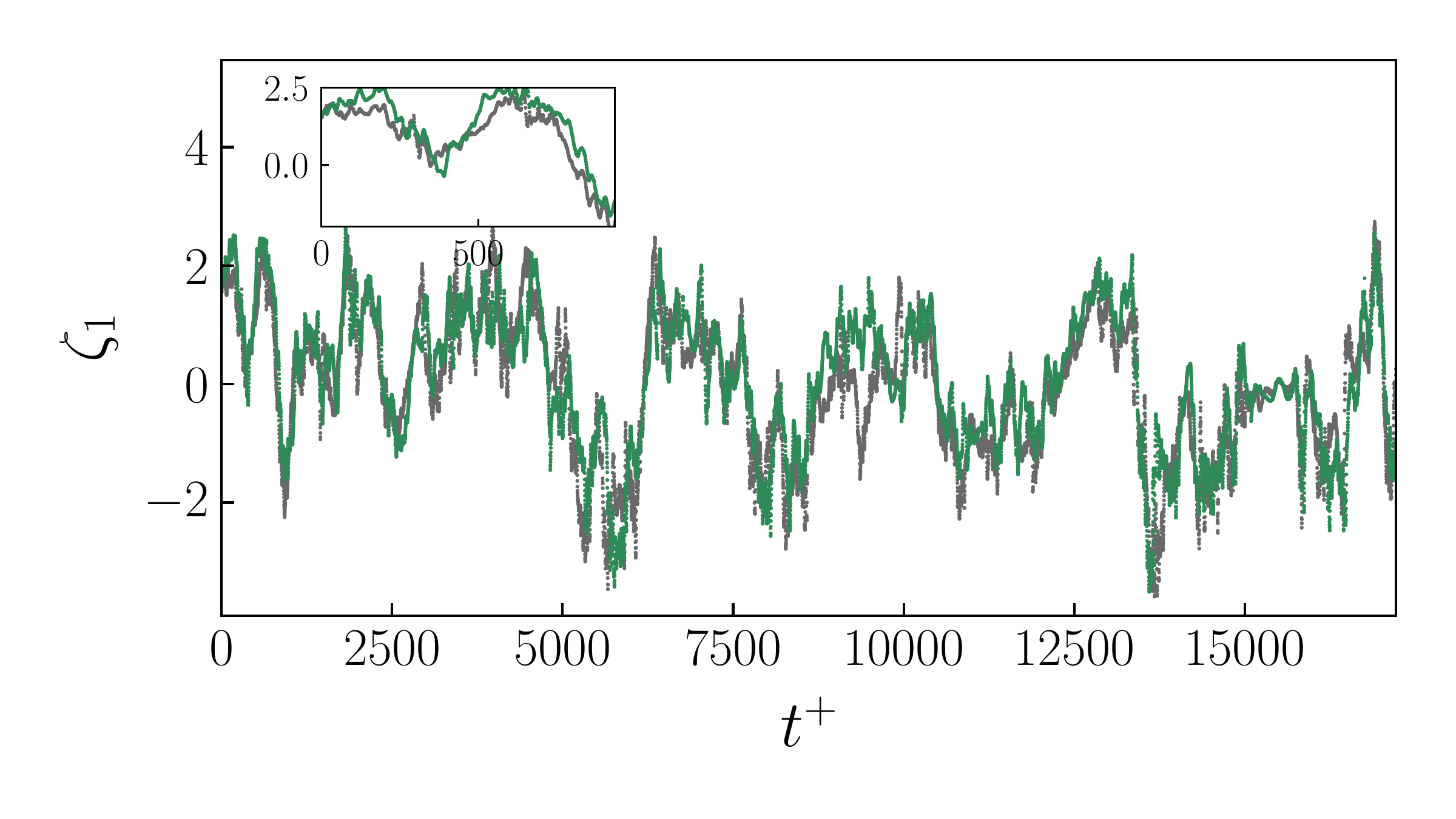}
        \includegraphics[width=0.3\linewidth]{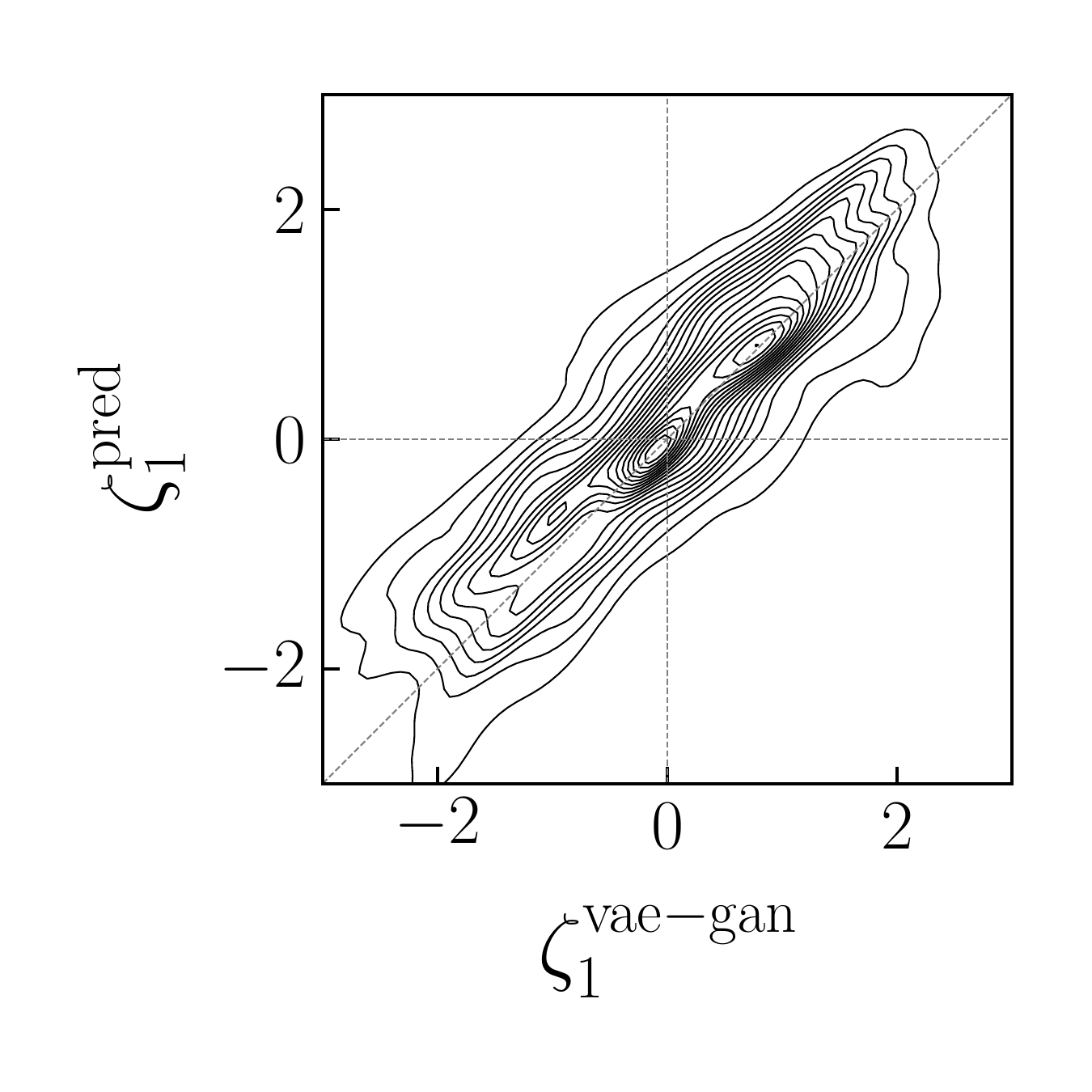}
    \end{minipage}
    \hfill
    \begin{minipage}[c]{\textwidth}
        \centering
        \includegraphics[width=0.58\linewidth]{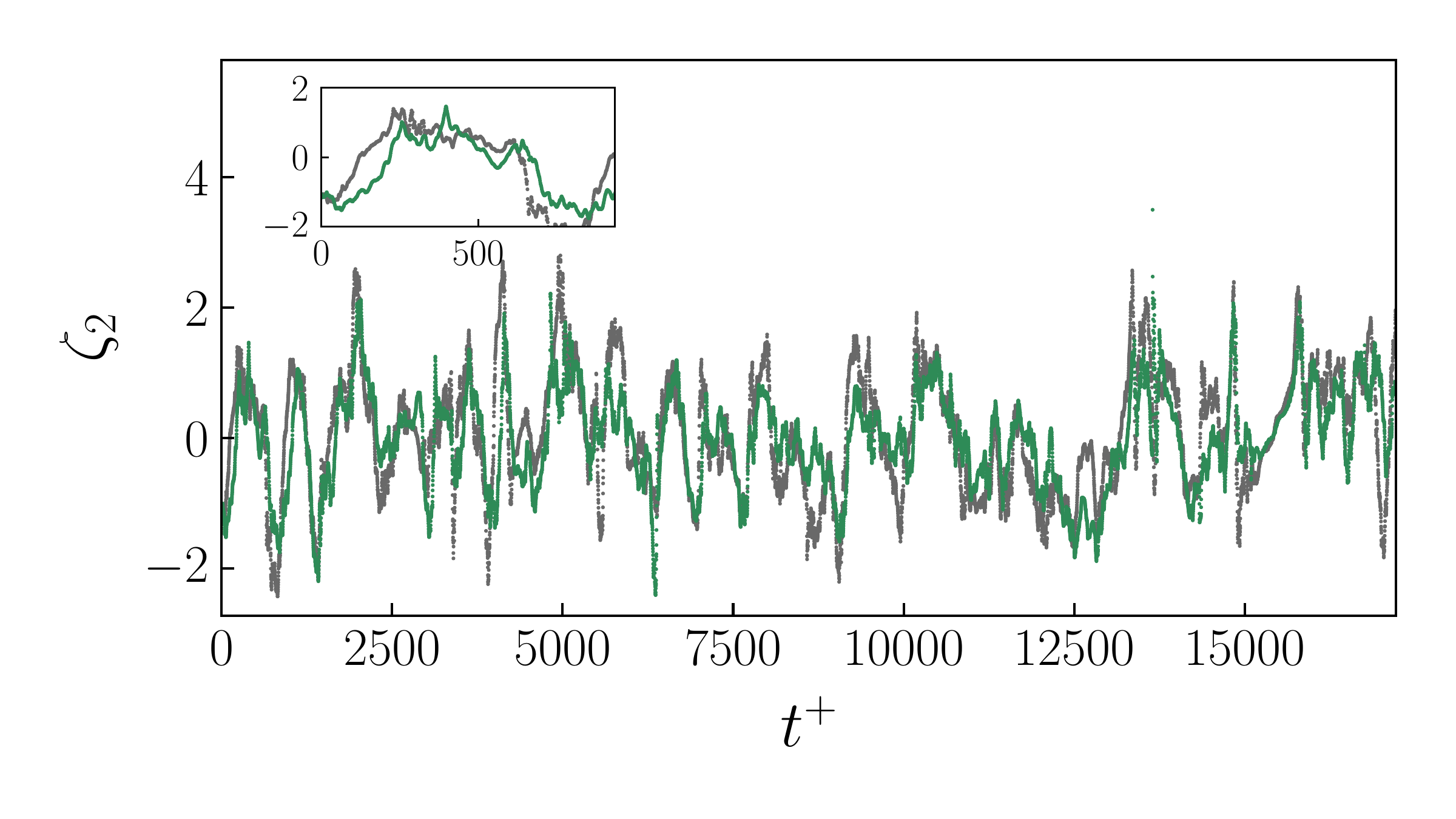}
        \includegraphics[width=0.33\linewidth]{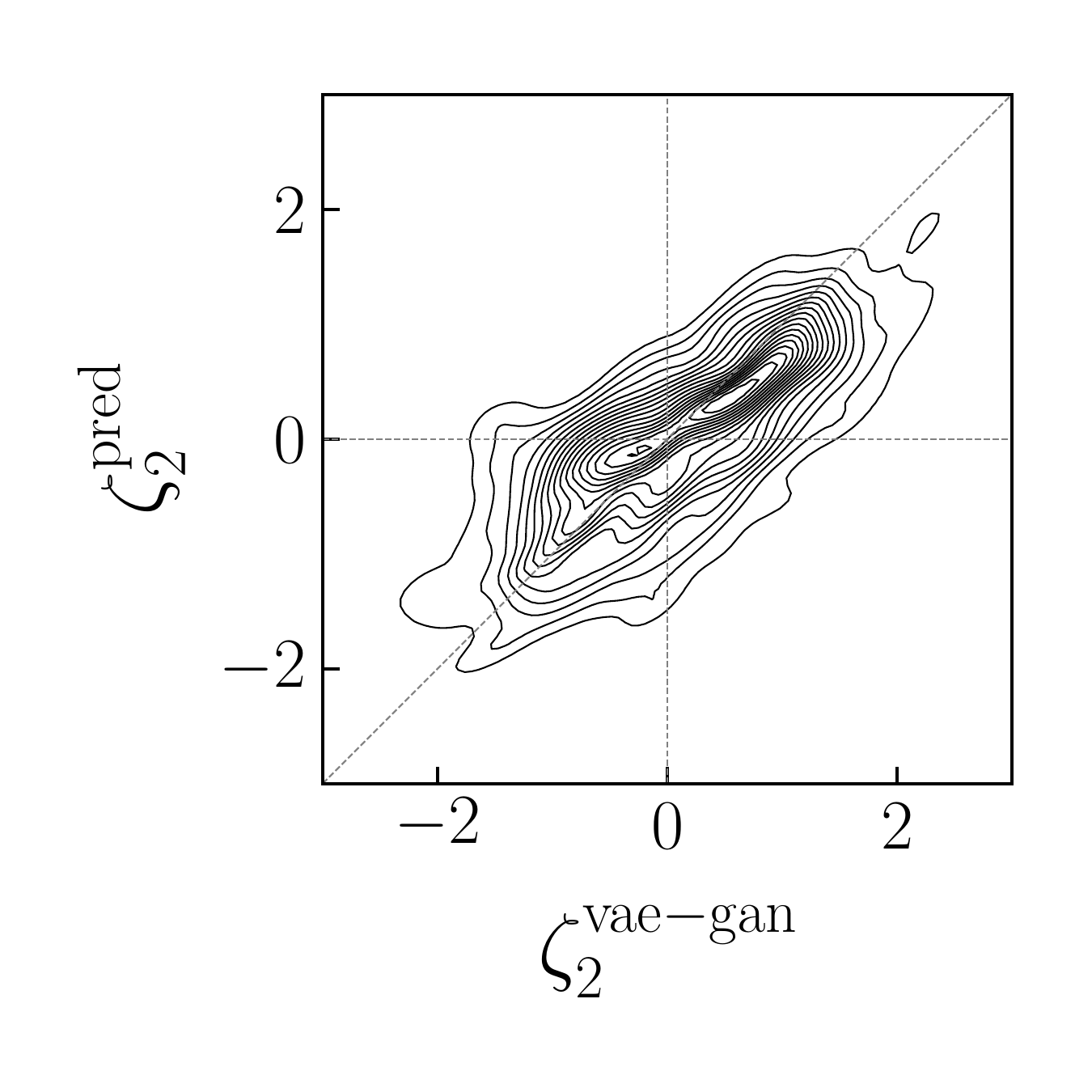}
    \end{minipage}
    \hfill
    \begin{minipage}[c]{\textwidth}
        \centering
        \includegraphics[width=0.58\linewidth]{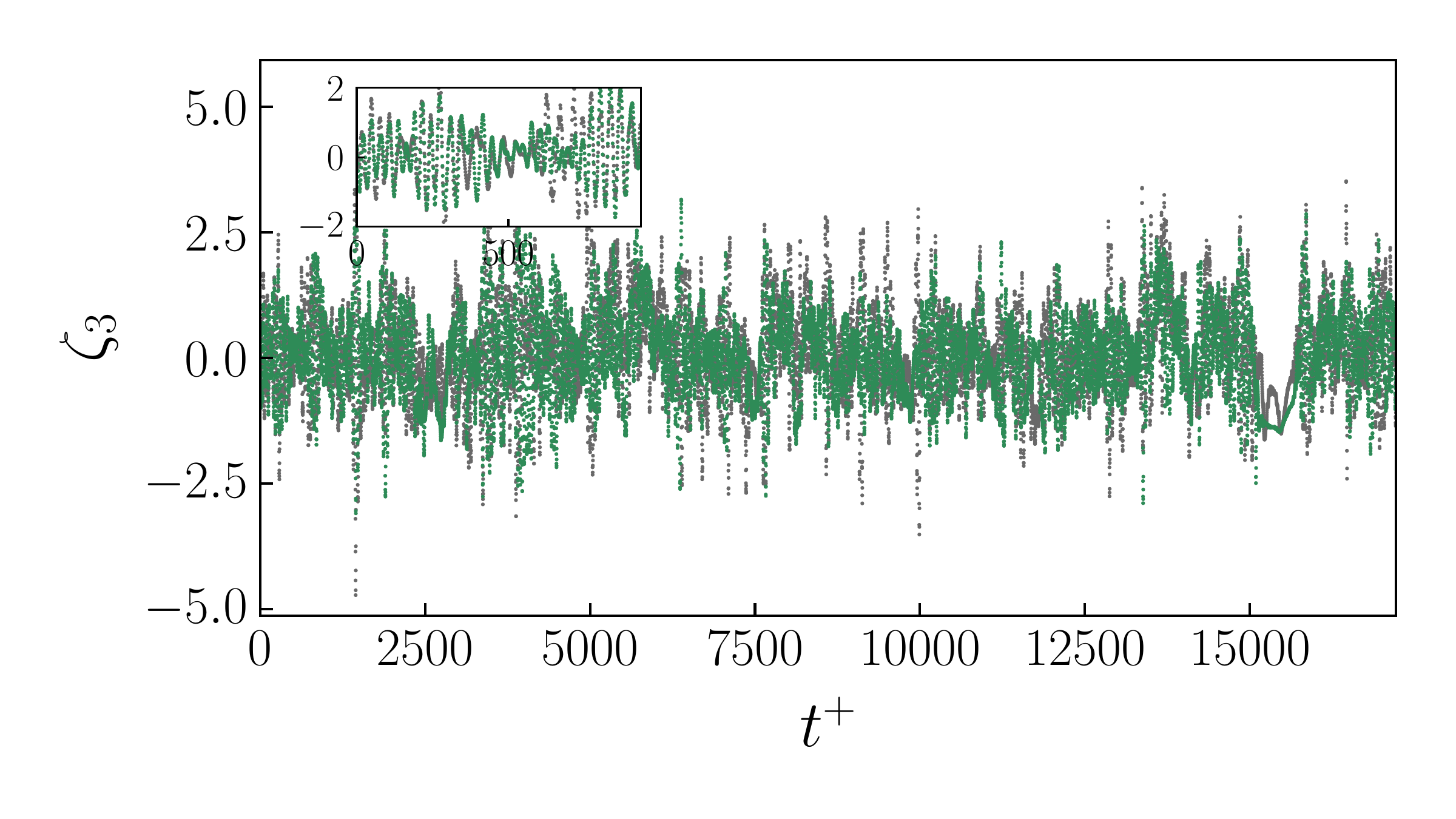}
        \includegraphics[width=0.33\linewidth]{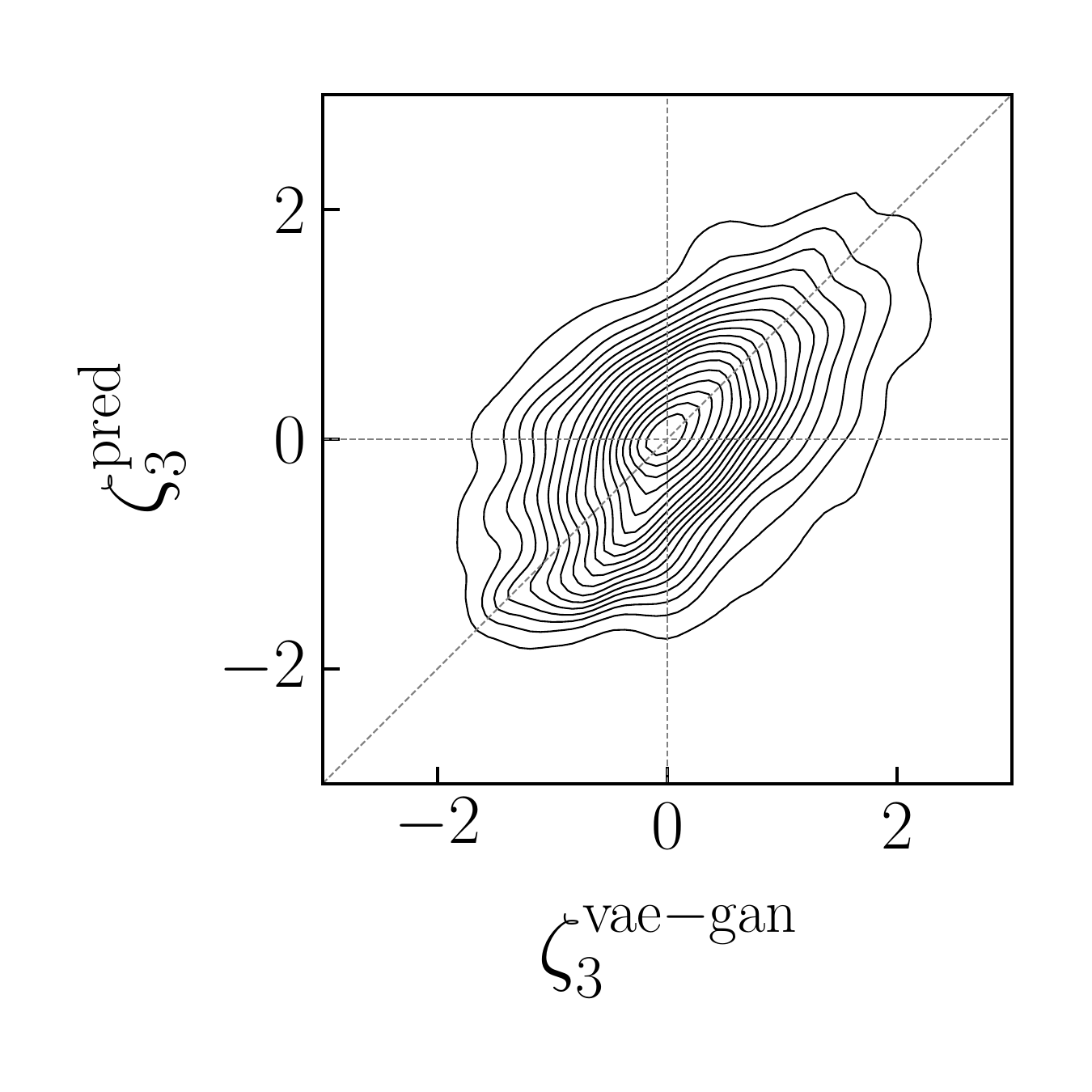}
    \end{minipage}
    \hfill
    \begin{minipage}[c]{\textwidth}
        \centering
        \includegraphics[width=0.58\linewidth]{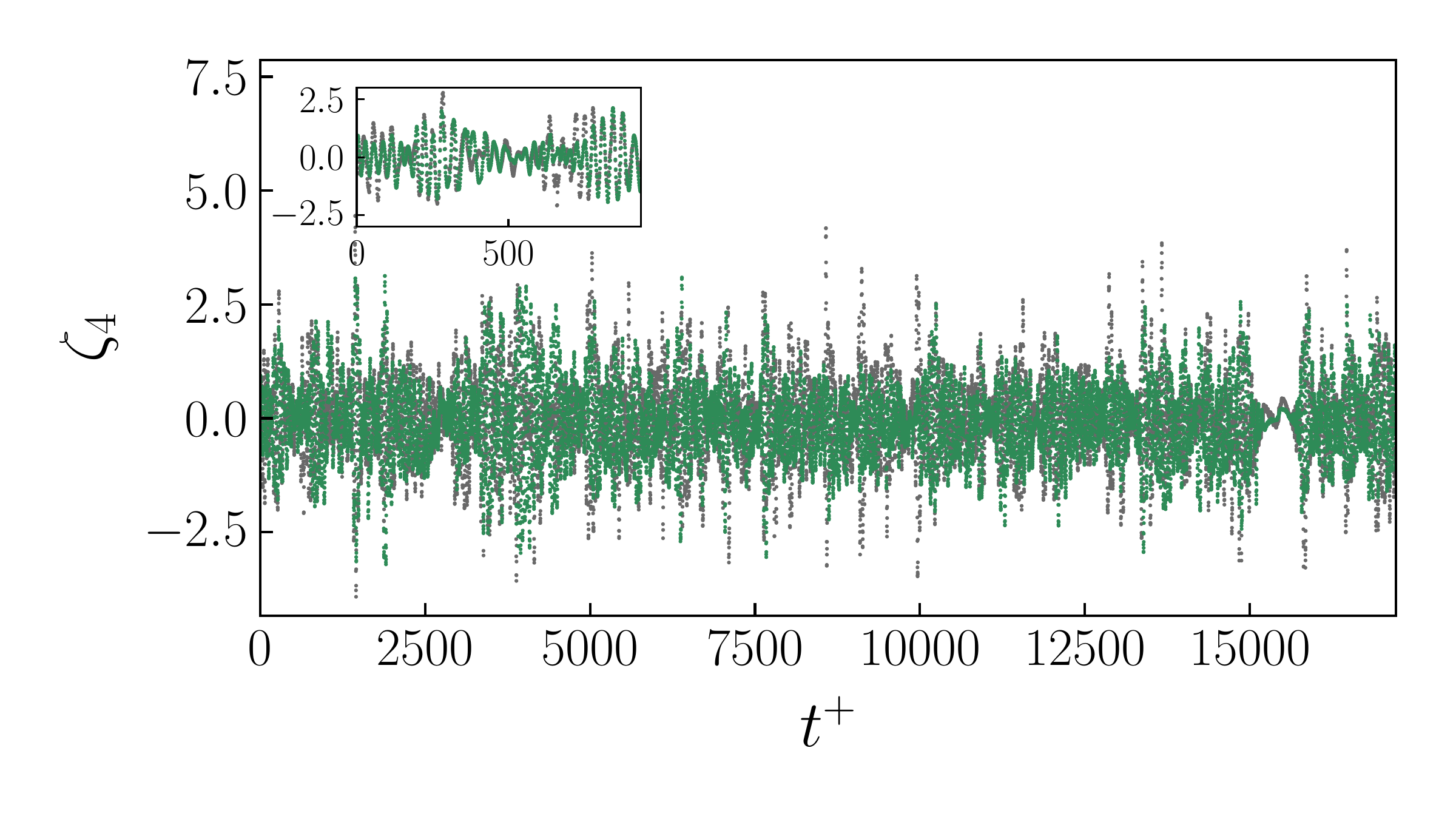}
        \includegraphics[width=0.33\linewidth]{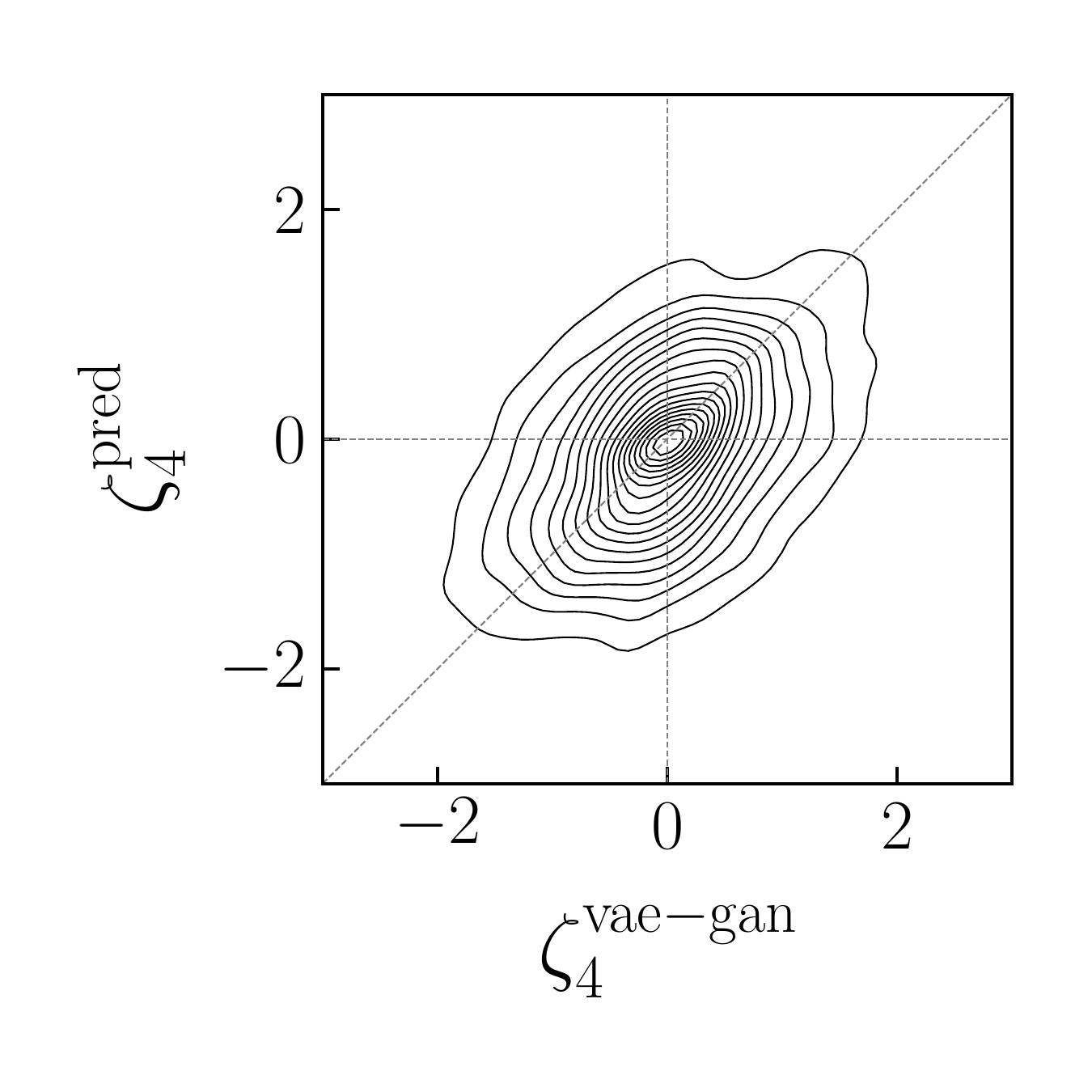}
    \end{minipage}
    \caption{Temporal evolution of the four latent variables over an extended autoregressive horizon. Left panels: time-series comparison between the predicted (sea-green) and reference (gray) trajectories. Right panels: corresponding joint probability density functions illustrating temporal alignment and dispersion error.}
    \label{fig:04.2_time_series_and_joint_pdf}
\end{figure}

The temporal fidelity of the predictions is examined in Figure~\ref{fig:04.2_time_series_and_joint_pdf}, which presents, for each latent variable, the time series of the predicted and reference signals (left panels) alongside the joint PDF of $\hat{\zeta}_{t,j}$ against $\zeta_{t,j}$ (right panels). The latter quantifies point-wise temporal alignment: a perfectly accurate forecast concentrates all probability mass along the main diagonal, whilst dispersion away from the diagonal indicates forecasting error. The variables $\zeta_1$ and $\zeta_2$ exhibit irregular, low-frequency trajectories, whilst $\zeta_3$ and $\zeta_4$ display higher-frequency, amplitude-modulated behaviour, consistent with their spectral content and statistical distributions analysed subsequently. Despite an autoregressive horizon of approximately $17{,}288\,t^+$ initialised from a context window of only ${\approx}\,128\,t^+$, the predicted trajectories maintain visual alignment with the reference. The joint PDFs for the low-frequency dimensions $\zeta_1$ and $\zeta_2$ are concentrated along the diagonal, confirming that their slowly evolving dynamics are reproduced with minimal dispersion error. In contrast, the joint PDFs for the high-frequency dimensions $\zeta_3$ and $\zeta_4$ exhibit broader scatter, consistent with the intrinsically harder task of predicting rapid fluctuations over long horizons. Notably, the autoregressive rollout successfully tracks localised temporal phenomena, such as the quiescent latent period at ${\approx}\,15{,}000\,t^+$, which is examined in detail at the end of \S\,\ref{sec:04.3_full_framework}.

\begin{figure}[htbp]
	\centering
	\begin{minipage}[c]{0.85\textwidth}
		\centering
		\includegraphics[width=0.48\linewidth]{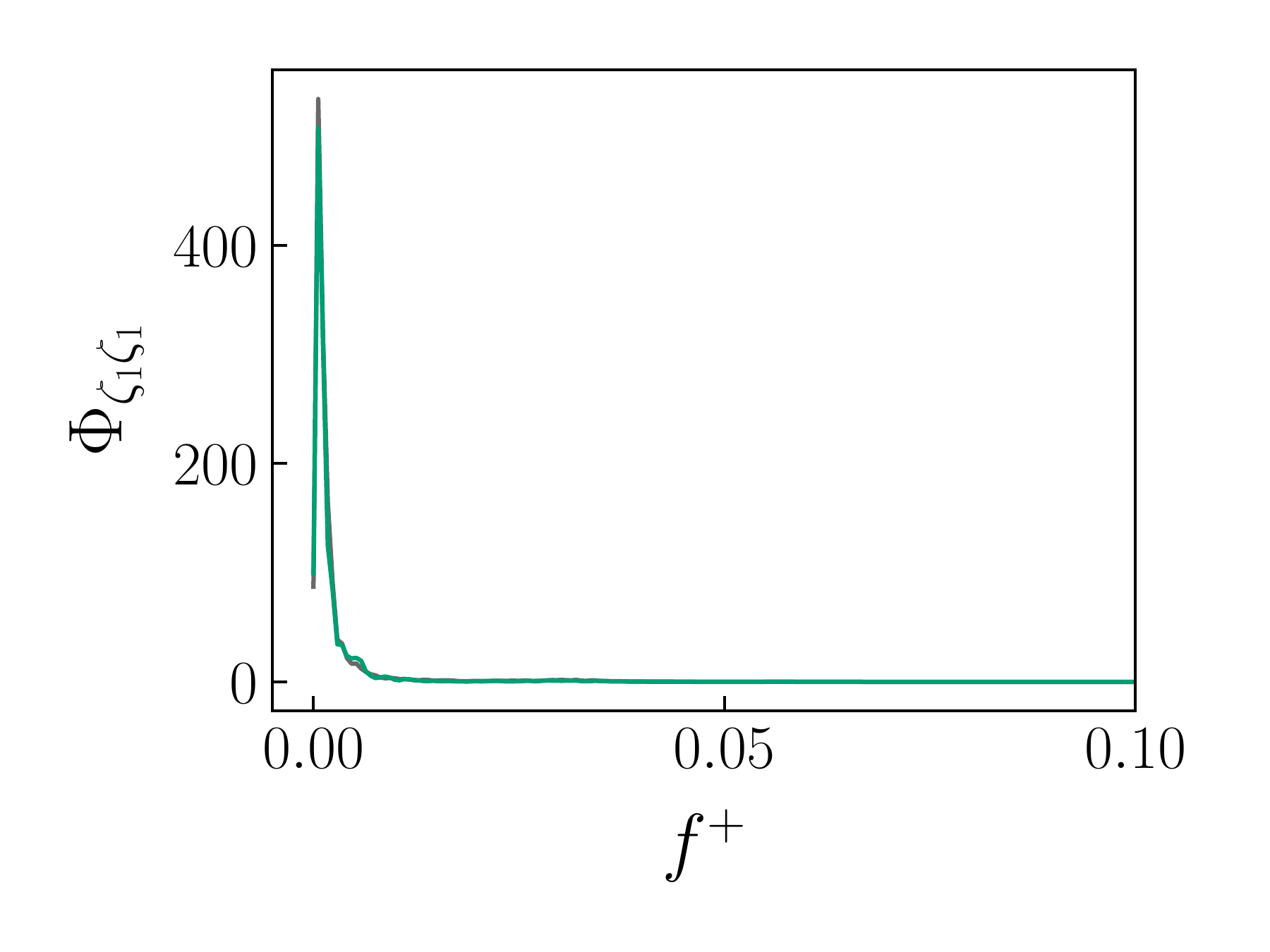}
        \includegraphics[width=0.48\linewidth]{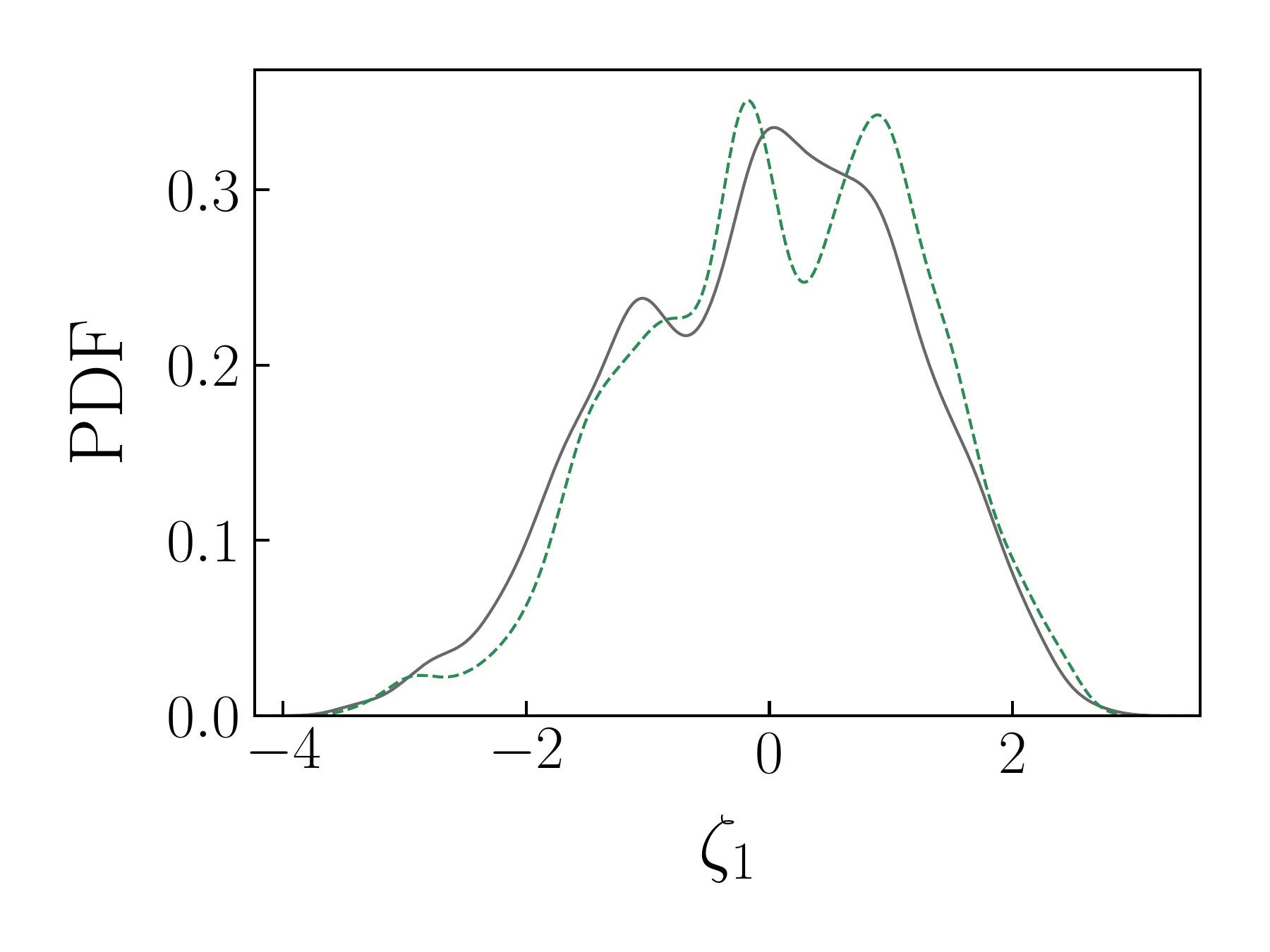}
	\end{minipage}
	\hfill
	\begin{minipage}[c]{0.85\textwidth}
		\centering
		\includegraphics[width=0.48\linewidth]{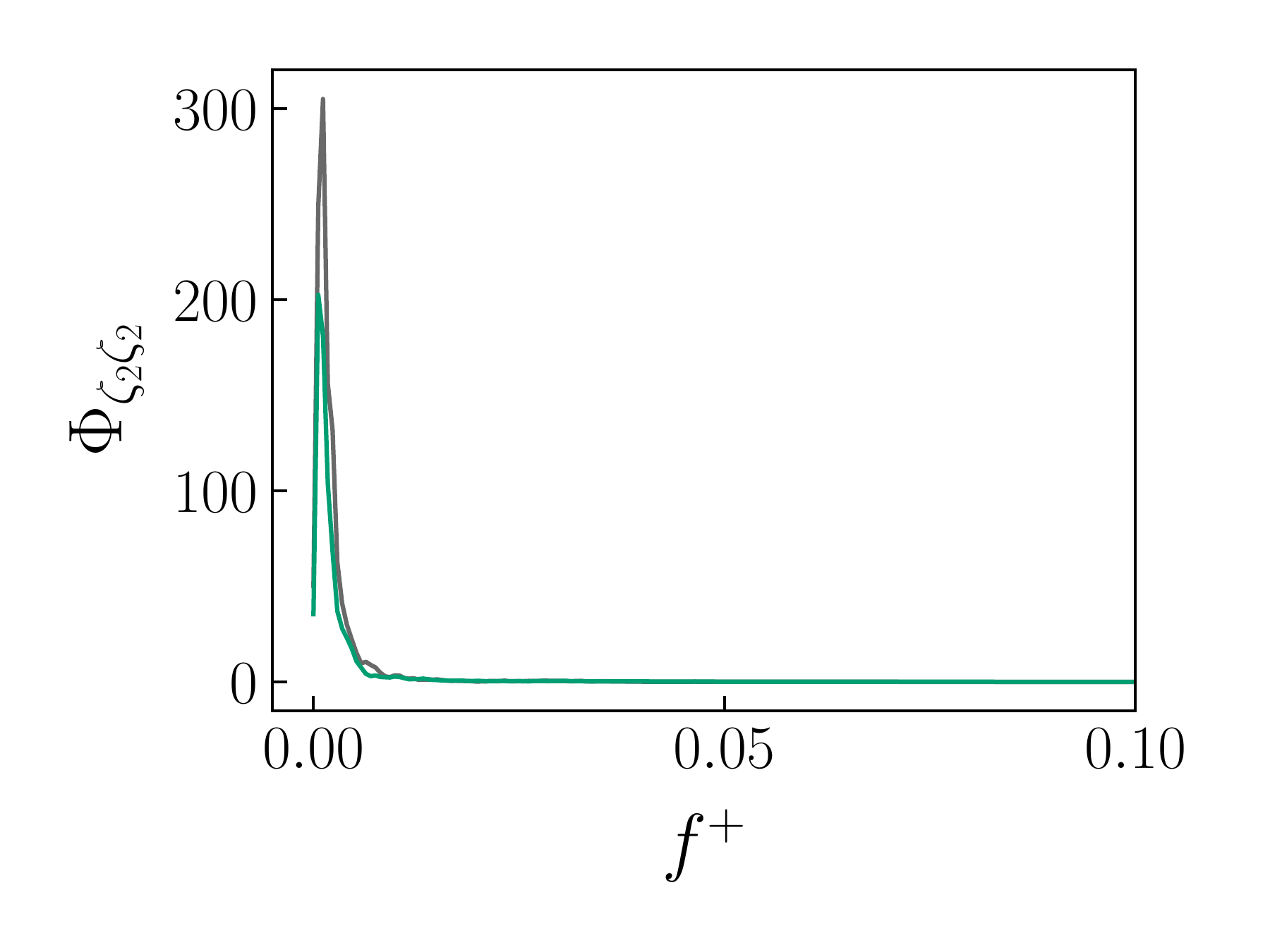}
        \includegraphics[width=0.48\linewidth]{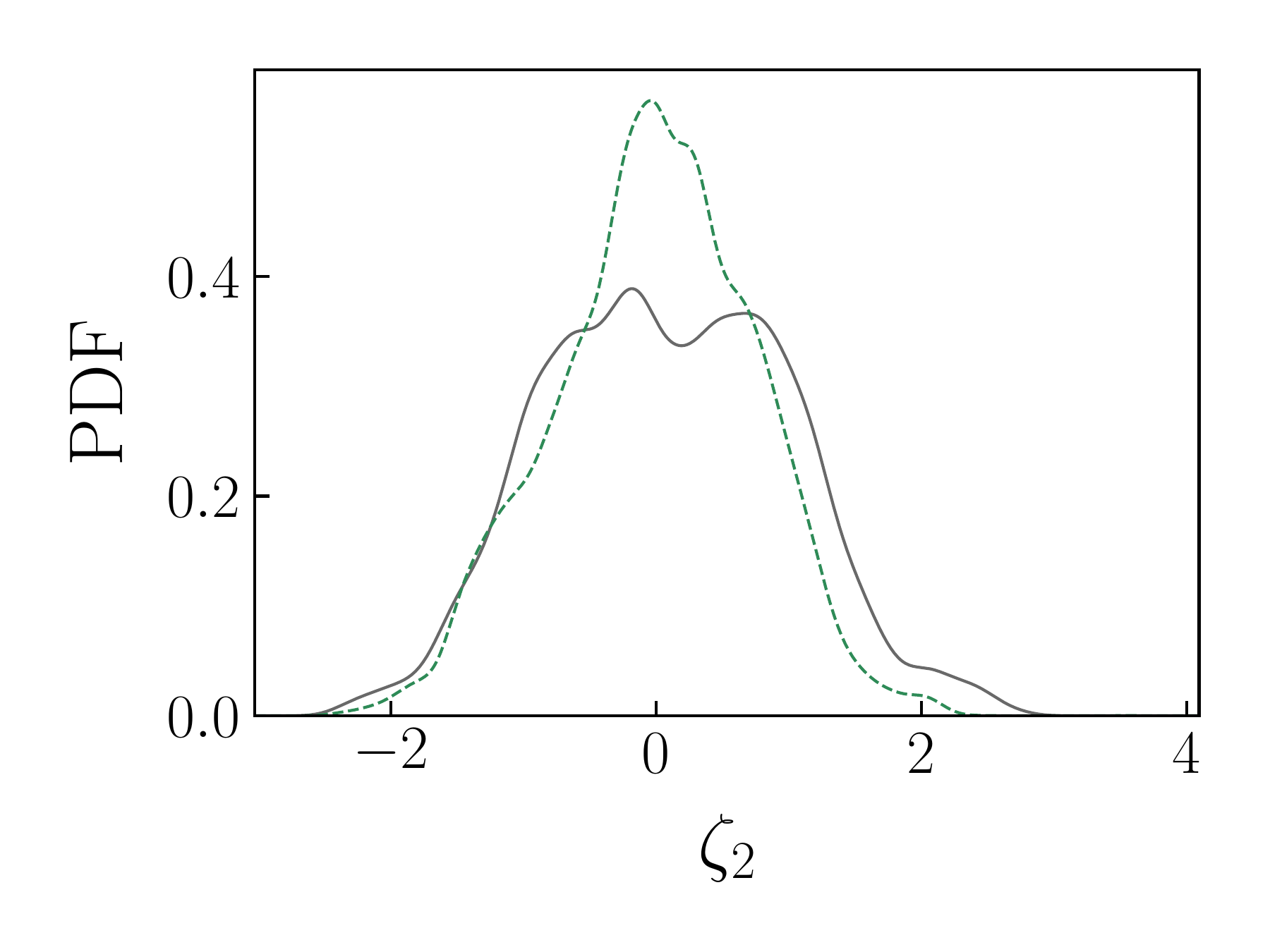}
	\end{minipage}
	\hfill
	\begin{minipage}[c]{0.85\textwidth}
		\centering
		\includegraphics[width=0.48\linewidth]{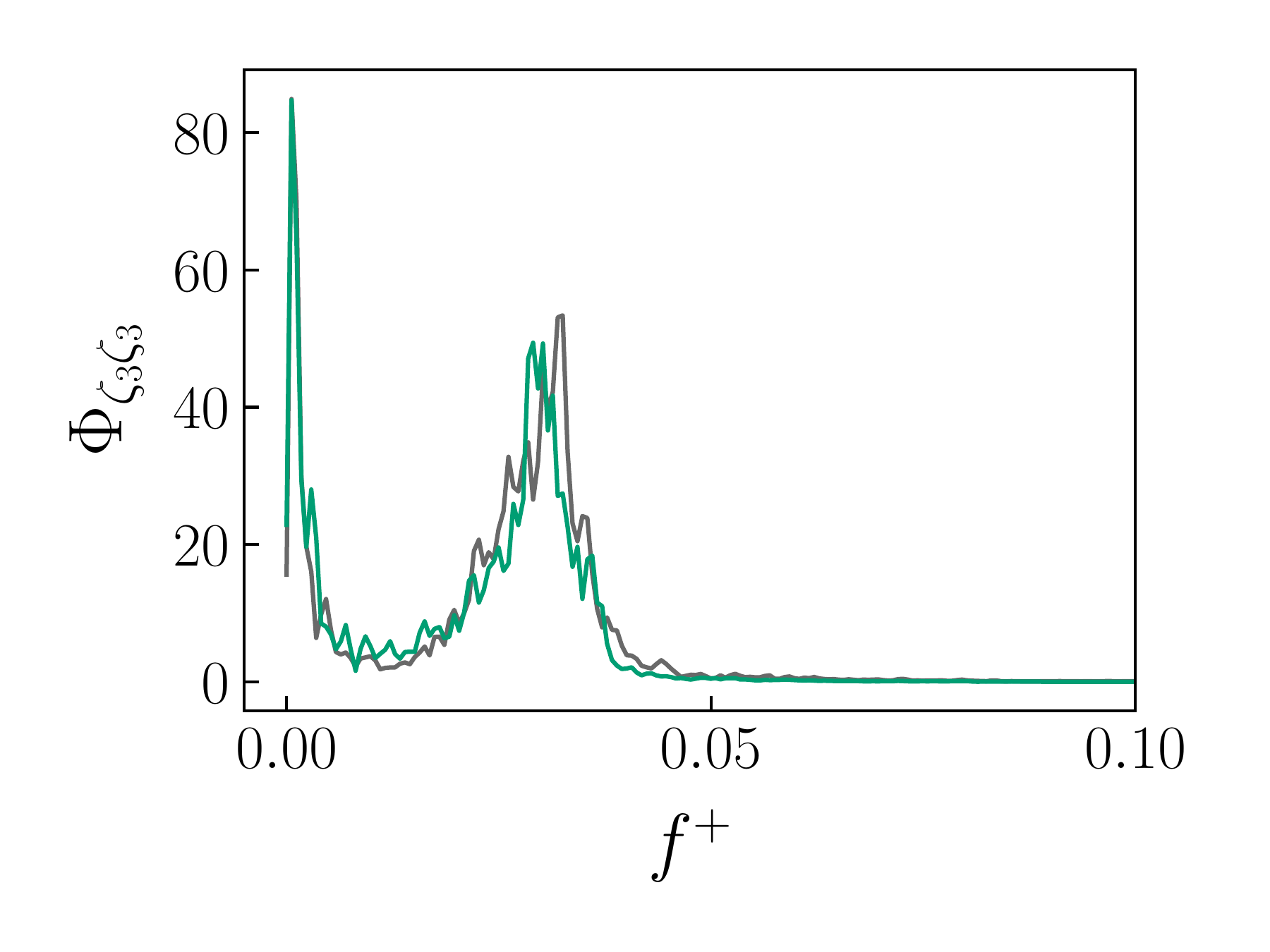}
        \includegraphics[width=0.48\linewidth]{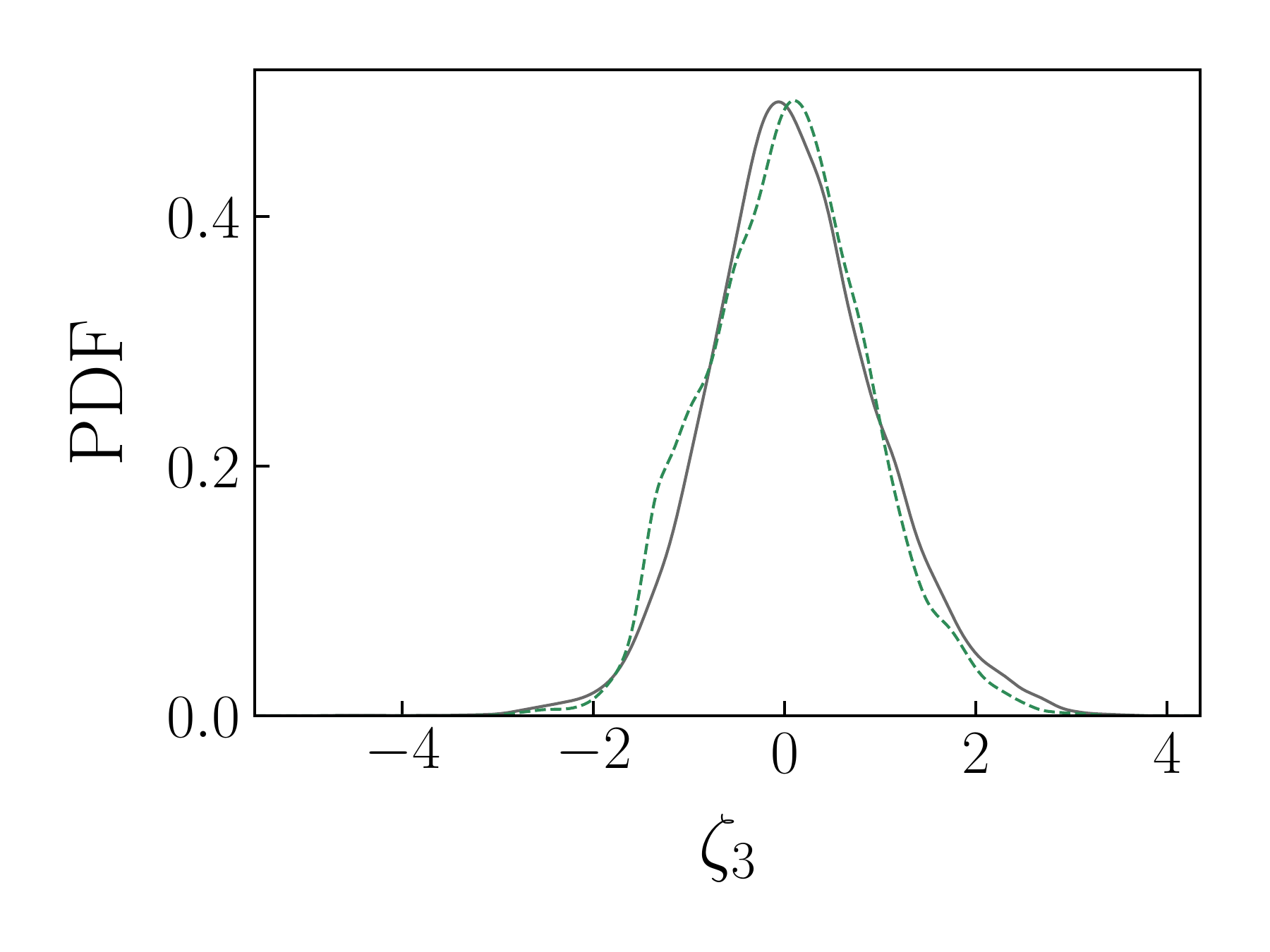}
	\end{minipage}
	\hfill
	\begin{minipage}[c]{0.85\textwidth}
		\centering
		\includegraphics[width=0.48\linewidth]{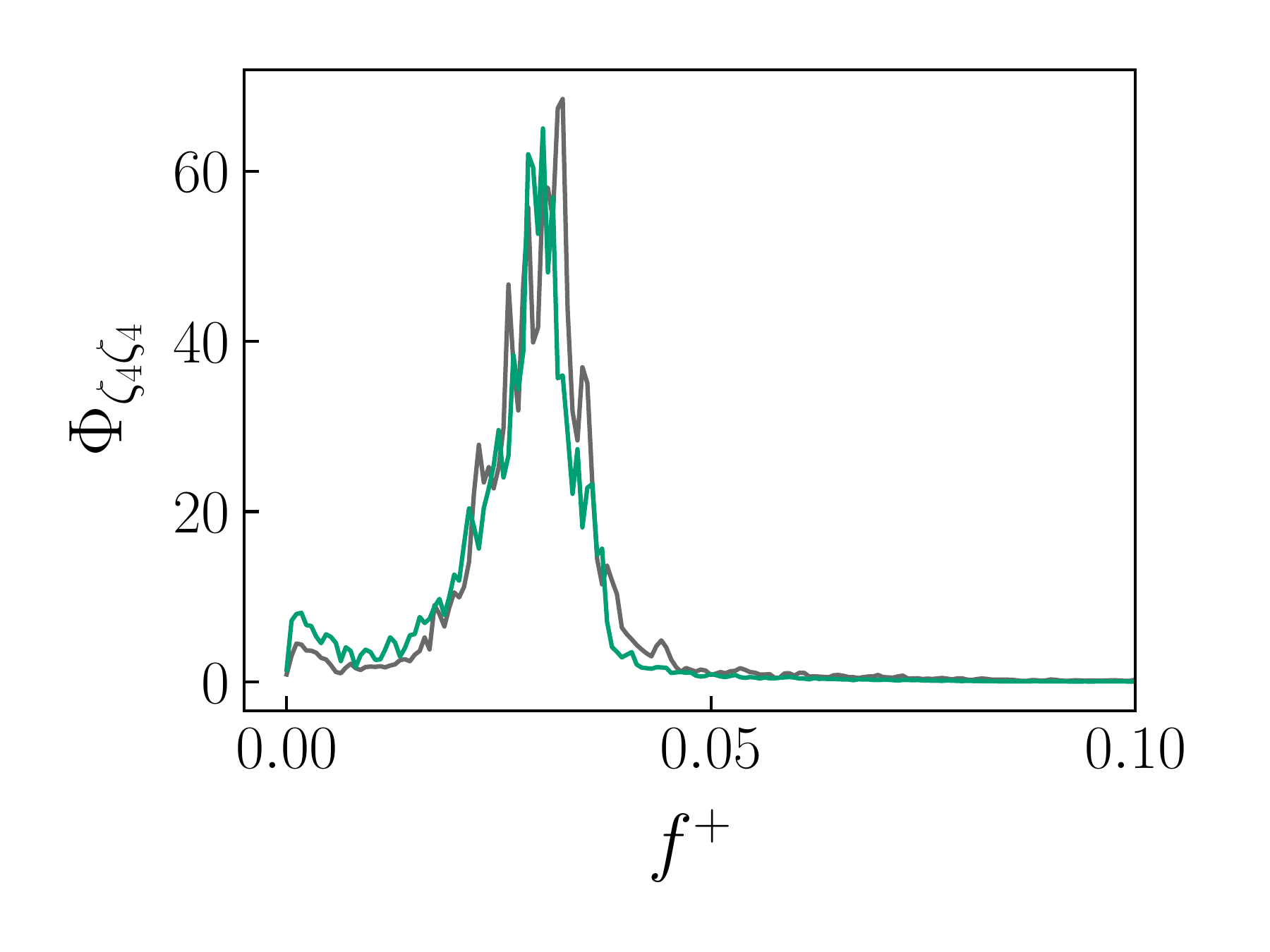}
        \includegraphics[width=0.48\linewidth]{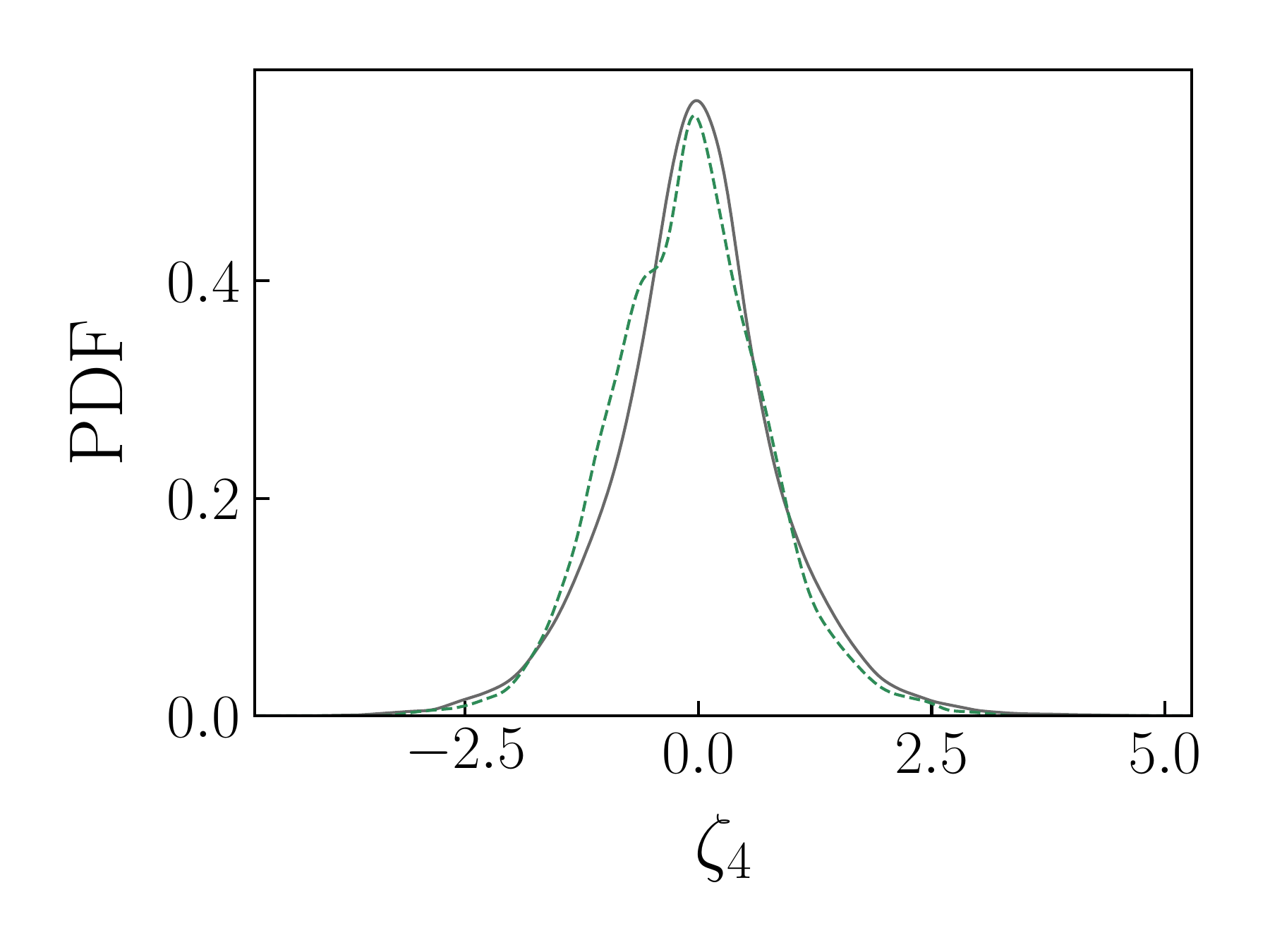}
	\end{minipage}
	\caption{Spectral and statistical analysis of the four latent variables. Left panel: power spectral density (PSD) of the predicted (sea-green) and reference (grey) latent trajectories, plotted against the frequency in wall units. Right panels: marginal probability density functions (PDFs) of the predicted (sea-green) and reference (grey) latent variables.}
	\label{fig:04.2_psd_analysis}
\end{figure}

The qualitative distinction between low- and high-frequency dimensions observed in the time series is confirmed and quantified by the PSD analysis shown in the left panels of Figure~\ref{fig:04.2_psd_analysis}. The PSDs are computed using Welch's method, partitioning the temporal signal into segments of 1/10 of the test dataset length with a 50\% overlap and a standard Hann window. Spectral peaks are reported in viscous units $f^+= f\,h/(u_\tau\,Re_\tau)$, as well as the corresponding viscous period $T^+ = 1/f^+$. The results show a clear spectral hierarchy: $\zeta_1$ and $\zeta_2$ exhibit dominant low-frequency peaks at $f^+ = 5.8 \times 10^{-4}$ ($T^+ \approx 1724$) and $f^+ = 1.16 \times 10^{-3}$ ($T^+ \approx 862$), respectively, whilst $\zeta_3$ and $\zeta_4$ display higher-frequency peaks near $f^+ = 3 \times 10^{-2}$ ($T^+ \approx 31$), with $\zeta_3$ also retaining the low-frequency peak at $T^+ \approx 1724$. The Transformer successfully reconstructs this spectral separation, capturing the primary peak frequencies and energy distributions of the reference manifold. These distinct spectral signatures indicate that the $\beta$-VAE-GAN autonomously decouples distinct aspects of the flow dynamics into separate latent dimensions, a direct consequence of explicitly promoting disentanglement during training, consistent with findings reported in \cite{jacobsen2022disentangling}. The physical origin of these spectral signatures is examined 
at the end of this section.

\newpage 

The marginal PDFs of each latent variable, shown in the right panels of Figure~\ref{fig:04.2_psd_analysis}, further characterise the statistical fidelity of the predictions. The low-frequency dimensions $\zeta_1$ and $\zeta_2$ present skewed distributions, consistent with their irregular temporal evolution, and exhibit the largest prediction discrepancies. For $\zeta_2$, the reference distribution is broad and slightly bimodal; the prediction captures the correct support and variance but over-concentrates probability at the centre, producing a more unimodal profile. For $\zeta_1$, the reference is strongly asymmetric, concentrated at positive values with a pronounced left tail, a feature that the model only partially reproduces. In both cases, the discrepancy is consistent with the slow timescales encoded by these dimensions: the finite training dataset contains comparatively few statistically independent realisations of the dominant dynamical states of $\zeta_1$ and $\zeta_2$, limiting the model's ability to learn their full amplitude and residence-time distributions. In contrast, the high-frequency dimensions $\zeta_3$ and $\zeta_4$ exhibit symmetric PDFs that are reproduced with high fidelity: $\zeta_4$ presents a near-Gaussian distribution recovered accurately, whilst $\zeta_3$, despite its left-skewed asymmetry, is also well captured with only a marginal amplitude difference at the peak.

Across the temporal, spectral, and statistical analyses presented above, a consistent separation emerges between the low- and high-frequency latent dimensions. The following analysis investigates the physical origin of this separation by comparing the latent spectra against those of representative DNS flow quantities, after briefly recalling the key dynamical properties of the MFU that provide the necessary physical context.

\begin{figure}[htbp]
	\centering
	\begin{minipage}[c]{0.48\textwidth}
		\centering
		\includegraphics[width=\linewidth]{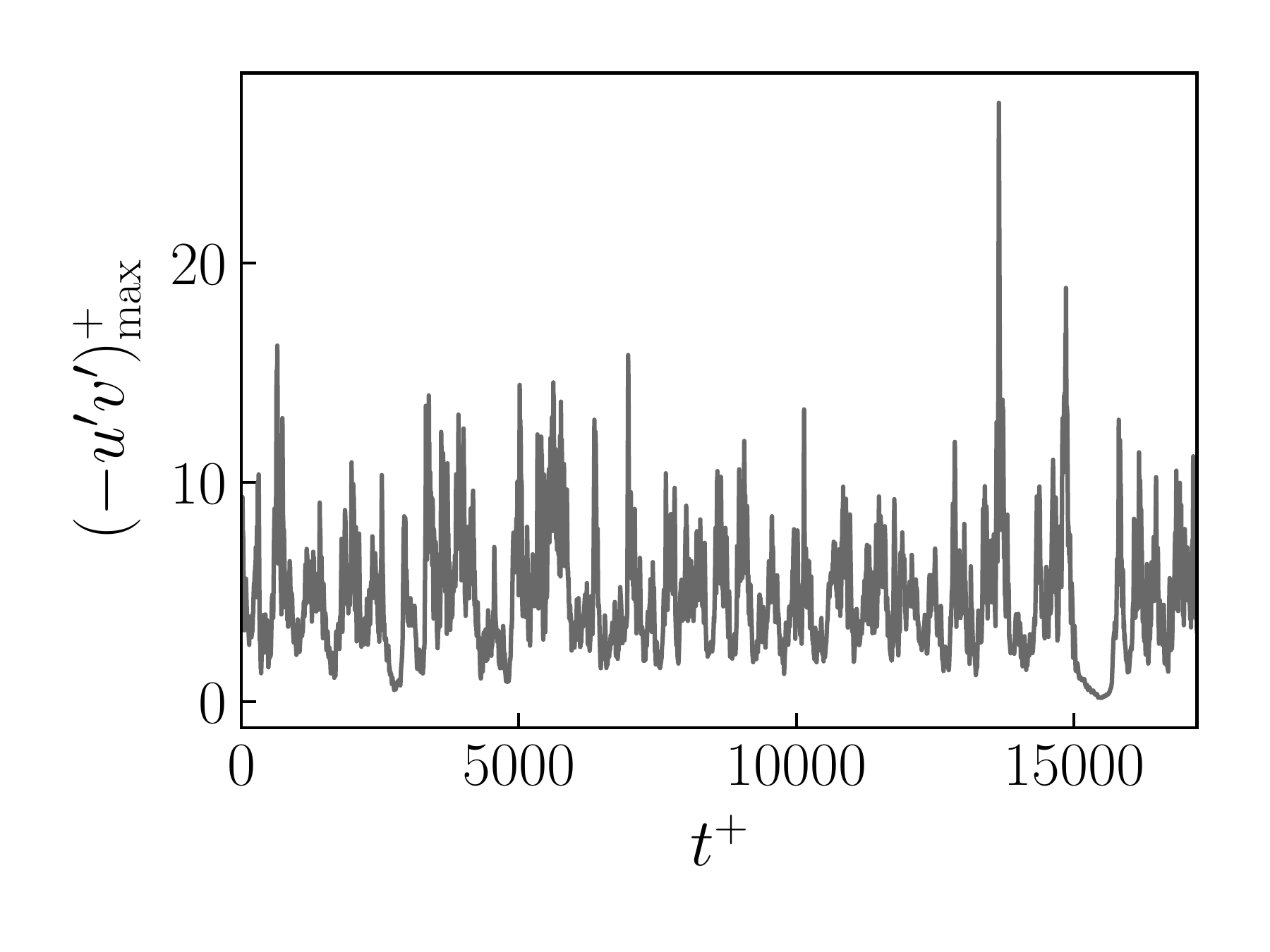}
	\end{minipage}
	\hfill
	\begin{minipage}[c]{0.48\textwidth}
		\centering
		\includegraphics[width=\linewidth]{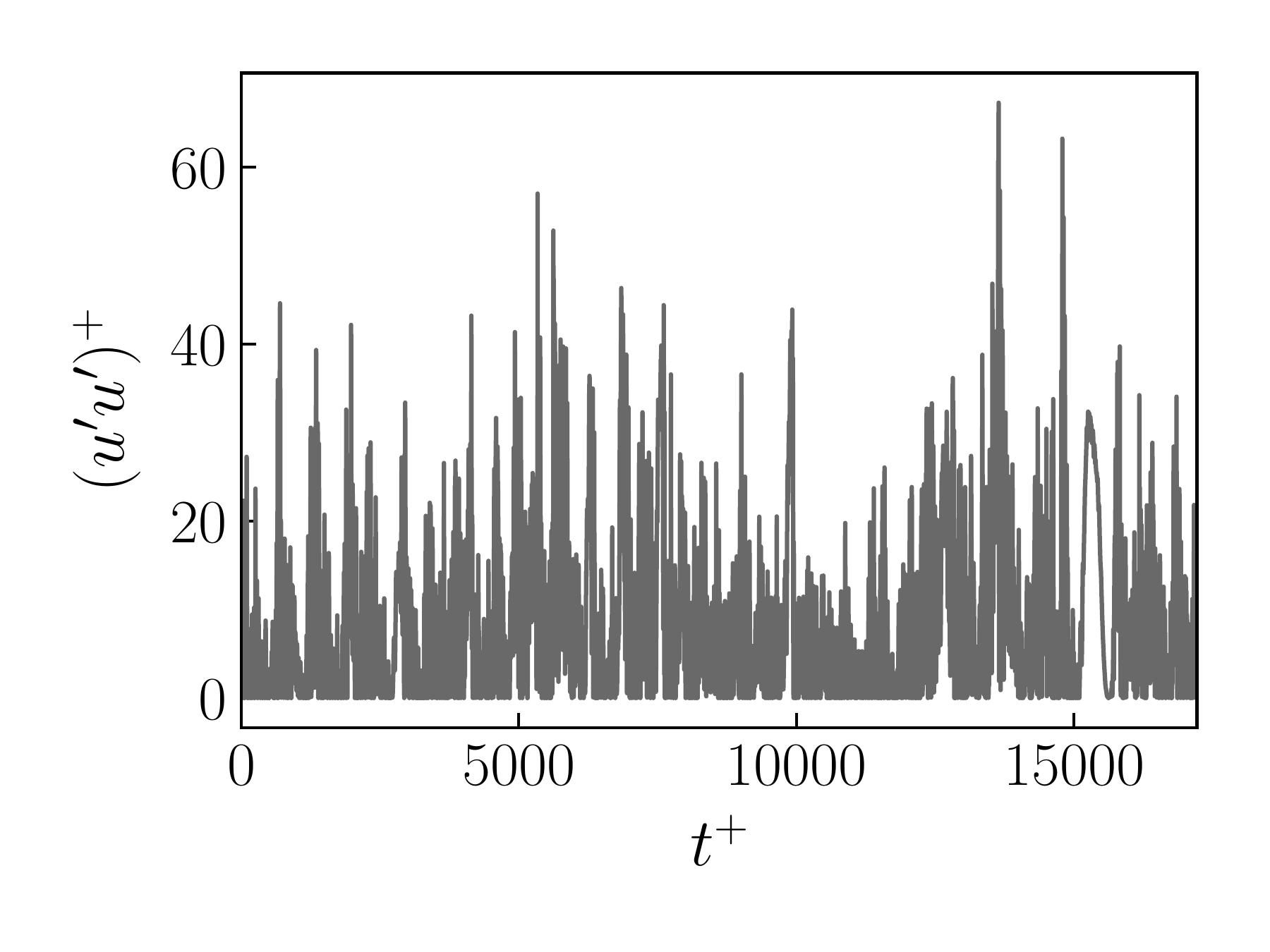}
	\end{minipage}
	\hfill
	\begin{minipage}[c]{0.48\textwidth}
		\centering
		\includegraphics[width=\linewidth]{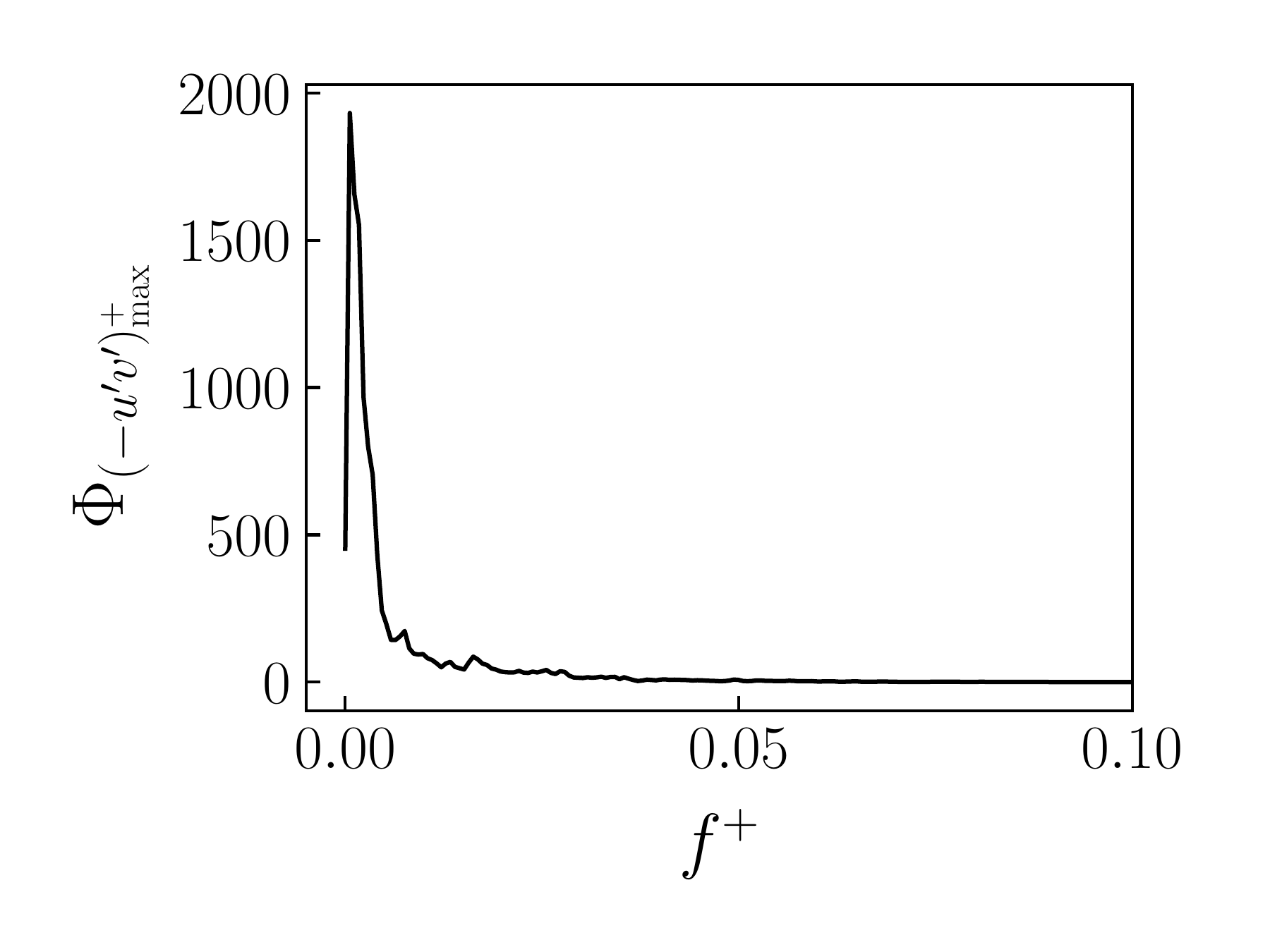}
	\end{minipage}
	\hfill
	\begin{minipage}[c]{0.48\textwidth}
		\centering
		\includegraphics[width=\linewidth]{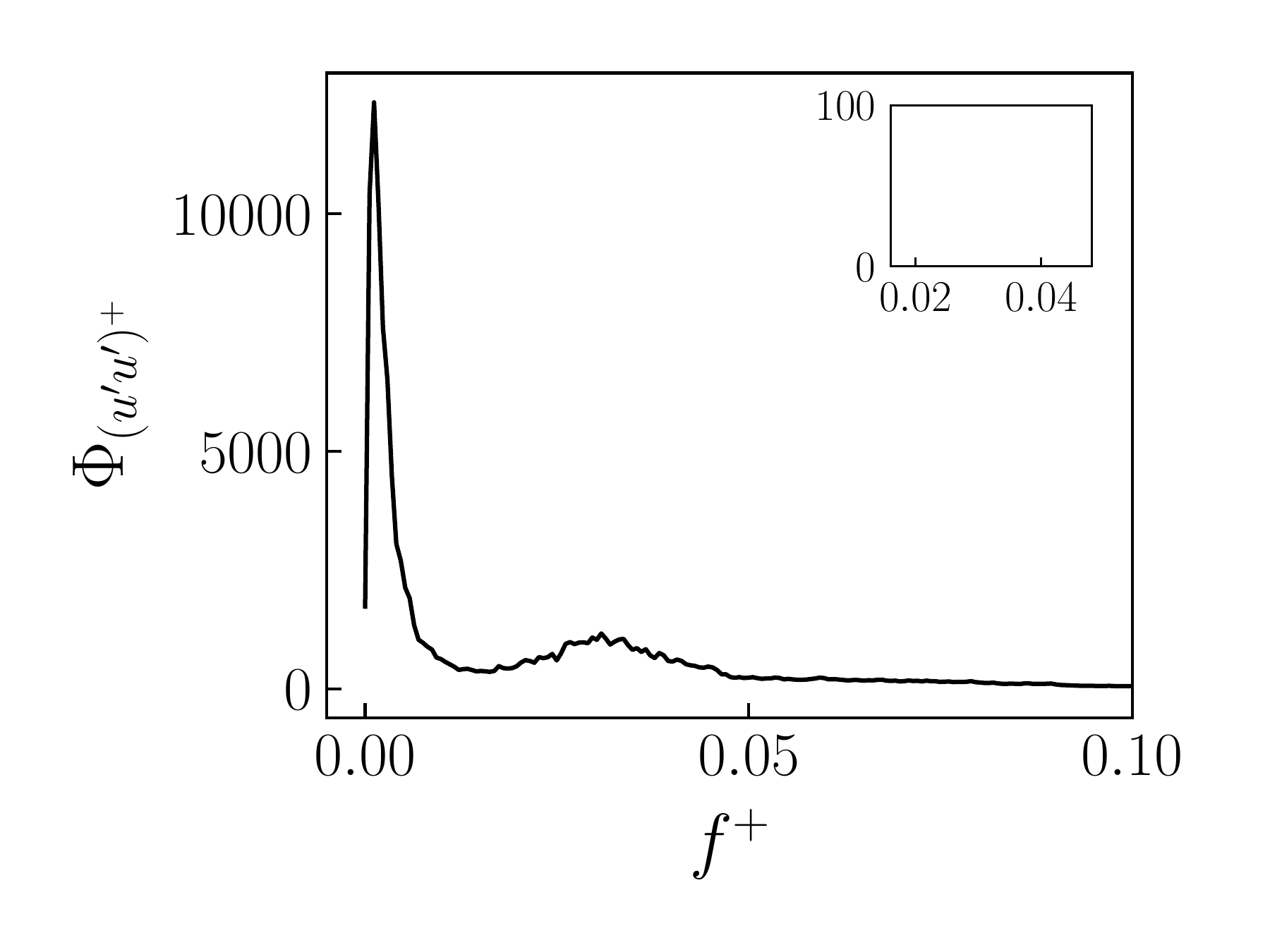}
	\end{minipage}
	\caption{Temporal evolution and associated power spectral densities (PSDs) of selected physical flow statistics. Left panels: time series (top) and corresponding PSD (bottom) of the spatial maximum of the instantaneous Reynolds shear stress, $(-u^\prime v^\prime)^+_{\max}$. Right panels: time series of the streamwise Reynolds normal stress, $(u^\prime u^\prime)^+$, sampled at the domain centre (top), and the temporal PSD of $(u^\prime u^\prime)^+$ averaged across all spatial points in the domain (bottom).The dominant spectral peaks of both physical quantities correspond directly to those identified in the latent space (Figure~\ref{fig:04.2_psd_analysis}), confirming the physical interpretability of the learnt representation.}
	\label{fig:04.2_psd_and_signal_of_uv_and_uu}
\end{figure}

The MFU domain confines a single near-wall low-speed streak and its associated quasi-streamwise vortices, whose dynamics are governed by an intermittent regeneration cycle \cite{jimenez1991minimal}, manifesting as global temporal bursts alternating with relatively quiescent periods. The period of this cycle scales inversely with the Reynolds number and is of the order of $\mathcal{O}(10^3)\,t^+$ at $Re_\tau = 200$ \cite{jimenez1991minimal}, in close agreement with the dominant latent peak at $T^+ \approx 1724$. To establish this link quantitatively, the spatial maximum of the instantaneous Reynolds shear stress is extracted at each time step:

\begin{equation}
    (-u^\prime v^\prime)^+_{\max}(t) = \max_{x^+,z^+}  \left( -u^\prime v^\prime \right)^+,
\label{eq:04.2_max_shear}
\end{equation}

\noindent as this is the observable employed by Jiménez and Moin \cite{jimenez1991minimal} to track the regeneration cycle. Its PSD (Figure~\ref{fig:04.2_psd_and_signal_of_uv_and_uu}, left column) exhibits a dominant peak at $T^+ \approx 1724$, matching the spectral signatures of $\zeta_1$ and $\zeta_3$. The spatially averaged PSD of the streamwise Reynolds stress $(u^\prime u^\prime)^+$ (Figure~\ref{fig:04.2_psd_and_signal_of_uv_and_uu}, right column) reveals a primary peak at $T^+ \approx 862$ and a secondary broad energetic hump near $T^+ \approx 31$, aligning closely with the dominant spectral peaks of $\zeta_2$ and $\zeta_4$, respectively. The secondary hump presents as a broad energetic feature rather than a discrete spike, indicating that it reflects turbulent content at these scales rather than a box-periodicity artefact. This one-to-one correspondence demonstrates that the $\beta$-VAE-GAN autonomously organises the characteristic timescales of the flow into distinct latent dimensions: $\zeta_1$ and $\zeta_3$ encode the slow regeneration-cycle dynamics ($T^+ \approx 1724$), $\zeta_2$ captures the intermediate timescale of the streamwise normal stress ($T^+ \approx 862$), and $\zeta_4$ reflects the higher-frequency streamwise turbulent fluctuations ($T^+ \approx 31$). 

Inspection of the signals in the vicinity of $t^+ \approx 15{,}000$ (Figure~\ref{fig:04.2_psd_and_signal_of_uv_and_uu}) reveals a dynamically distinct interval: $(-u^\prime v^\prime)^+_{\max}$ drops sharply towards zero and remains suppressed over a prolonged period, whilst $(u^\prime u^\prime)^+$, after an initial attenuation, rises to values of order 30 before $(-u^\prime v^\prime)^+_{\max}$ eventually recovers. During the quiescent phase of the regeneration cycle, the quasi-streamwise vortices weaken and the vortex-driven pumping that sustains active ejection and sweep events ceases, consistent with the near-zero $(-u^\prime v^\prime)^+_{\max}$. The low-speed streak, however, may persist as a passive structure near the measurement point even in the absence of active vortex forcing, maintaining a local streamwise velocity deficit and hence a non-negligible $(u^\prime u^\prime)^+$ signal. This quiescent interval corresponds directly to the suppressed latent activity visible in Figure~\ref{fig:04.2_time_series_and_joint_pdf} at the same time period, confirming that the latent representation responds to the collapse of the organised near-wall motions. Crucially, the Transformer successfully tracks this event despite its rarity in the dataset and the extended forecasting horizon. Importantly, the joint PDF analysis demonstrates that the model tracks the temporal evolution of $\zeta_1$, with reasonable fidelity; the imperfect reproduction of the marginal distribution shape is therefore not a forecasting failure in the dynamical sense, but rather a consequence of the scarcity of such quiescent events in the training data, which limits the model's ability to learn their full residence-time statistics.

Whilst the one-to-one spectral correspondences established here provide a clear picture of how the $\beta$-VAE-GAN organises the flow timescales, spectral matching alone is insufficient for a precise attribution of each latent dimension to specific flow structures. Whether these dimensions encode streak instability dynamics, quasi-streamwise vortex activity, or their nonlinear coupling remains an open question that motivates dedicated spatio-temporal correlation analysis in future work.

\newpage

\subsection{End-to-End Framework Performance}
\label{sec:04.3_full_framework}

In this final stage of evaluation, the forecasting capabilities of the end-to-end predictive framework are analyzed. The system operates in full inference mode: latent variables are generated autoregressively by the Transformer conditioned on the sensor signals, and are subsequently decoded to physical space by the $\beta$-VAE-GAN. To isolate the errors introduced by temporal forecasting from those inherent to spatial compression, all results are compared against the baseline compression mode, in which physical fields are decoded directly from the encoded latent variables of the test data, constituting a performance upper bound for the architecture. Results for the DNS test data, the compression baseline, and the end-to-end inference mode are denoted by the superscripts $(\cdot)^{\text{test}}$, $(\cdot)^{\text{vae-gan}}$, and $(\cdot)^{\text{pred}}$, respectively. All metrics reported in this section exclude the 128 initialisation time steps. The analysis proceeds from global energy balance and spatial structure, through the statistics of the Reynolds stress mechanisms, to the temporal fidelity of the regeneration cycle.

The quantitative performance of the end-to-end framework is summarised in Table~\ref{tab:04.3_performance_metrics}, which reports the reconstructed energy coefficient $E_{k^+}$ (Eq.~\ref{eq:04.1_tke_reconstruction_coefficient}) and the global entropy $H$ (Eq.~\ref{eq:02.2_global_entropy}), the former measuring the fraction of turbulent kinetic energy recovered and the latter quantifying the retention of spatio-temporal complexity. The energy recovery coefficient for the inference mode is $E_{k^+}^{\text{pred}} = 0.82$, closely tracking the compression baseline of $E_{k^+}^{\text{vae-gan}} = 0.87$. This $\approx 5\%$ relative drop confirms that the temporal forecasting module preserves the dominant energetic content of the flow. The entropy results reinforce this picture: the principal reduction occurs at the compression stage, where $H$ drops from $H^{\text{test}} = 0.6975$ to $H^{\text{vae-gan}} = 0.6159$, with the subsequent transition to full autoregressive forecasting incurring a negligible further loss of less than 2\%. Both results imply that the primary source of information loss in the framework is the bottleneck constraint of the $\beta$-VAE-GAN, rather than the Transformer.

\begin{table}[htbp]
    \centering
    \small
    \begin{tabular}{l c c}
        \toprule
        \textbf{Configuration} & $E_{k^+}$ & $H$ \\
        \midrule
        Test Data $(\cdot)^{\text{test}}$ & -- & 0.6975 \\
        Baseline $(\cdot)^{\text{vae-gan}}$ & 0.85 & 0.6159 \\
        Inference $(\cdot)^{\text{pred}}$ & 0.82 & 0.6041 \\
        \bottomrule
    \end{tabular}
    \caption{Summary of performance metrics: reconstructed energy coefficient ($E_{k^+}$) and global entropy ($H$).}
    \label{tab:04.3_performance_metrics}
\end{table}

\begin{figure}[htbp]
	\centering
	\begin{minipage}[c]{0.48\textwidth}
		\centering
		\includegraphics[width=\linewidth]{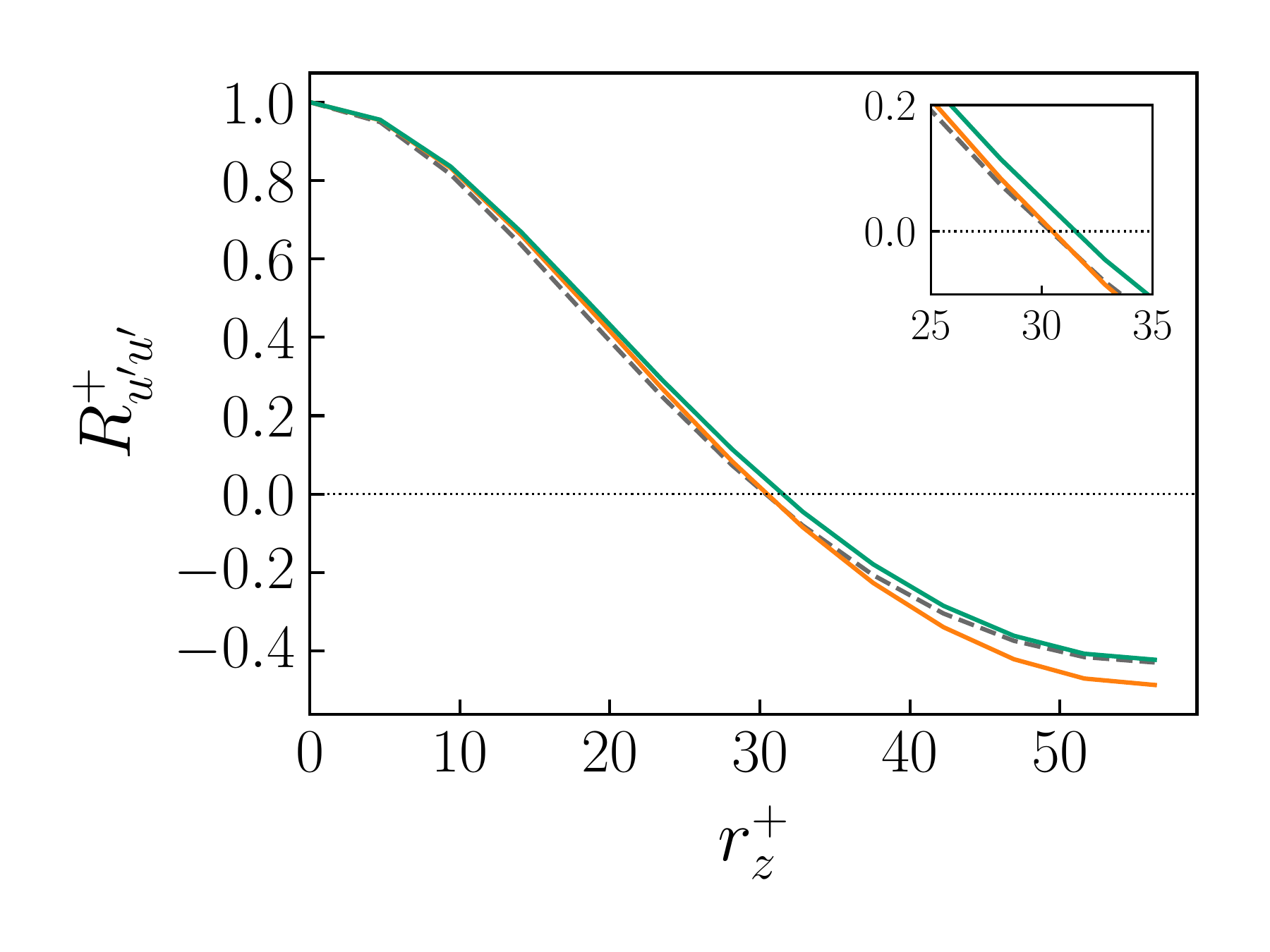}
	\end{minipage}
	\hfill
	\begin{minipage}[c]{0.48\textwidth}
		\centering
		\includegraphics[width=\linewidth]{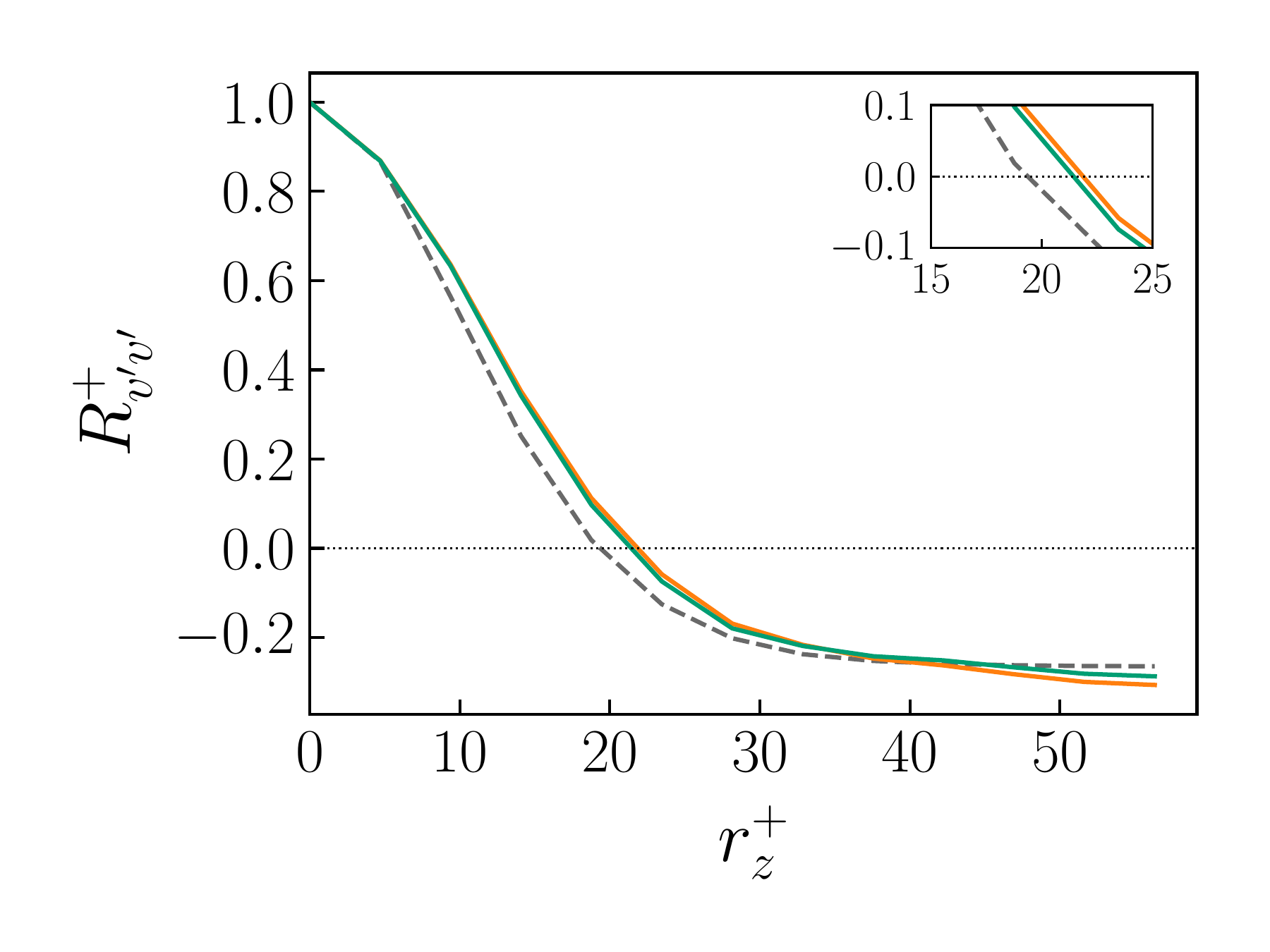}
	\end{minipage}
	\hfill
	\begin{minipage}[c]{0.48\textwidth}
		\centering
		\includegraphics[width=\linewidth]{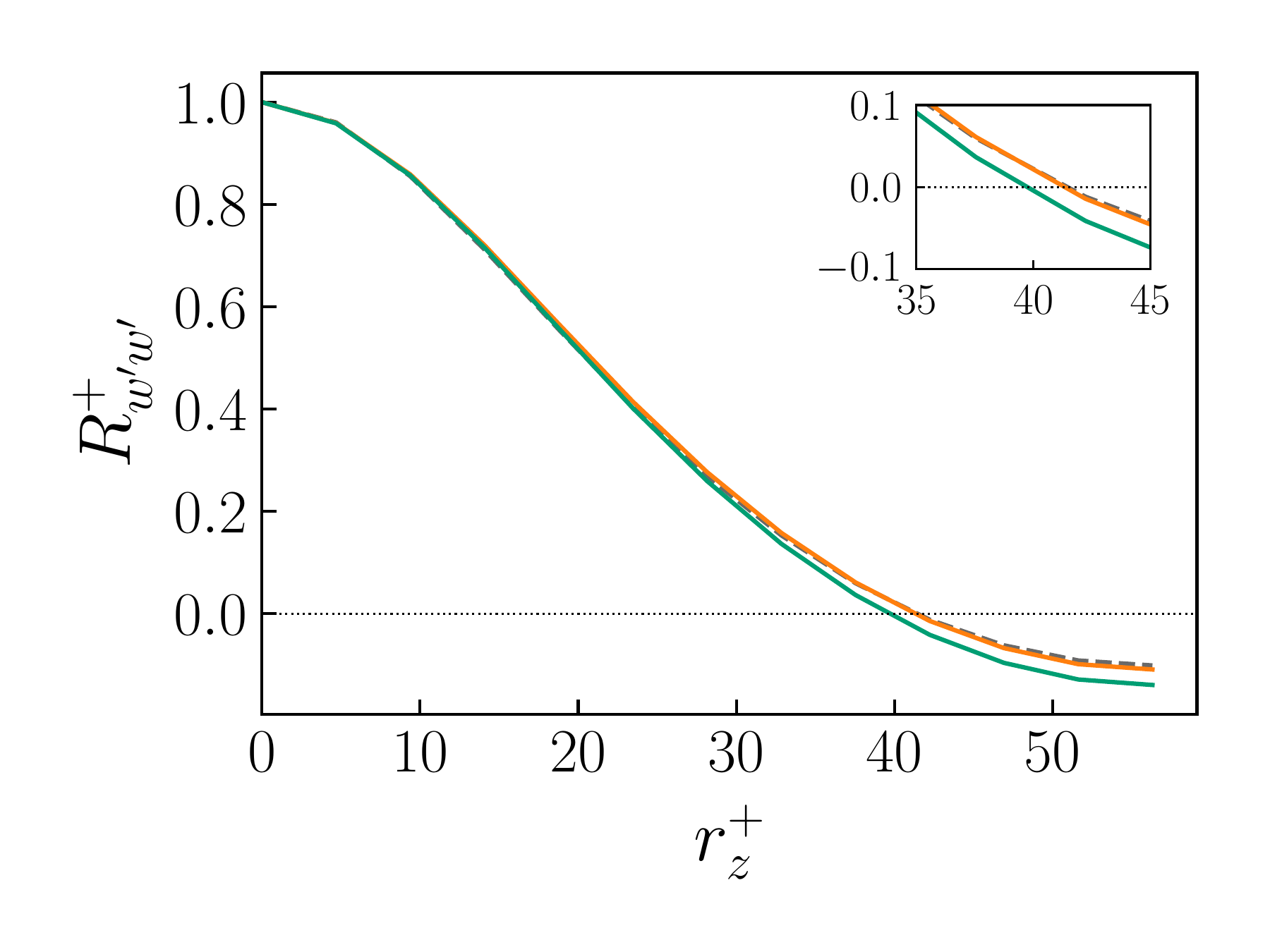}
	\end{minipage}
	\caption{Two-point spatial autocorrelation of the streamwise ($u$), wall-normal ($v$), and spanwise ($w$) velocity fluctuations for the DNS test data (gray), the compression baseline (orange), and the end-to-end inference framework (sea-green). The inset in the upper right of each panel provides a magnified view of the region near the zero-crossing.}
	\label{fig:04.3_two_point_correlation}
\end{figure}

To analyse the spatial organisation of the predicted flow fields, Figure~\ref{fig:04.3_two_point_correlation} reports the two-point spatial autocorrelation profiles for the three velocity components. The predicted profiles (sea-green) are in close agreement with the DNS reference (grey) and the compression baseline (orange), with only a minor departure near the zero-crossing. The most pronounced discrepancy arises in the wall-normal velocity component $v$, for which the predicted and baseline profiles are nearly coincident whilst both deviate from the DNS. This collapse corroborates the conclusion drawn from the global metrics: the spatial error stems from the $\beta$-VAE-GAN bottleneck rather than the Transformer. 

\begin{figure}[htbp]
	\centering
	\includegraphics[width=0.48\linewidth]{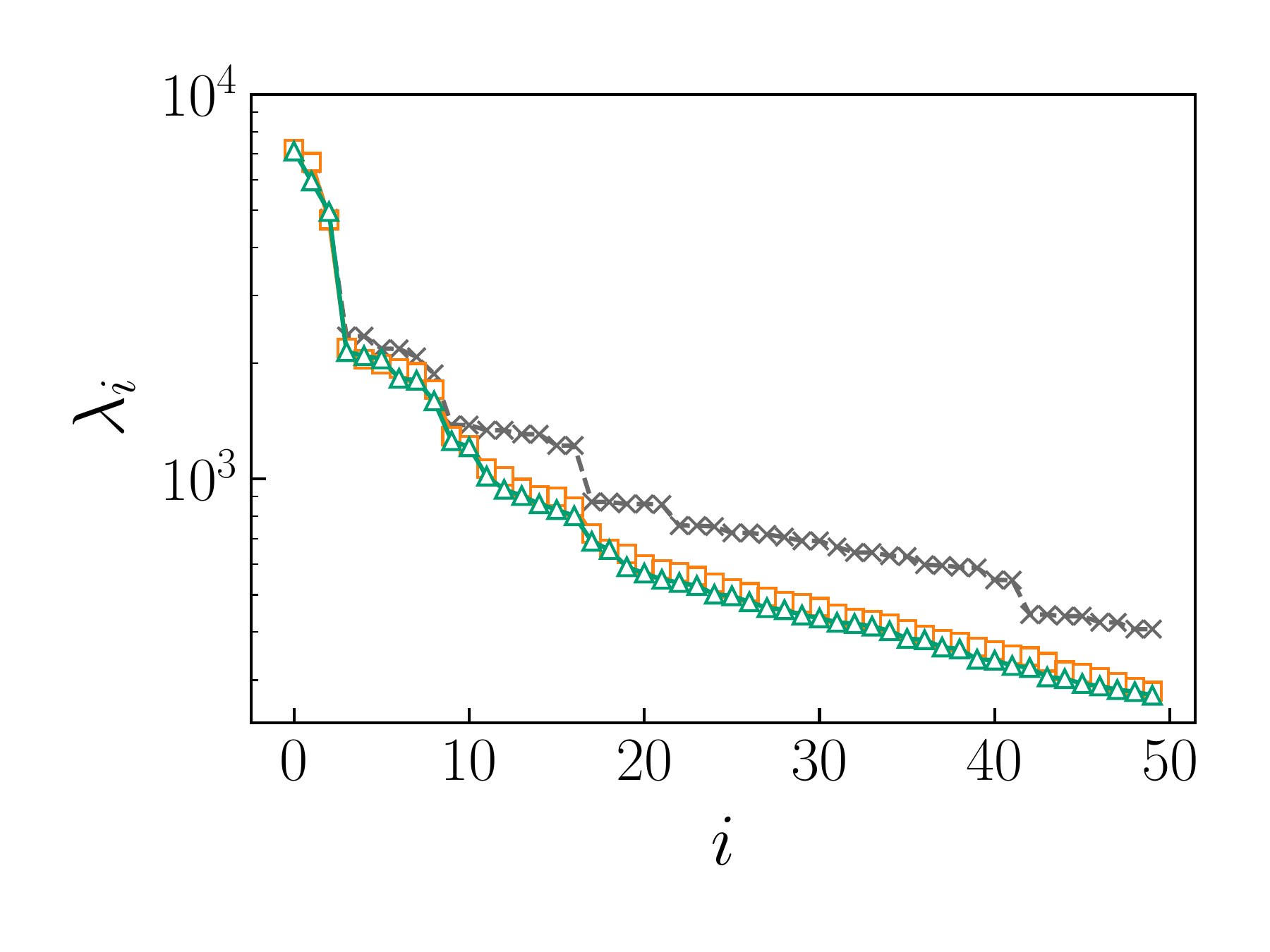}
	\caption{Singular value decay of the POD modes for the DNS test data (gray), the compression baseline (orange), and the end-to-end inference framework (sea-green).}
	\label{fig:04.3_pod_decay}
\end{figure}

This regularisation acts preferentially on the less energetically dominant fluctuation components, as further evidenced by the POD singular value decay shown in Figure~\ref{fig:04.3_pod_decay}. The first three singular values of the reconstructed and predicted fields nearly coincide with the DNS reference, confirming that the dominant spatial structures are preserved. Beyond the first ten modes, however, a steeper drop-off is observed, reflecting the attenuation of smaller, lower-energy structures by the compression stage.

\begin{figure}[htbp]
	\centering
	\begin{minipage}[b]{0.48\textwidth}
		\centering
		\includegraphics[width=\linewidth]{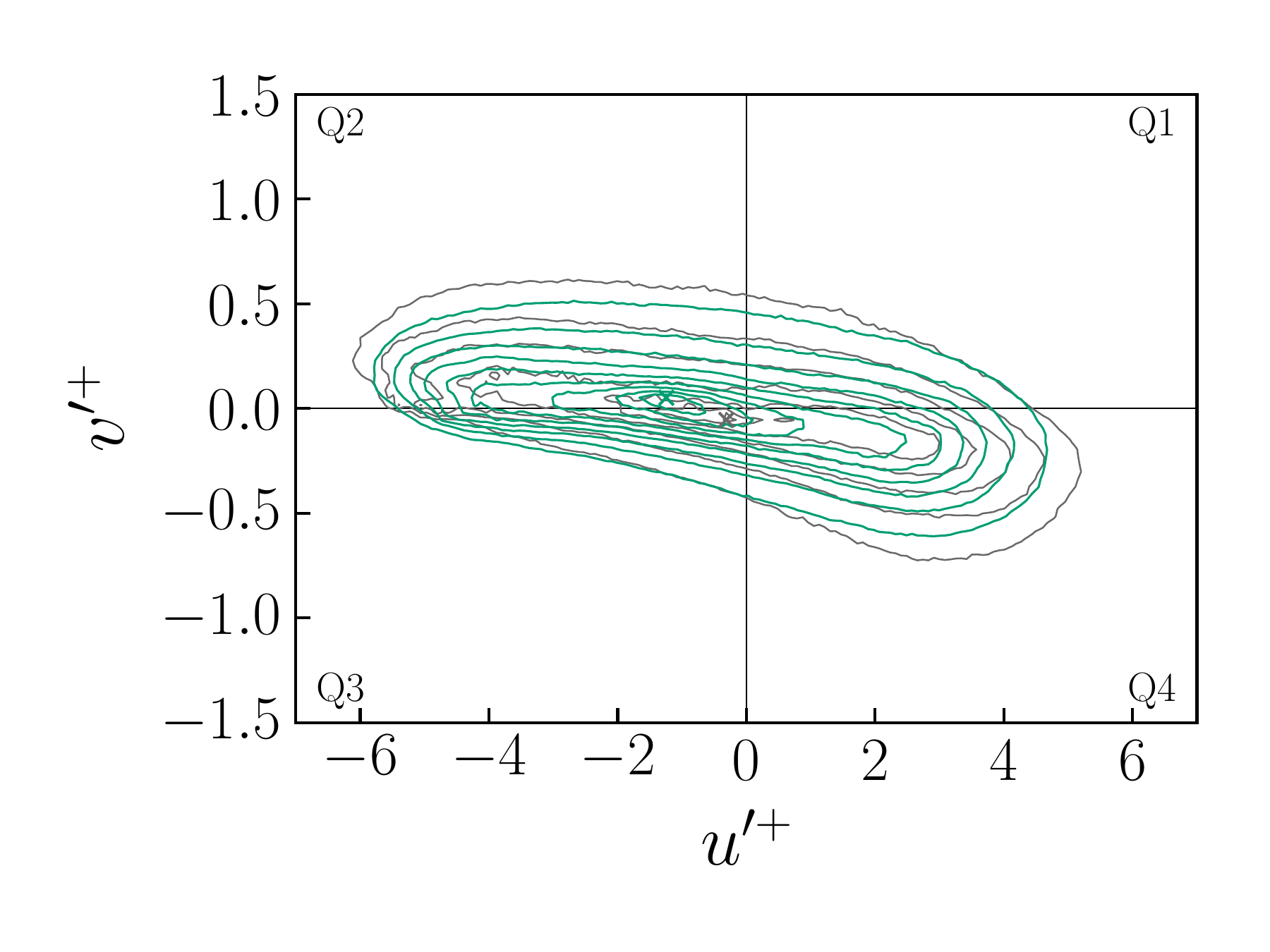}
	\end{minipage}
	\hfill
	\begin{minipage}[b]{0.48\textwidth}
		\centering
		\includegraphics[width=\linewidth]{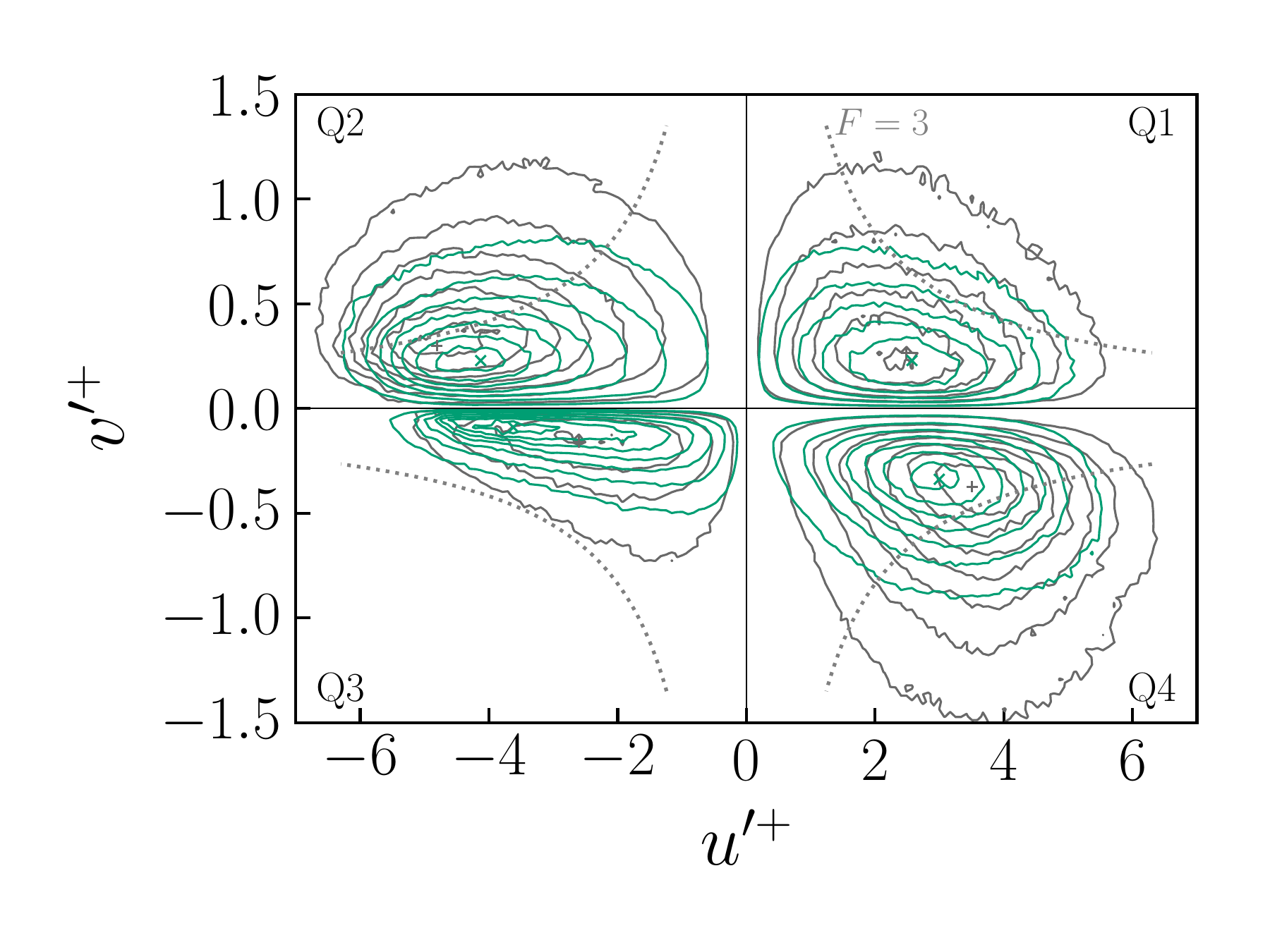}
	\end{minipage}
	\hfill
	\begin{minipage}[b]{0.48\textwidth}
		\centering
		\includegraphics[width=\linewidth]{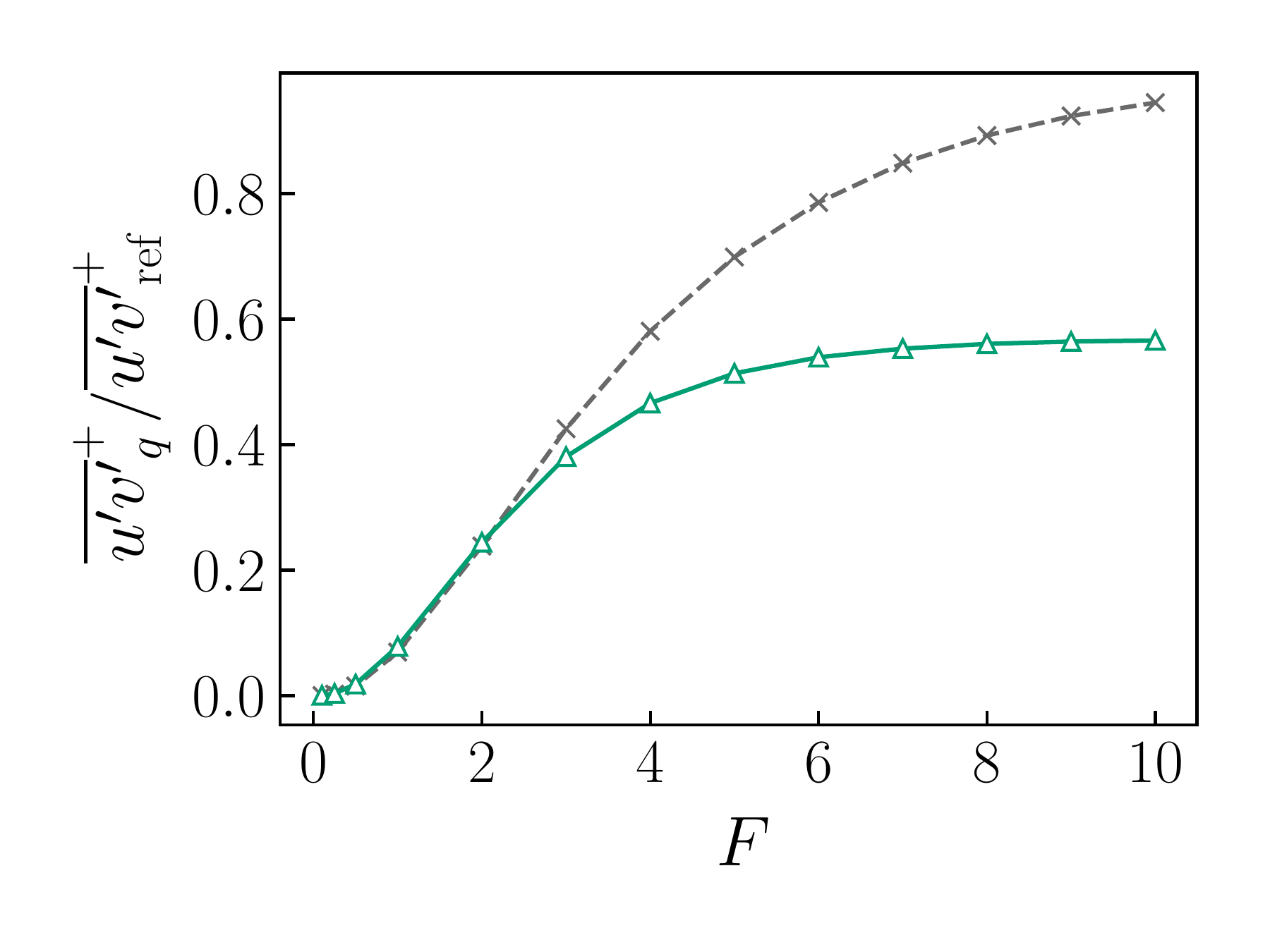}
	\end{minipage}
	\hfill
	\begin{minipage}[b]{0.48\textwidth}
		\centering
		\includegraphics[width=\linewidth]{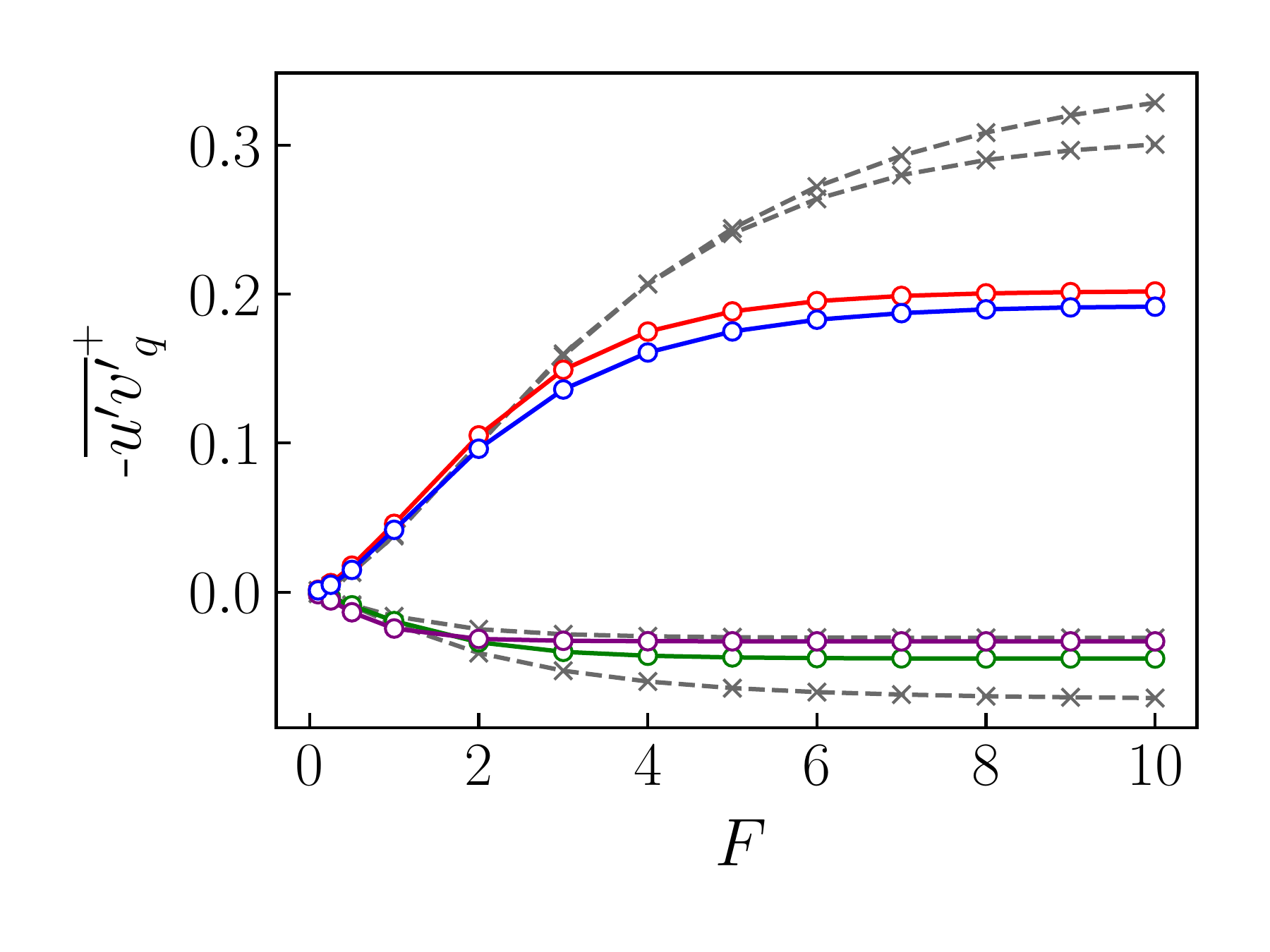}
	\end{minipage}
	\caption{Quadrant analysis of the Reynolds shear stress. Top row: joint probability density function $P(u^{\prime +},v^{\prime +})$ (left) and covariance integrand $u^{\prime +} v^{\prime +} P(u^{\prime +},v^{\prime +})$ (right), with hyperbolic hole-boundary curves overlaid for $F = 3$. Bottom row: hole-filtered contributions to the Reynolds shear stress as a function of $F$, combined (left) and fractional by quadrant (right). DNS reference data are denoted by grey lines/contours and end-to-end framework predictions by sea-green lines/contours. In the bottom-right panel, coloured lines indicate the fractional contribution of each quadrant for the predicted flow: outward interactions (Q1, green), ejections (Q2, red), inward interactions (Q3, purple), and sweeps (Q4, blue). Combined contributions are normalised by the reference value $\overline{u^\prime v^\prime}^+_\mathrm{ref}$.}
	\label{fig:04.3:quadrant_analysis}
\end{figure}

The ability of the model to reproduce the physical mechanisms responsible for Reynolds shear stress generation is assessed next via quadrant analysis \cite{wallace2016quadrant}, shown in Figure~\ref{fig:04.3:quadrant_analysis}. The top row displays the joint probability density function $P(u^{\prime +},v^{\prime +})$ (left) and the covariance integrand $u^{\prime +} v^{\prime +} P(u^{\prime +},v^{\prime +})$ (right), the latter relating to the total Reynolds shear stress via:

\begin{equation}
    \overline{u^{\prime} v^{\prime}}^+ = \int_{-\infty}^\infty \int_{-\infty}^\infty u^{\prime +} v^{\prime +} P(u^{\prime +}, v^{\prime +}) \, \mathrm{d}u^{\prime +} \, \mathrm{d}v^{\prime +}.
\end{equation}

\noindent In the $u^{\prime +}$--$v^{\prime +}$ planee, the events contained in each quadrant represent distinct types of turbulent motions. The Q2 ($u^{\prime +}<0,\,v^{\prime +}>0$) and Q4 ($u^{\prime +}>0,\,v^{\prime +}<0$) quadrants correspond to ejections and sweeps, respectively: gradient-type motions that transport low-momentum fluid away from the wall (Q2) or high-momentum fluid towards it (Q4), constituting the dominant contribution to the Reynolds shear stress. Conversely, Q1 ($u^{\prime +}>0, v^{\prime +}>0$) and Q3 ($u^{\prime +}<0, v^{\prime +}<0$) represent counter-gradient inward and outward interactions. The predicted joint PDF correctly reproduces the characteristic elongation along the Q2--Q4 axis, demonstrating that the most frequently occurring ejection and sweep events are successfully captured. The covariance integrand (top right), however, reveals that the most intense and intermittent events are not fully represented: the predicted $u^{\prime +} v^{\prime +} P(u^{\prime +},v^{\prime +})$ exhibits a contraction in the $v^{\prime +}$ direction relative to the DNS reference. The primary origin of this attenuation resides in the information bottleneck of the $\beta$-VAE-GAN. By compressing the flow into a 
low-dimensional latent space of dimension $d_\zeta = 4$, the encoder prioritises the dominant, statistically recurrent structures of the flow. Rare, extreme-amplitude events contribute marginally to the training loss and are consequently attenuated in the reconstruction. A remedy would be to increase the latent dimension $d_\zeta$, which would allow the encoder to represent a broader range of flow 
states. However, this comes at the cost of a higher-dimensional latent space for the Transformer to forecast, possibly increasing the risk of error accumulation over long autoregressive horizons. The optimal latent dimension therefore represents an application-dependent trade-off between latent space compactness and forecasting tractability.

To quantifies this attenuation more precisely, a hole-filtered extension of the quadrant analysis \cite{willmarth1972structure} is presented in the bottom row of Figure~\ref{fig:04.3:quadrant_analysis}. In this analysis, hyperbolic isolines of constant $|u^{\prime +} v^{\prime +}|$ are defined via a magnitude-filtering parameter $F$ as:

\begin{equation}
    |u^{\prime +} v^{\prime +}| < F \left|\overline{u^{\prime} v^{\prime}}^+_\mathrm{ref}\right|,
\end{equation}

\noindent where $\overline{u^{\prime} v^{\prime}}^+_\mathrm{ref}$ is taken as the total DNS Reynolds shear stress. The fractional contribution of each quadrant $\overline{u^{\prime} v^{\prime}}^+_q$ is then computed by only integrating samples inside the threshold. The combined contribution (bottom left) follows the reference up to $F = 3$, beyond which a discrepancy emerges. Examination of the fractional contributions by quadrant (bottom right) confirms that this divergence arises primarily from the under-prediction of extreme-amplitude sweep and ejection events in Q4 and Q2, respectively. The model thus reproduces the bulk of the Reynolds stress budget with high fidelity, whilst the most intermittent, large-amplitude events, which correspond to the tail of the $u^{\prime +} v^{\prime +}$ distribution, are attenuated in the reduced-order representation.

\begin{figure}[htbp]
	\centering
	\includegraphics[width=0.5\linewidth]{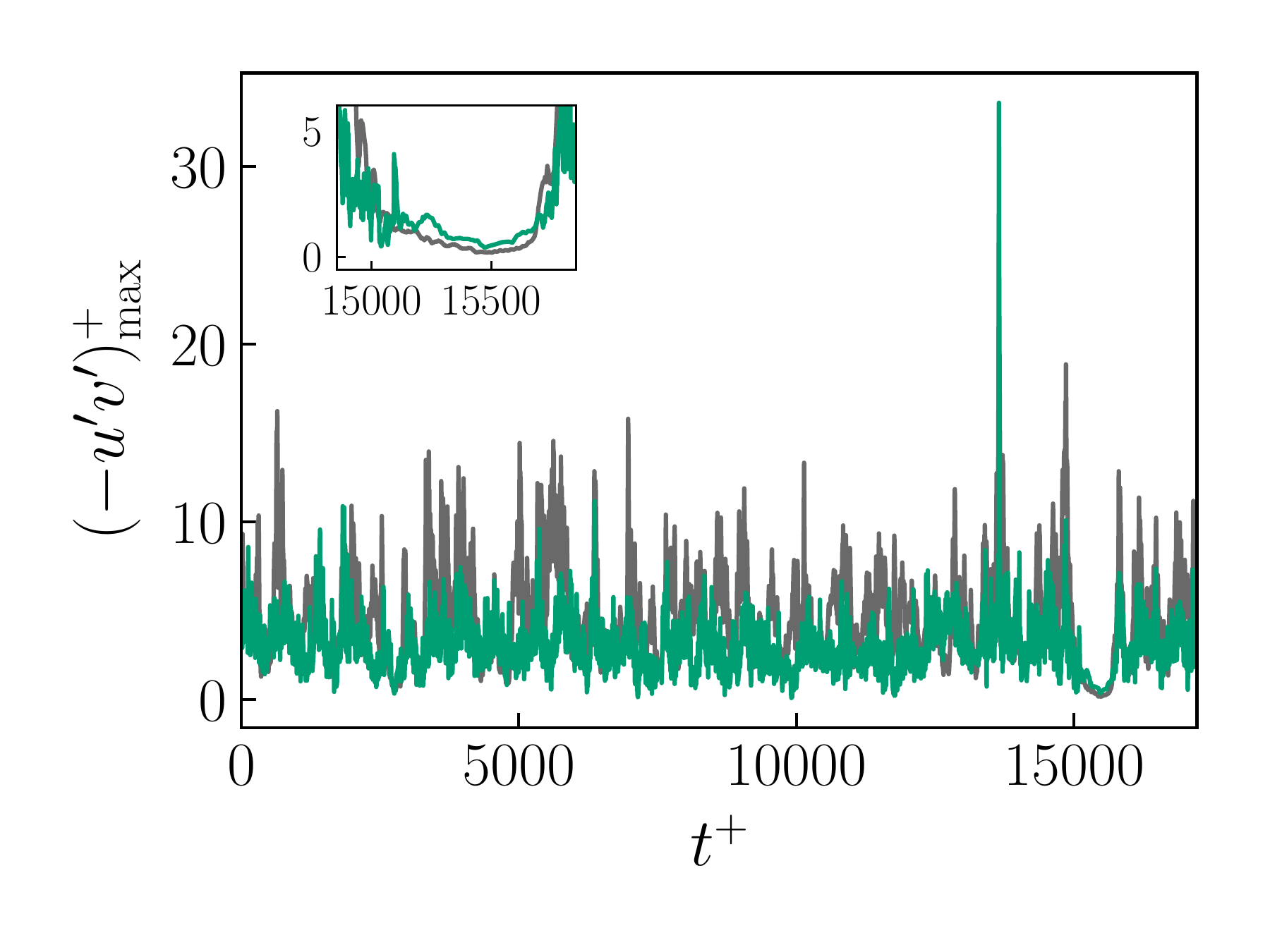}
	\caption{Temporal evolution of the spatial maximum of the instantaneous Reynolds shear stress, $(-u^\prime v^\prime)^+_{\max}$, for the DNS reference data (gray) and the end-to-end framework prediction (sea-green).}
	\label{fig:04.3_max_uv_comparison}
\end{figure}

Whilst the quadrant analysis demonstrates that the statistical extremes of the Reynolds stress are attenuated, a more fundamental question concerns whether the model preserves the temporal organisation of the flow, in particular the intermittent regeneration cycle whose timescale was identified in \S\ref{sec:04.2_latent_prediction} as $T^+ \approx 1724$ and shown to be directly encoded in the latent dimensions $\zeta_1$ and $\zeta_3$. Figure~\ref{fig:04.3_max_uv_comparison} conveys the temporal evolution of $(-u^\prime v^\prime)^+_{\max}$, the spatial maximum of the instantaneous Reynolds shear stress, for both the predicted and reference DNS fields. The absolute peak amplitudes of the most extreme bursting events are, as anticipated from the quadrant analysis, truncated by the model. However, the alternating sequence of active and quiescent phases is reproduced with satisfactory fidelity over the full extent of the forecasting horizon. This observation is consistent with the spectral analysis of \S\ref{sec:04.2_latent_prediction}: the model reproduces the correct cyclical pacing of the regeneration cycle. Of particular interest is the extended quiescent period around $15{,}000\,t^+$. Both the DNS reference and the model prediction drop to near-zero values during this interval, confirming that the latent variables capture the quiescent state of the flow. The predicted signal exhibits a slightly earlier recovery than the DNS reference. This minor temporal offset reflects the difficulty of the autoregressive model in precisely locating the transition from quiescent to active phase, consistent with the limited number of such events in the training dataset. Importantly, this quiescence was also directly observable in the latent space, as a distinct and concurrent attenuation of $\zeta_1$ and $\zeta_3$ towards zero (cf. \S\ref{sec:04.2_latent_prediction}), reinforcing the conclusion that the learnt manifold encodes dynamical phases of the flow.

\section{Conclusion}
\label{sec:05_conclusion}
This study introduces a reduced-order modelling framework for constructing surrogate models of wall-bounded turbulent flows. The approach is equation-free and data-driven, combining two machine-learning components: a $\beta$-VAE-GAN for non-linear spatial dimensionality reduction and an Easy Attention Transformer, conditioned on sparse sensor signals via an adapted AdaLN-Zero modulation mechanism, for modelling the resulting low-order dynamics. The framework is assessed on the Minimal Flow Unit \cite{jimenez1991minimal} at $Re_\tau = 200$ and $y^+ = 14$, a domain that isolates the essential near-wall regeneration cycle whilst remaining computationally tractable.

At a latent dimension of $d_\zeta = 4$, the $\beta$-VAE-GAN recovers approximately $87\%$ of the test-set turbulent kinetic energy, compared with $76\%$ for the standard $\beta$-VAE at the same compression level, confirming the contribution of the adversarial training component. The two-point spatial autocorrelations confirm that the integral length scales of all three velocity components are faithfully preserved, and the singular value decay of the POD modes demonstrates that the dominant energetic structures are retained. Importantly, this gain in reconstruction fidelity is achieved without compromising the disentanglement of the latent representation, as evidenced by the low mean absolute off-diagonal Pearson correlation of the latent variables.

The power spectral density analysis reveals that the latent dimensions autonomously decouple the characteristic timescales of the MFU: $\zeta_1$ and $\zeta_3$ encode the slow regeneration-cycle period ($T^+ \approx 1724$), $\zeta_2$ captures the intermediate shear-stress timescale ($T^+ \approx 862$), and $\zeta_4$ reflects the higher-frequency streamwise velocity fluctuations ($T^+ \approx 31$). The one-to-one correspondence between latent spectral peaks and the DNS power spectra of $(-u^\prime v^\prime)^+_{\max}$ and $(u^\prime u^\prime)^+$ establishes that the $\beta$-VAE-GAN produces a physically interpretable latent manifold rather than a purely statistical compression.

The sensor-conditioned Easy Attention Transformer reproduces the latent dynamics over an autoregressive horizon of approximately $17{,}288\,t^+$ from an initialisation window of only ${\approx}\,128\,t^+$. The geometric and statistical properties of the predicted manifold are preserved: the joint probability density functions of the low-frequency dimensions $\zeta_1$ and $\zeta_2$ are tightly concentrated along the main diagonal, indicating minimal dispersion error, whilst the high-frequency dimensions $\zeta_3$ and $\zeta_4$ exhibit greater point-wise scatter, consistent with the harder task of predicting rapid temporal variations over long horizons.

When the two components are coupled in end-to-end inference, the framework recovers approximately $82\%$ of the turbulent kinetic energy, closely approaching the compression upper bound of $87\%$. The two-point spatial autocorrelations and the POD singular value decay confirm the preservation of the dominant spatial structures, with the residual discrepancy attributable to the compression stage rather than to the Transformer. The quadrant analysis demonstrates the model's ability to reproduce the most frequently occurring ejection and sweep events, as evidenced by the correct elongation of the predicted joint PDF along the Q2--Q4 axis. The under-prediction of rare, extreme-amplitude events, observed for hole-size parameter $F > 3$, arises from the information bottleneck: compressing the flow into four latent dimensions retains the most statistically recurrent structures at the expense of infrequent, large-amplitude events that contribute marginally to the training loss. Despite this attenuation, the model successfully reproduces the alternating sequence of active and quiescent phases that characterises the near-wall regeneration cycle, including an extended quiescent interval at $t^+ \approx 15{,}000$, confirming that the learnt manifold encodes the distinct dynamical phases of the flow and not merely its mean spectral content.

The principal limitation of the present framework concerns the attenuation of rare, extreme-amplitude events, a direct consequence of the bottleneck dimensionality. As discussed, this may be partially addressed by increasing the latent dimension $d_\zeta$, at the cost of a higher-dimensional forecasting problem for the Transformer; the optimal trade-off is application-dependent. More broadly, the one-to-one correspondence established here between the latent spectral peaks and the DNS flow frequencies opens questions regarding the precise attribution of each dimension to specific near-wall structures. The present evaluation is restricted to the Minimal Flow Unit at a single Reynolds number; extension to larger domains and higher $Re_\tau$ would be required to assess the scalability of the latent dimensionality with flow complexity.

Notwithstanding these open directions, the framework demonstrates the ability to compress wall-bounded turbulent flow fields into a physically interpretable low-dimensional manifold, to decouple the characteristic timescales of the near-wall regeneration cycle into distinct latent dimensions, and to track the alternating sequence of active and quiescent phases over extended autoregressive horizons. These properties establish its suitability as a surrogate model for wall-bounded turbulence and motivate its extension to more complex flow configurations.



\section*{Acknowledgments}
This work was supported by the French government program "Investissements d'Avenir" (EUR INTREE, Reference No. ANR-18-EURE-0010 and LABEX INTERACTIFS, Reference No. ANR-11-LABX-0017-01). N.T. acknowledges the support of HPC resources provided by GENCI--IDRIS  (Grant No. 20XX-AD011015409). L.A.'s contribution was supported by French National Research Agency (ANR) from the ANR Young Researchers (JCJC) programme under Grant No. ANR-23-CE46-0004.  M.S-A. acknowledges financial support from the EU Doctoral Network MODELAIR. R.V. acknowledges the financial support from ERC Grant No. 2021-CoG-101043998, DEEPCONTROL. Views and opinions expressed are however those of the author(s) only and do not necessarily reflect those of the European Union or the European Research Council. Neither the European Union nor the granting authority can be held responsible for them. 

\bibliographystyle{unsrtnat}
\bibliography{references}  






\end{document}